%% file: main.tex
\documentclass[sigconf, nonacm]{acmart}
\AtBeginDocument{%
  }

\usepackage{algorithm}
\usepackage[noend]{algpseudocode}
\usepackage{circledsteps}
\usepackage{amsmath}
\usepackage{mathtools}
\usepackage{amsthm}
\usepackage{dsfont}
\usepackage{multirow}
\usepackage{euscript}
\usepackage{array}
\usepackage{booktabs}
\usepackage{physics}
\usepackage{pifont}
\usepackage{capt-of}
\usepackage{cleveref}
\usepackage{spverbatim}
\usepackage{enumitem}
\usepackage{mdframed}
\usepackage{makecell}
\usepackage{float}
\usepackage{xparse}
\usepackage{xifthen} 

\Crefname{figure}{Fig.}{Figs.}
\usepackage[cal=boondox,scr=txupr]{mathalpha}

\usepackage{titlesec}

\setcopyright{acmlicensed}
\copyrightyear{2018}
\acmYear{2018}
\acmDOI{XXXXXXX.XXXXXXX}
\acmConference[Conference acronym 'XX]{Make sure to enter the correct
  conference title from your rights confirmation email}{June 03--05,
  2018}{Woodstock, NY}
\acmISBN{978-1-4503-XXXX-X/2018/06}

\begin{document}

\settopmatter{printfolios=true}  
\pagestyle{plain}     
\title{Cut Costs, Not Accuracy: \\ LLM-Powered Data Processing with Guarantees}

\newcommand{\name}{BARGAIN}

\newcommand{\techreport}[1]{#1}
\newcommand{\nontechreport}[1]{}

\newcommand{\rone}[1]{#1}
\newcommand{\rtwo}[1]{#1}
\newcommand{\rthree}[1]{#1}
\newcommand{\rall}[1]{#1}

\newcommand{\sep}[1]{\textcolor{blue}{sep:#1}}
\newcommand{\shreya}[1]{\textcolor{teal}{shreya:#1}}
\renewcommand{\mathfrak}[1]{\mathscr{#1}}
	
\definecolor{bargain}{rgb}{0.5859375, 0.01171875, 0.01171875}
\definecolor{cadmiumgreen}{rgb}{0.0, 0.42, 0.24}
\definecolor{burgundy}{rgb}{0.5, 0.0, 0.13}
\definecolor{pink_figma}{rgb}{0.8828125, 0.4765625, 0.84765625}
\settopmatter{printacmref=false}
\pagestyle{plain}

\techreport{
\author{Sepanta Zeighami}
\affiliation{%
  \institution{UC Berkeley}
}
\email{zeighami@berkeley.edu}

\author{Shreya Shankar}
\affiliation{%
  \institution{UC Berkeley}
}
\email{shreyashankar@berkeley.edu}

\author{Aditya Parameswaran}
\affiliation{%
  \institution{UC Berkeley}
}
\email{adityagp@berkeley.edu}
}

\setlength{\abovedisplayskip}{2.5pt}
\setlength{\belowdisplayskip}{2.5pt}
\setlength{\abovedisplayshortskip}{2.5pt}
\setlength{\belowdisplayshortskip}{2.5pt}

\setlength{\textfloatsep}{2pt}
\setlength{\floatsep}{2pt}
\setlength{\intextsep}{2pt}
\setlength{\dbltextfloatsep}{2pt}
\setlength{\dblfloatsep}{2pt}
\setlength{\abovecaptionskip}{2pt}
\setlength{\belowcaptionskip}{2pt}
\titlespacing*{\section}{0pt}{1ex}{0.7ex}
\titlespacing*{\subsection}{0pt}{0.9ex}{0.7ex}
\titlespacing*{\subsubsection}{0pt}{0.8ex}{0.7ex}


\NewDocumentEnvironment{revcomment}{m o}
{
  \par\vspace{0.5em}
  \IfValueT{#2}{%
    \IfEqCase{#2}{%
      {true}{\noindent\rule{\linewidth}{0.4pt}}
    }[\PackageWarning{revcomment}{Unknown flag `#2', no rule added.}]
  }
  \noindent\textbf{#1. }\itshape
}
{
  \par
}

\begin{abstract}
\vspace{-0.1cm}
Large Language Models (LLMs) are being increasingly used as a building block in data systems to process large text datasets. To do so, LLM model providers offer multiple LLMs with different sizes, spanning various cost-quality trade-offs when processing text at scale. Top-of-the-line LLMs (e.g., GPT-4o, Claude Sonnet) operate with high accuracy but are prohibitively expensive when processing many records. To avoid high costs, more affordable but lower quality LLMs (e.g., GPT-4o-mini, Claude Haiku) can be used to process records, but we need to ensure that the overall accuracy does not deviate substantially from that of the top-of-the-line LLMs. The model cascade framework provides a blueprint to manage this trade-off, by using the confidence of LLMs in their output (e.g., log-probabilities) to decide on which records to use the affordable LLM. However, existing solutions following this framework provide only marginal cost savings and weak theoretical guarantees because of poor estimation of the quality of the affordable LLM's outputs. We present \name{}, a method that judiciously uses affordable LLMs in data processing to significantly reduce cost while providing strong theoretical guarantees on the solution quality.  \name{} employs a novel adaptive sampling strategy and statistical estimation procedure that uses data and task characteristics and builds on recent statistical tools to make accurate estimations with tight theoretical guarantees. Variants of \name{} can support guarantees on accuracy, precision, or recall of the output. Experimental results across 8 real-world datasets show that \name{} reduces cost, on average, by up to 86\% more than state-of-the-art, while providing stronger theoretical guarantees on accuracy of output, with similar gains when guaranteeing a desired level of precision or recall.   
\end{abstract}

\maketitle

\input{common_pars}

\input{intro}

\input{background}
\input{newer_precision}

\input{extension}
\input{discussion}

\input{exp}

\input{related_work}

\input{conclusion}
\begin{acks}
 We acknowledge support from grants DGE-2243822, IIS-2129008, IIS-1940759, and IIS-1940757 awarded by the National Science Foundation, funds from the State of California, an NDSEG Fellowship, funds from the Alfred P. Sloan Foundation, as well as EPIC lab sponsors: Adobe, Google, G-Research, Microsoft, PromptQL, Sigma Computing, and Snowflake. Compute credits were provided by Azure, Modal, NSF (via NAIRR), and OpenAI.
\end{acks}

\nontechreport{\clearpage}

\bibliographystyle{ACM-Reference-Format}
\bibliography{references}

\nontechreport{\clearpage}
\appendix

\input{appendix/overviews}
\input{appendix/appendix_overview}
\input{appendix/appendix_additional_results}
\input{appendix/appendix_proofs}
\input{appendix/multi_proxy}
\input{appendix/candidate_set}

\input{appendix/taskprompts}
\input{appendix/additional_exp}

\end{document}

%% file: common_pars.tex
\newcommand{\fighyperparam}{fig:hyperparam_AT_rebuttal}
\newcommand{\tabchernoff}{tab:hoef_vs_chernoff_rebuttal}
\newcommand{\chernoffall}{tab:avg_results_rebuttal}
\newcommand{\variance}{tab:avg_variance_rebuttal}
\newcommand{\naivevariant}{tab:naive_vs_naive_witheq9_rebuttal}

\newcommand{\chernoff}{We note that it is possible to use Chernoff's inequality to define the estimation function instead of Hoeffding's. In Appx.~\ref{appx:chernoff} we discuss how the bound can be applied and present results comparing the use of Chernoff's and Hoeffding's inequality. We saw marginal differences between the two and thus present only Hoeffding's inequality here because it is simpler to apply.}

\newcommand{\chernoffappx}{In our naive approach, we can use Chernoff's bound instead of Hoeffding's inequality. This impacts our estimation function, modifying Eq.~\ref{eq:naive_est}. Applying Chernoff's bound~\cite{mitzenmacher2017probability}, we obtain:

\begin{align*}
\hspace*{-10pt}
\mathscr{E}^{\text{Chernoff}}(S, T, \rho, \alpha)=\mathds{I}\big[\mathfrak{P}_S(\rho)\geq T+\sqrt{\frac{2(1-T)\log(1/\alpha)}{|S^\rho|}}\big]. 
\end{align*}
Appx.~\ref{appx:chernoff_prof} shows the derivation of $\mathscr{E}^{\text{Chernoff}}$ form Chernoff's bound~\cite{mitzenmacher2017probability} (which is less straightforward than using Hoeffding's) and its statistical guarantees. 

We next empirically compare this estimator with $\mathscr{E}^{\text{naive}}$ derived from Hoeffding's inequality. To use $\mathscr{E}^{\text{Chernoff}}$, we follow the same naive approach as before (see Eq.~\ref{eq:rho_S_naive}) but replace $\mathscr{E}^{\text{naive}}$ with $\mathscr{E}^{\text{Chernoff}}$. In the following we use Chernoff to refer to the naive method using $\mathscr{E}^{\text{Chernoff}}$ and Hoeffding to the navie methd using $\mathscr{E}^{\text{naive}}$ as presented in Sec.~\ref{sec:pt:naive}. We present the result of the experiments across different quality metrics averaged across all datasets from Sec.~\ref{sec:exp}, for targets $T=0.7$ and $0.9$ in Table~\ref{\tabchernoff}. (Recall that for AT, the utility is cost saved, for PT it is the recall and for RT it is the  precision, similar to Table~\ref{tab:all_res}.) We see that when quality target is 0.9, Chernoff outperforms Hoeffding, but not at target 0.7. Indeed, comparing $\mathscr{E}^{\text{Chernoff}}$ with $\mathscr{E}^{\text{naive}}$ from Eq.~\ref{eq:naive_est}, we observe that $\mathscr{E}^{\text{Chernoff}}$ provides a tighter bound when $T>\frac{3}{4}$ while Eq.~\ref{eq:naive_est} is tighter when $T<\frac{3}{4}$, meaning Chernoff's bound is sharper only when true mean of observations is close to 1, which implies that its application is only beneficial when $T$ is close to 1, whereas Hoeffding's inequality is sharper when $T$ is smaller. Nonetheless, even when $T$ is large (and Chernoff's bound is sharper), \name{} significantly outperforms both bounds as we see in \Cref{\chernoffall} which presents utility of different methods averaged across all datasets from Sec.~\ref{sec:exp} at $T=0.9$. Superior performance of \name{} is not only due to our sharper bounds using \citet{waudby2024estimating}, but also because of our adaptive sampling and selection methods. We note that Sec. 2.3 of \citet{waudby2024estimating} discusses why their proposed estimation method is sharper than Chernoff/Hoeffding bounds, and we provide an informal discussion on why it is so in our Sec.~\ref{sec:stat_tools}.
}

\newcommand{\proofoverview}{Alg.~\ref{alg:iterative_sample_precision} fails to meet the target if $\mathscr{E}^{\text{\name{}}}$ wrongly estimates a threshold meets the target at any iterations of the algorithm. 
To bound the probability of this event, we ensure that (1) the estimates for any threshold $\rho$ can be wrong with a bounded probability and use this result to show (2) the total probability of making a wrong estimate across all thresholds is bounded by $\delta$. To show (1), observe that every estimate by $\mathscr{E}^{\text{\name{}}}$ for a threshold $\rho$ uses samples taken uniformly from $D^{\rho}$. Thus, the probability that a single estimate is incorrect is bounded by Lemma~\ref{lemma:lb}. Furthermore, the repeated estimation while sampling for the same threshold is accounted for through \cite{waudby2024estimating} which shows that the estimation procedure in Lemma~\ref{lemma:lb} is \textit{anytime valid}, that is, the estimation can be performed repeatedly during sampling, while still providing the same bound on the overall probability of making a wrong estimate (see Lemma~\ref{lemma:_anytime_replace_original}). Then, to show (2), we use Lemma~\ref{lemma:select_eta} which uses the union bound to account for the total probability of making a wrong estimate across all thresholds.}

\newcommand{\naivemain}{We additionally performed experiments using an alternative naive variant that uses Eq.~\ref{eq:select_eta} (with $\eta=0$). This variant performed similarly to the one in Sec.~\ref{sec:pt:naive}; results are presented in Appx.~\ref{exp:additional_naive}.}

\newcommand{\naiveappx}{Here, we consider the alternative naive approach of using Eq.~\ref{eq:select_eta} with $\eta$=0 for threshold selection, instead of Eq.~\ref{eq:naive_sel}. We implemented this approach and observed that it performs almost identically to using Eq.~\ref{eq:naive_sel}. We show the results in \Cref{\naivevariant} which contains the recall for PT queries with $T=0.9$ averaged across all datasets presented in Table~\ref{tab:all_res}. As Fig.~\ref{fig:prec_candidate} shows, precision is, in practice, monotonic in proxy scores, so Eq.~\ref{eq:select_eta} and Eq.~\ref{eq:naive_sel} often end up choosing the same threshold. Overall, \Cref{\naivevariant} shows that modifying the selection procedure alone does not improve utility; the sampling and estimation also need to be improved, as done by \name{}.}

\newcommand{\naivevariance}{Here, we present the standard deviation of the utility of the methods across 50 different runs (runs whose average was presented in Table~\ref{tab:all_res}). Table~\ref{\variance} shows the results, where we compute the standard deviation of the utility of each method for each dataset (i.e., standard deviation corresponding to values in Table~\ref{tab:all_res}), and present the average across datasets in the table. Table~\ref{\variance} shows that \name{} has lower standard deviation than SUPG. To put this in the context of our broader experimental results, we note that Table~\ref{tab:all_res} showed that \name{} often provides an order of magnitude better utility than SUPG; \Cref{\variance} shows that it provides such high utility \textit{with lower variance} than SUPG. }

\newcommand{\calibration}{Note that the use of proxy scores to decide whether to use the proxy or not is beneficial only if the model is well-calibrated \cite{wang2023calibration}, i.e., there is positive correlation between the value of proxy score and probability of correctness of the proxy. Although our guarantees on output quality always hold even when the models are not calibrated, in such cases, the cascade framework is unlikely to provide high utility, e.g., for PT queries, one obtains low recall even though the precision is guaranteed to meet the precision target.}

\newcommand{\paramsettingmain}{Here, we discuss the role of system parameters in \name{}.

\textbf{Number of Candidate Thresholds, $M$}. $M$ controls the number of candidate thresholds \name{}$_A$ and \name{}$_P$ consider. Increasing $M$ (i.e., having more candidate thresholds) increases the chance of finding a threshold with high utility; however, considering many thresholds requires more samples and could be wasteful if most do not improve utility. To understand the trade-offs, note that there are most $\frac{n}{M}$ records between any two consecutive candidate thresholds in $\mathscr{C}_M$ (by definition, see Eq.~\ref{eq:c_M_def}), and that \name{} returns the $i$-th threshold $\rho_i\in \mathscr{C}_M$ when the $i+1$-th threshold, $\rho_{i+1}$ is estimated to not meet the target. If the precision or accuracy of the proxy monotonically increases in the proxy scores (which, according to Fig.~\ref{fig:prec_candidate}, typically holds in practice), then increasing $M$ can lead to choosing a threshold between $\rho_i$ and $\rho_{i+1}$, but not smaller. This means that, for AT queries (similar argument holds for PT queries, see Appx.~\ref{sec:candidate_set}), the fraction of records processed by the proxy can improve by $\frac{1}{M}$ at most, showing diminishing returns as $M$ increases (e.g., any $M>20$ leads to at most $5\%$ increase in utility over $M=20$). So large values of $M$ are unlikely to provide significantly better utility and the cost incurred by sampling more records to evaluate many thresholds when $M$ is large may offset any such gains. We show this empirically in Appx.~\ref{appx:hyperparams} and recommend $M=20$ as the default value. 

\textbf{Minimum Number of Samples per Threshold, $c$}. The parameter $c$ controls the minimum number of samples taken by \name{}$_A$ variants at a threshold before it decides a threshold does not meet the target. If $c$ is too small, the algorithm might prematurely decide a threshold does not meet the target (before having observed sufficient samples), while if $c$ is too large it might spend too many samples at a threshold that does not meet the target, thus wasting samples. Setting $c$ as a small constant fraction of data size ensures that not too many samples are wasted during estimation relative to total processing cost. This is supported by our experiments (see Appx.~\ref{appx:hyperparams}) that show setting $c$ to 1\% to 5\% of data size performs well across datasets, and in general, utility of \name{} is not very sensitive to $c$ as long as it is not very large compared with data size.

\textbf{Tolerance Parameter, $\eta$}. As \name{}$_A$ and \name{}$_P$ variants iterate through candidate thresholds, they terminate after they consider $\eta$ thresholds that do not meet the target (see Eq.~\ref{eq:select_eta}). $\eta$ is a tolerance parameter that can be set based on whether the quality metric is expected to be monotonic or not. That is, whether, after estimating that a threshold $\rho$ does not meet the target, thresholds smaller than $\rho$ are expected to also not meet the target. In real-world datasets, and as discussed in Sec.~\ref{sec:prism_pu:thresh}, this monotonicity property is expected to hold, and as such we set $\eta=0$ by default. Experiments in Appx.~\ref{appx:hyperparams} validate this.}

\newcommand{\systemparamsexp}{Here, we study the impact of hyperparameters, $M, c, \eta$ on \name{}$_A$-A (we observe similar results on other variants) across various datasets. The results are presented in Figure~\ref{\fighyperparam}.

\textbf{Varying $M$}. Results for varying $M$ are shown in Fig.~\ref{\fighyperparam} (a). The results show that very small values of $M$ lead to low utility, while after $M$ reaches a value of around $20$, the utility stabilizes and increasing $M$ further leads to similar utilities. This is in line with the discussion in Sec.~\ref{sec:hyperparam} which shows diminishing returns for increasing $M$ as $M$ gets larger. Low utility for small $M$ is due to the approach not trying enough thresholds, e.g., when $M=1$, \name{} considers only a single threshold, which in many datasets may not meet the accuracy target and thus \name{} has to use the oracle to process all the records. 

\textbf{Varying $c$}. The results for varying $c$ is presented in Fig.~\ref{\fighyperparam} (b). As discussed in Sec.~\ref{sec:hyperparam}, we see that utility improves initially when $c$ increases, by allocating more samples to each threshold to ensure accurate estimates. However, the trend reverses as $c$ increases further, since spending too many samples on thresholds that does not meet the target causes wasting samples. Nonetheless, Fig.~\ref{\fighyperparam} (b) shows that a large range of values of $c$ set as a small fraction of data size provide good utility across datasets.

\textbf{Varying $\eta$}. We next study the impact of $\eta$ from Eq.~\ref{eq:select_eta}. Note that Alg.~\ref{alg:iterative_sample_accurcay} presented for \name{}$_A$-A only considered $\eta=0$ for ease of presentation (since, as we sill, $\eta=0$ is the best performing variant). We extend the algorithm to support $\eta>0$ in Alg.~\ref{alg:iterative_sample_accurcay_w_etha}, where, based on Eq.~\ref{eq:select_eta}, we modify it to consider $\eta$ thresholds that miss the target before returning the final threshold. The results of running this algorithm with various $\eta$ values is presented in Fig.~\ref{\fighyperparam} (c). As the figure shows, the utility decreases with $\eta$, showing that $\eta=0$ provides the highest utility across datasets. As discussed in Sec.~\ref{sec:hyperparam}, if the accuracy of the proxy monotonically decreases as proxy scores deceases, the thresholds smaller than the first threshold that misses the target will also miss the target, so that there is no benefit having $\eta>0$. Meanwhile, $\eta>0$ means each application of the estimation function needs to be more conservative (see Lemma~\ref{lemma:select_eta}), thus providing worse utility.}

\newcommand{\stattools}{To solve the cascade threshold problem through sampling, we need to estimate, based on observed samples, whether a specific threshold meets the quality target. This estimation can be done using classic concentration bounds such as Hoeffding's or Chernoff's inequality. We instead use recent results by \citet{waudby2024estimating} that provide tighter bounds (as discussed in \cite{waudby2024estimating} and empirically validated in our results). Here, we provide an informal overview of the result by \cite{waudby2024estimating} used in our paper. Formal statement of results are presented in Appx.~\ref{sec:statements}.

For a set of i.i.d random variables, $X=\{X_1, ..,. X_k\}$ whose true mean is $\mu$, we would like to estimate, using $X$, whether the true mean is more than a threshold $m$ or not. Theorem 3 of \cite{waudby2024estimating} provides a hypothesis testing approach for this estimate. It specifies a boolean function $\mathscr{T}(m, X, \alpha)$, which, with high probability, is 1 when $\mu$ is at least $m$ and  0 otherwise. \cite{waudby2024estimating} shows that whenever $\mu<m$,  $\mathds{P}\big(\mathscr{T}(m, X, \alpha)=1\big)\leq \alpha$. That is, $\mathscr{T}$ is unlikely to wrongly estimate the mean is more than $m$ when it is not.  
\begin{lemma}[Informal and Simplified Statement of Theorem 3 by \cite{waudby2024estimating}]\label{lemma:hypothesis_test}
    Consider the set of i.i.d random variables $X$ with mean $\mu$. 
    For a confidence parameter $\alpha\in[0, 1]$, and any $\mu<m$, we have
    \begin{align}\label{eq:betting_lb1}
        \mathds{P}\big(\mathscr{T}(m, X, \alpha)=1\big)\leq \alpha,\,\text{where}
    \end{align}    
    \begin{align}\label{eq:betting_lb2}
        \mathscr{T}(m, X, \alpha)\approx\mathds{I}\big[\mathscr{K}(m, X)\geq\frac{1}{\alpha}\big],
    \end{align}    
    \begin{align}\label{eq:k_new}
        \mathscr{K}(m, X)\approx\Pi_{i=1}^{k}\Big(1+\frac{(X_i-m)}{\hat{\sigma}_{i-1}}\sqrt{\log(1/\alpha)}\Big),
    \end{align}
    \begin{align*}
        \hat{\sigma}_i^2=\frac{1/4+\sum_{j=1}^i(X_j-\hat{\mu}_j)^2}{i+1}, \;\hat{\mu}_i = \frac{1/2+\sum_{j=1}^iX_j}{i+1}.
    \end{align*} 
\end{lemma}
Note that $\approx$ means \textit{we have dropped some of the terms from the definition} to convey high-level intuition (see Appx.~\ref{sec:statements} for exact formulas). Above, $\hat{\sigma}_i^2$ (resp., $\hat{\mu}_i$) is a term analogous to empirical variance (resp., mean). $\mathscr{K}(m, X)$, informally, quantifies whether the sequence of $X_i$ consistently exceeds $m$, when normalized by the variance $\hat{\sigma}_i^2$; larger values show observations exceed $m$ more frequently (thus suggesting $\mu>m$). The use of empirical variance $\hat{\sigma}_i^2$ contrasts with Hoeffding's inequality that only relies on the empirical mean. Taking standard deviation into account significantly improves the bounds when standard deviation is small, as our experiments show (see Sec.~\ref{sec:prec:unif} for comparison).}

\newcommand{\overview}{Our main proposed \name{} variants for each of the queries are shown in Table~\ref{tab:prism_variants}. The variants in the table perform adaptive sampling; for ease of exposition and comparison purposes, we also present other variants (not listed in the table) that perform uniform sampling. All variants take the quality target, $T$, and confidence parameter, $\delta$, as input, while \name{}$_P$-A and \name{}$_R$-A additionally require the oracle budget, $k$, upfront. All methods additionally use other parameters which need not be set by the users and modifying their default values have limited impact on utility. We discuss how these parameters are set in Sec.~\ref{sec:hyperparam}. 

\textbf{Use-cases of Variants.} \name{}$_A$ is suitable for multi-class classification, as well as binary classification without a maximum oracle budget constraint but when the goal is to minimize the total budget used. \name{}$_P$ and \name{}$_R$ on the other hand respect a maximum oracle budget and thus are useful when user wants to stay within a budgetary constraint. Moreover,  \name{}$_P$ and \name{}$_R$ are useful for binary classification tasks with class imbalance, e.g., when there is expected to be few records in the positive class, while \name{}$_A$ is additionally applicable when classes are balanced. Finally, \name{}$_A$-A and \name{}$_A$-M differ in that \name{}$_A$-A chooses a single cascade threshold for {\em all} output classes while \name{}$_A$-M chooses a cascade threshold {\em per class}. As such, \name{}$_A$-M is beneficial if the proxy is differently calibrated for different output classes, which may occur if it is more difficult to correctly estimate one class but not another.}

%% file: intro.tex
\vspace{-0.25cm}
\section{Introduction}
\vspace{-0.1cm}
LLMs are being increasingly used as a building block in data systems that process large text datasets for tasks such as extraction, filtering, summarization, and question answering \cite{shankar2024docetl, patel2024lotus, liu2024declarative, anderson2024design, urban2024eleet, arora2023language, narayan2022can}. For example, given a set of legal contracts, a lawyer might want to find those that relate to a specific law. To do so, a system can iterate over the contracts and, for each contract, ask an LLM to decide if it relates to the specific law. To obtain the most accurate results possible, users want to use \textit{top-of-the-line LLMs} (e.g., GPT-4o, Claude Sonnet). However, such LLMs are prohibitively expensive at scale. Even a single scan of a few thousand-page long documents by an LLM can cost hundreds of dollars---and gets more expensive with more documents or when multiple scans are needed to address complex or iterative information needs. To support users with budgetary constraints, LLM companies often also provide \textit{affordable LLMs} (e.g., GPT-4o-mini, Claude Haiku) that are much cheaper but can be less accurate; for instance, GPT-4o-mini is more than 15$\times$ cheaper than GPT-4o \cite{openaipricing}. 
However, using the affordable model can reduce answer quality relative to the top-of-the-line model. In such cases, users can often tolerate some marginal quality degradation as long as we can reduce cost substantially, e.g., if the system is guaranteed to match the top-of-the-line LLM's output 90\% of the time but at half the cost. The system then needs to decide when to use which LLM to minimize cost while guaranteeing this desired answer quality.

\begin{figure}
\begin{minipage}{0.34\textwidth}
    \centering
    \includegraphics[width=1\linewidth]{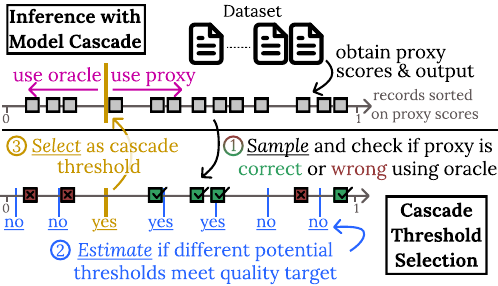}
    \caption{\rthree{Overview of Model Cascade}}
    \label{fig:cascade_overview}
\end{minipage}
\hfill
\end{figure}

A common paradigm to decide when to use the affordable model and when the more expensive model is \textit{model cascades}~\cite{kang2017noscope, anderson2019physical, kang2022tasti, russo2023accelerating, Kang2021AcceleratingAA}. In this paradigm, the more expensive LLM is called the \textit{oracle} while the affordable LLM is called the \textit{proxy}. 
In model cascades, we additionally have access to  \textit{proxy scores} 
which quantify, for each data record, how confident the proxy is in its output\footnote{Proxy scores are provided as part of prediction, e.g., log-probability of LLM outputs }. Proxy scores help decide whether to use the proxy or the oracle to process a record based on a \textit{cascade threshold}, $\rho$ (see Fig.~\ref{fig:cascade_overview}, top): the proxy model answer is used for records whose proxy score is more than $\rho$, and the remaining records are processed with the oracle. The smaller the cascade threshold is, the more frequently the proxy is used to process the records. This means cheaper data processing at potentially worse quality, leading to a classic cost/quality trade off. 

A central problem in model cascades is setting the cascade threshold based on users' desired answer quality. In the general case, users provide an \textit{accuracy target}, e.g., outputs should match the oracle 90\% of the time, while minimizing the number of oracle invocations. We refer to this problem as \textit{accuracy target (or AT) queries}, where the number of oracle invocations avoided is the \textit{utility} of the approach. Alternatively, in filtering tasks, users may be interested in a desired \textit{recall target} while maximizing precision, or a \textit{precision target} while maximizing recall, both settings introduced by \citet{kang2020approximate}, and respectively referred to as RT and PT Queries. We refer to the achieved precision for RT queries and recall for PT queries as the \text{utility} for these two queries. Moreover, we collectively refer to the user-specified accuracy, precision, and recall targets in AT, PT, and RT queries as \textit{quality targets}. Similar to \cite{kang2020approximate}, the user provides a failure probability, $\delta$, and algorithms must meet quality targets with probability at least $1-\delta$ while maximizing utility. \rtwo{We focus on using LLMs for classification, specifically multiclass classification for AT queries and binary classification for PT and RT queries.}

To solve any of AT, PT and RT queries, a common approach \cite{kang2020approximate, kang2017noscope, lu2018accelerating, patel2024lotus, kang13blazeit} is to (1) \textit{sample} and label a subset of records using the oracle, (2) \textit{estimate} if various cascade thresholds meet the desired quality target and (3) \textit{select} a cascade threshold among the ones estimated to meet the quality target, as illustrated in Fig.~\ref{fig:cascade_overview} (bottom). However, most existing methods \cite{kang2017noscope, lu2018accelerating, kang13blazeit} do not provide \textit{any} guarantees on meeting the quality target. Such methods miss the quality target frequently; e.g., \cite{kang2017noscope, lu2018accelerating} frequently achieve precision below 65\%  when given a target of 90\% \cite{kang2020approximate}. To provide some guarantees, SUPG \cite{kang2020approximate} (also used by \cite{patel2024lotus}), employs Central Limit Theorem (CLT) to estimate if different thresholds meet the quality target from labeled samples. 
Due to the use of CLT, SUPG meets the quality target \textit{only asymptotically as sample size goes to infinity}. Since the cascade threshold is usually set with small sample sizes, there are cases where SUPG frequently misses the quality target, like other work that don't claim to provide guarantees.
Besides weak guarantees, SUPG relies on worst-case analysis that ignores data characteristics. This leads to
\textit{inaccurate threshold estimates} that may incorrectly exclude high-utility thresholds that meet the target yielding poor utility. This problem is compounded by SUPG's use of importance sampling (sampling based on proxy scores) that ignores the quality target and label distribution. 
Ignoring such characteristics leads to \textit{ineffective samples} that don't provide useful information for identifying high-utility and high-quality thresholds.

\begin{figure}
\setlength{\tabcolsep}{0pt}
\begin{minipage}{0.17\textwidth}
    \centering
    \includegraphics[width=1\linewidth]{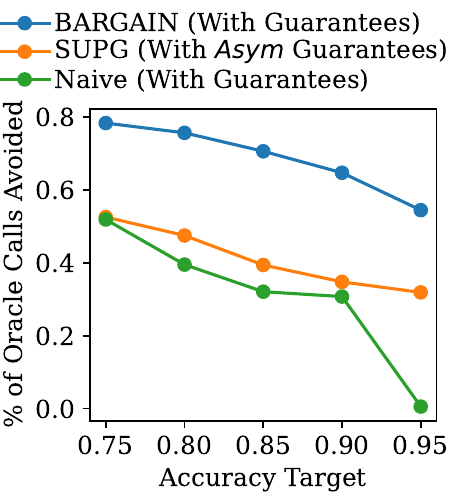}
    \caption{\rthree{Summary of AT Query Results}}
    \label{fig:avg_results}
\end{minipage}
\hfill
\begin{minipage}{0.29\textwidth}
    \centering
    \includegraphics[width=1\linewidth]{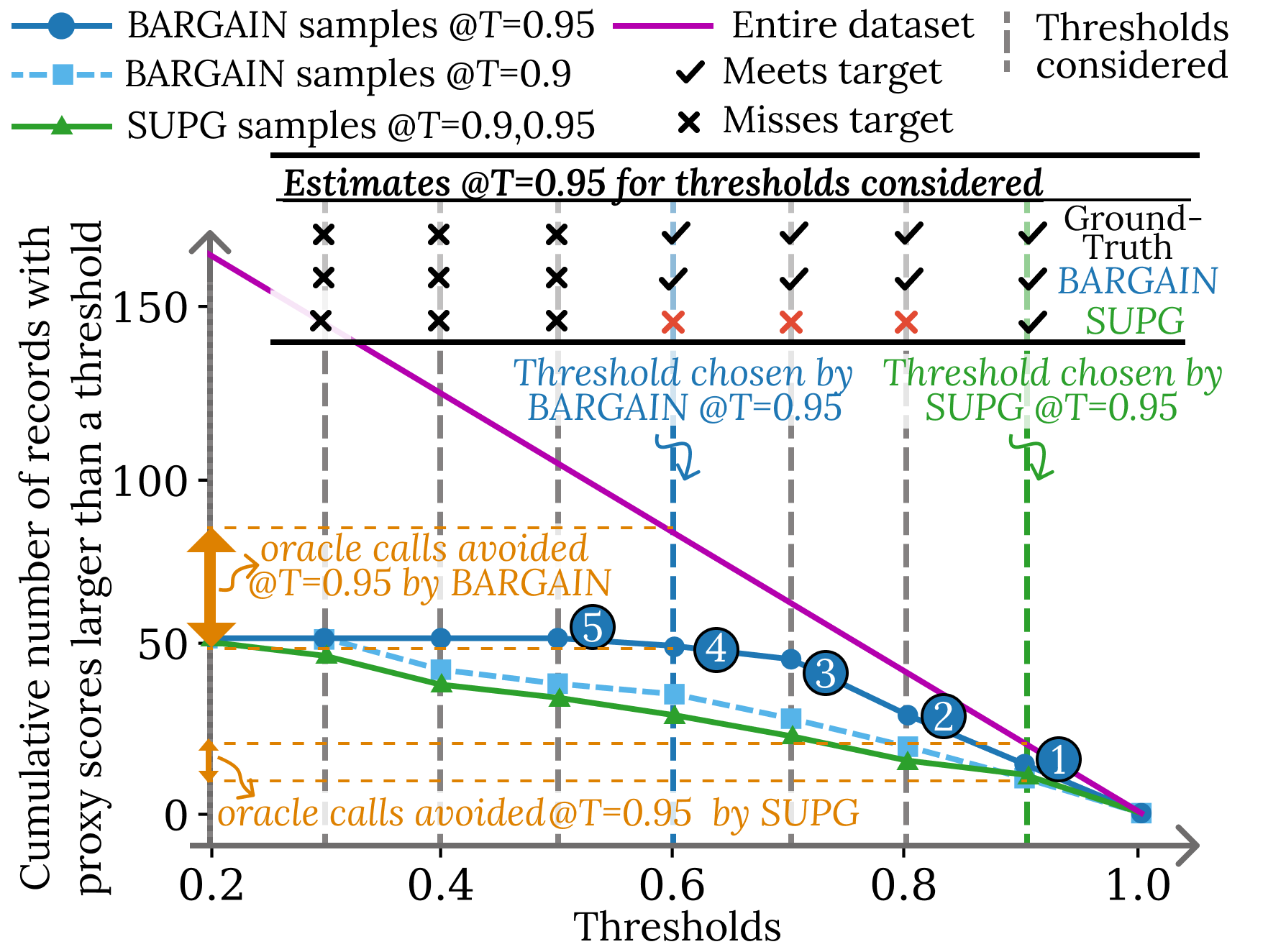}
    \caption{\rthree{Example \name{} workflow for AT queries}}
    \label{fig:ex_prism}
\end{minipage}
\end{figure}

In this paper, we present \textit{BARG\textcolor{bargain}{AI}N}, an approach for \textit{\textcolor{bargain}{AI}-powered} \textit{data} \textit{processing using model cascade with tight theoretical guarantees}. 
We present novel theoretical and algorithmic insights that use \textit{data and task characteristics} to \textit{sample records effectively} and make \textit{accurate threshold estimations}, thus improving utility while providing rigorous theoretical guarantees. \techreport{The benefits of \name{} are summarized in Table~\ref{tab:approaches}, which also shows a Naive approach that achieves the same strong theoretical guarantees as \name{}, but through black-box application of statistical tools (i.e., uniform sampling and Hoeffding's inequality) without considering data and task characteristics. To convey a sense of benefits, Fig.~\ref{fig:avg_results} shows that \name{} provides significantly better utility compared with both the Naive approach and SUPG (results averaged over eight real-world datasets, see Sec.~\ref{sec:exp}). }
\nontechreport{To convey a sense of benefits, Fig.~\ref{fig:avg_results} shows that \name{} provides significantly better utility compared with SUPG (results averaged over eight real-world datasets, see Sec.~\ref{sec:exp}), as well as a Naive approach that achieves the same strong theoretical guarantees as \name{}, but through black-box application of statistical tools (uniform sampling and Hoeffding's inequality) without considering task characteristics.  }
For AT queries, \textbf{\name{} reduces oracle usage by up to 86\% more than SUPG}. Similarly, \textbf{\name{} improves recall in PT queries by up to 118\% and precision in RT queries by up to 19\% over SUPG}. 
\name{} achieves these empirical benefits \textbf{while providing rigorous theoretical guarantees that SUPG lacks}. This rigor is significant since our results show cases where SUPG misses the target more than 75\% of the time when given a failure probability of only 10\%. 



\if 0\name{} views the cascade threshold problem as the problem of searching over a set of possible thresholds to find one with high utility that meets the quality target, where the quality of a threshold needs to be estimated through sampling. To achieve \textit{high sample effectiveness}, \name{} samples records adaptively and as it considers different thresholds. This ensures \name{} can make accurate estimates for every threshold considered.  sample considered  
\fi

\techreport{
\begin{figure}
\setlength{\tabcolsep}{2pt}
    \centering
    \small
    \begin{tabular}{>{\centering\arraybackslash}p{1.3cm} >{\centering\arraybackslash}p{1.4cm} >{\centering\arraybackslash}p{1.7cm} >{\centering\arraybackslash}p{1.7cm}}
    \toprule
        \multirow{2}*{\textbf{Method}} & \textbf{Quality Guarantee} & \textbf{Sample Effectiveness} & \textbf{Threshold Est. Accuracy} \\\midrule
        \textbf{SUPG}  & Asymptotic &  Medium & Medium\\\hline
        \textbf{Naive}  & Yes & Low & Low\\\hline
        \textbf{\name{}} & \textbf{Yes} & \textbf{High} &\textbf{High}\\\bottomrule
    \end{tabular}
    \captionof{table}{Overview of Approaches}
    \label{tab:approaches}    
\end{figure}
}

\textbf{\name{} Overview}. We use Fig.~\ref{fig:ex_prism} to discuss how \name{} achieves these benefits on a toy dataset for AT queries (other queries are similar).  
The blue and green lines in the figure (ignore the large numbered circles, pink line, and orange markings for now) show, for each proxy score, the cumulative number of the samples taken with scores greater than that proxy score by \name{} and SUPG respectively, and for accuracy targets $T=0.9$ and $T=0.95$. Both approaches take a total of 50 samples in all settings (as shown by the y-intercept), and SUPG has the same distribution of samples for both $T=0.9$ and $T=0.95$. The figure also shows the cascade thresholds chosen by the approaches using the samples for $T=0.95$: 0.6 by \name{} and 0.9 by SUPG. The cascade thresholds are chosen based on each approach's estimates, shown in the table embedded in Fig.~\ref{fig:ex_prism}. SUPG wrongly estimates that thresholds between 0.6-0.8 miss the target, but \name{} correctly identifies that those thresholds meet the target. Making correct estimates enables \name{} to find a better cascade threshold and improve utility. 
To see the utility, consider the pink line that shows the cumulative number of records in the entire dataset with scores greater than a given threshold. Note that all samples taken by the approaches are labeled by the oracle. Thus, the \textit{total number of oracle calls avoided} by each approach is the number of records in the entire dataset with a proxy score more than the cascade threshold that are not already sampled. In Fig.\ref{fig:ex_prism}, we visualize this as the difference between the number of cumulative records (pink line) and the cumulative number of samples taken (blue or green line respectively) marked by the orange double arrows. SUPG's cascade threshold results in almost all records being processed by the oracle. \name{} instead avoids a significant number of oracle calls. 


\name{} achieves this, in part, by using adaptive sampling (Step 1 in Fig.~\ref{fig:cascade_overview}) to ensure \textit{high sample effectiveness}. \name{} iterates through different candidate thresholds (in decreasing order) and, for each threshold, samples sufficient records to accurately estimate if it meets the target before moving to the next. \name{} stops sampling once it reaches thresholds estimated to miss the target. This is shown in Fig.~\ref{fig:ex_prism} for $T=0.95$ where circles show the iteration at which a threshold was considered. 
In contrast with SUPG \cite{kang2020approximate} that samples records based on their proxy scores independent of $T$, \name{}'s sampling adapts to the label distribution and quality target. In Fig.~\ref{fig:ex_prism}, for $T=0.95$ (dark blue line), \name{} takes enough samples with proxy score larger than 0.8 to correctly estimate that threshold 0.8 meets the target, while SUPG takes fewer such samples and ends up making an incorrect estimate. Both SUPG and \name{} sample the same number of records in total (50), but SUPG takes many uninformative samples with proxy scores less than 0.6 that aren't useful for estimating the quality of thresholds larger than 0.6. 
Fig.~\ref{fig:ex_prism} also shows samples taken by \name{} when $T=0.9$ (dashed blue line). \name{} adapts to this lower target. 

Furthermore, \name{}'s estimation and selection approach (Steps 2 \& 3 in Fig.~\ref{fig:cascade_overview}) is optimized based on task and data characteristics to achieve \textit{high threshold estimation accuracy} when using the samples. We show that the variance of our samples decreases as the quality of a threshold increases; we take advantage of this low variance to improve our estimation significantly for high-quality thresholds. We use a hypothesis-testing formulation for estimating whether a threshold meets a target and solve it through recent statistical results by \citet{waudby2024estimating} to obtain significantly more accurate estimates than the Naive approach that uses Hoeffding's inequality \cite{hoeffding1994probability} with the same guarantees---unlike SUPG's use of CLT with weaker guarantees. 
Furthermore, we perform an in-depth analysis of statistical events during threshold selection to incorporate useful dataset characteristics into the analysis rather than relying on loose worst-case bounds. This analysis with corresponding modifications to the threshold selection algorithm helps \name{} significantly improve utility on real-world datasets. We apply the above intuitions separately to each of AT, PT, and RT queries. For RT queries, we prove a negative result that when the number of true positives on a dataset is small, no approach can achieve high utility while guaranteeing it meets the target. In light of this result, to achieve high utility, we allow users to opt for a relaxation of our quality guarantee designed to match the original guarantees in realistic settings but not in worst-case datasets.

\textbf{Contributions}. To summarize, our contributions are as follows.
\begin{itemize}
    \item We present \name{}, the first practical method to perform AT, PT and RT queries with rigorous theoretical guarantees.
    \item We show how to perform task and data-aware sampling, estimation, and threshold selection, each designed to take advantage of the characteristics of different quality metrics and the dataset to provide high utility.
    \item We present tight theoretical analyses for our algorithms, presenting novel theoretical insights and showing how recent statistical tools can be used in the analysis. 
    \item We perform extensive empirical evaluation, showing that \name{} reduces oracle usage by up to 86\% for AT queries,  improves recall by 118\% for PT queries, and precision by 19\% for RT queries over the state-of-the-art, SUPG.
\end{itemize}

\techreport{
We present the necessary background in Sec.~\ref{sec:background}, \name{} for PT queries in Sec.~\ref{sec:pt}, and other queries in Sec.~\ref{sec:discussion}, with Sec.~\ref{sec:hyperparam} discussing how to set the parameters of the methods. We present our empirical results in Sec.~\ref{sec:exp}, and present related work in Sec.~\ref{sec:rel_work}. 
}

%% file: background.tex
\section{Background}\label{sec:background}

\subsection{Problem Definition}
\textbf{Setup}. Consider a dataset $D=\{x_1, ..., x_n\}$ containing $n$ records. The user wants to process these records with an expensive AI model, e.g., to apply the same prompt to each document using an LLM such as GPT-4o. We call this expensive AI model an oracle $\mathcal{O}$, and let $\mathcal{O}(x_i)$ denote its output for the $i$-th record, given a fixed prompt. The user also has access to a cheaper AI model, e.g., a smaller LLM such as GPT-4o-mini. We call this model a proxy model $\mathcal{P}$, where $\mathcal{P}(x_i)$ is the output of this model on the $i$-th record, given a fixed prompt (we discuss extensions to multiple proxies in \Cref{sec:other_extesions}). If the user is interested in filtering  (or binary classification of the records in) the dataset, then the outputs of both models are in $\{0, 1\}$; we refer to 0 and 1 as the negative and positive class, respectively. In general, model outputs can be arbitrary. 

For now, we assume the oracle is expensive while the cost of the proxy is negligible compared to the oracle (\Cref{sec:other_extesions} discusses extensions when considering the proxy cost). This often holds in practice where there is more than an order of magnitude cost difference between large and small LLMs (such as for OpenAI and Claude models \cite{openaipricing}). Ideally, the user wants to process the dataset $D$ using the oracle, but performing oracle invocations on the entire dataset is expensive. Instead, the user can tolerate some deviation in output quality relative to the oracle to reduce cost, as long as the output does not deviate \textit{too much} from the oracle. 
We consider accuracy, precision, or recall quality guarantees on the output. 


\textbf{Accuracy Target Queries}. In Accuracy Target (AT) queries, the user specifies a desired accuracy target, and the goal is to minimize the number of times the oracle is used while meeting the desired accuracy target. Formally, consider an algorithm $A$ that processes records in $D$ and provides an answer $\hat{y}_i$ for the $i$-th record $x_i$. $\hat{y}_i$ is either $\mathcal{P}(x_i)$ or $\mathcal{O}(x_i)$, depending on whether $A$ uses the proxy or the oracle on $x_i$. Let $C$ be the number of records where $A$ uses the oracle; $C$ is the \textit{cost} of $A$.  Furthermore, consider the answer set $\hat{Y}=\{\hat{y}_1, ..., \hat{y}_n\}$, and define its accuracy as $\mathfrak{A}(\hat{Y})=\sum_{i\in[n]}\frac{\mathds{I}[\mathcal{O}(x_i)=\hat{y}_i]}{n}$. Then, given an accuracy target $T$, and failure probability $\delta$, an AT query is the problem of returning an answer set $\hat{Y}$ using minimum number of oracle calls, $C$, while guaranteeing the accuracy target is met with probability at least $1-\delta$, that is, $\mathds{P}(\mathfrak{A}(\hat{Y})\geq T)\geq 1-\delta$, where the probability is over runs of the algorithm. 

\textbf{Precision/Recall Target Queries}. Precision Target (PT)  queries and Recall Target (RT) queries were formalized by \cite{kang2020approximate} as follows. In Precision (resp., Recall) Target queries, the user specifies a precision (resp., recall) target and an oracle budget, and our algorithm needs to meet the precision (resp., recall) target and maximize recall (resp., precision). Here, unlike AT queries, the oracle budget is fixed, and the goal is to maximize recall given precision (or vice versa). PT/RT queries only apply to binary classification or filtering where $\mathcal{O}(x_i)\in\{0, 1\}$ for all $i\in[n]$; so our goal is to find the subset of $D$ with positive labels. Formally, given an oracle budget $k$, consider an algorithm $A$ that performs at most $k$ oracle calls and returns a set of data indexes $\hat{Y}=\{i_1, ..., i_r\}$ for some integer $r$, where $i_j\in[n]$. $\hat{Y}$ is the set of indexes of records in $D$ that $A$ estimated to be labeled positive. Precision and recall of this set are defined, respectively, as
\begin{align*}
    \mathfrak{P}(\hat{Y})&=\sum_{i\in \hat{Y}}\frac{\mathcal{O}(x_i)}{r},\; \text{where}\;r=|\hat{Y}|,\,\text{and}\\
    \mathfrak{R}(\hat{Y})&=\sum_{i\in \hat{Y}}\frac{\mathcal{O}(x_i)}{n^+},\; \text{where}\;n^+=\sum_{j\in[n]}\mathcal{O}(x_j).
\end{align*}
Then, given an oracle budget $k$, precision target $T$, and probability of failure $\delta$, a PT query is the problem of returning an answer set $\hat{Y}$ that maximizes recall, $\mathfrak{R}(\hat{Y})$, while using at most $k$ oracle calls and guaranteeing the precision target is met with a probability at least $1-\delta$, that is, $\mathds{P}(\mathfrak{P}(\hat{Y})\geq T)\geq 1-\delta$, where the probability is over runs of the algorithm. RT queries are defined analogously but with precision and recall swapped in the problem definition. 

\textbf{Quality Constraints and Utility.} In the remainder of this paper, we collectively refer to \textit{quality constraints} in AT, PT and RT queries as the constraints on accuracy, precision and recall in the respective problem definition. We use the letter $\mathfrak{F}$ to refer the quality constraints for all problems (e.g., $\mathfrak{F}(Y)\geq T$ means $\mathfrak{A}(Y)\geq T$, $\mathfrak{P}(Y)\geq T$ and $\mathfrak{R}(Y)\geq T$ depending on the situation). We collectively refer to \textit{utility} of the solution for AT, PT and RT queries as the objective optimized in their respective problem definitions. That is, for AT queries, the utility is measured in terms of cost, for PT in terms of recall and for RT queries in terms of precision.

\subsection{Model Cascade Framework}\label{sec:background:cascade}
To decide when to use the proxy model or the oracle, we follow the model cascade framework \cite{kang2017noscope, anderson2019physical, kang2022tasti, russo2023accelerating, Kang2021AcceleratingAA}. Recall that in  PT and RT queries, we are given a fixed oracle budget, while in AT queries, our goal is to minimize the number of oracle calls. Thus, the cascade framework is used differently in each case. Here, we describe the framework for RT and PT queries, following the formalization by \cite{kang2020approximate}, and defer the discussion of AT queries to Sec.~\ref{sec:at}. 

For PT and RT queries, model cascade relies on \textit{proxy scores} $\mathcal{S}(x)\in[0, 1]$ that quantify the proxy model's confidence in the record $x$ being positive (recall that for PT/RT queries model outputs are binary). Such scores are typically produced by the proxy model as part of inference (i.e., the output tokens' probability for LLMs). The framework uses the fixed oracle budget, $k$, to determine a \textit{cascade threshold} $\rho$ on the proxy scores so that the records with proxy scores more than $\rho$ are deemed positive and those below $\rho$ are deemed negative. More formally, first define, for a set of records $S$, $S\subseteq D$, and a threshold $\rho\in[0, 1]$, $S^\rho=\{x; x\in S, \mathcal{S}(x)>\rho\}$. In our framework, a subset $S$, $S\subseteq D$, of size at most $k$ is labeled and used to determine the cascade threshold, $\rho$. Using $\rho$, we estimate the set of records with positive labels as $D^\rho$. In practice, $D^\rho$ can additionally be augmented with the observed positive labels in $S$. 

Thus, performing PT queries boils down to finding a cascade threshold, $\rho$, by labeling a subset $S\subseteq D$ of size $k$ with the oracle, such that $\mathfrak{P}(D^\rho)$ meets the precision target, that is $\mathds{P}(\mathfrak{P}(D^\rho)\geq T)\geq 1-\delta$, and maximizes recall, $\mathfrak{R}(D^\rho)$. The problem is analogously defined for RT queries. For convenience of notation, we define $\mathfrak{P}_D(\rho)=\mathfrak{P}(D^\rho)$ and $\mathfrak{R}_D(\rho)=\mathfrak{R}(D^\rho)$. 
We further abuse notation and, for any set $S\subseteq D$, define:
\begin{align*}
    \mathfrak{R}_S(\rho)=\frac{\sum_{x\in S^\rho}\mathds{I}[\mathcal{O}(x)=1]}{\sum_{x\in S}\mathds{I}[\mathcal{O}(x)=1]},
\mathfrak{P}_S(\rho)=\frac{\sum_{x\in S^\rho}\mathds{I}[\mathcal{O}(x)=1]}{|S^\rho|}.
\end{align*}
$\mathfrak{P}_S(\rho)$ (resp. $\mathfrak{R}_S(\rho)$) is the precision (resp. recall) with respect to $S$ if $S^\rho$ is estimated as the set of records in $S$ with positive labels. 

\rone{\calibration{}}

\textbf{Notation and Terminology}. For a random sample $S$ of $D$, we refer to metrics calculated on the sample as \textit{observed} metrics (e.g., $\mathfrak{P}_S(\rho)$ is the \textit{observed precision} at $\rho$), and refer to the metrics on the entire dataset as \textit{true} metrics (e.g., $\mathfrak{P}_D(\rho)$ is the \textit{true precision}). When sampling points, the oracle labels every sampled point, so we use the terms \textit{sampling} and \textit{labeling} interchangeably. We use $[i]$ to denote the set $\{1, ..., i\}$ for any integer $i$.

\begin{table}[t]
\vspace{-0.6cm}
\rthree{
    \small
    \begin{tabular}{ll}
    \toprule
    \textbf{Notation} & \textbf{Description} \\
    \midrule
    $D, S$ & Dataset of all records and a subset of the records\\
    $\mathcal{O}(x)$ & Oracle output for a record $x$ \\
    $\mathcal{P}(x), \mathcal{S}(x)$ & Proxy output and score for a record $x$ \\
    $\rho$ & Cascade threshold on proxy scores \\
    $\mathfrak{A}_S(\rho)$ & Accuracy on a set $S$ at threshold $\rho$ \\
    $\mathfrak{P}_S(\rho), \mathfrak{R}_S(\rho)$ & Precision and recall on a set $S$ at threshold $\rho$ \\
    $S^\rho$ & $\{x \in S : \mathcal{S}(x) > \rho\}$ for any set of records, $S$\\
    $S_+$ & $\{x;x\in S, \mathcal{O}(x)=1\}$ for any set of records, $S$\\
    \multirow{2}{*}{$\mathfrak{E}(S, T, \rho, \alpha)$} & Estimation function to test if a threshold, $\rho$,  \\
    & meets $T$ given a sample $S$ with confidence $\alpha$ \\
    $\mathscr{C}$ & Set of candidate cascade thresholds to choose from\\
    $k$ & Oracle budget \\
    $T$ & Target precision, recall, or accuracy \\
    $\delta$ & Allowed probability of failure to meet the target\\
    \bottomrule
    \end{tabular}
\caption{\rthree{Summary of mathematical notation}}\label{tab:notation}
\vspace{-0.2cm}
}
\end{table}


%

\subsection{\rall{Statistical Tools}}\label{sec:stat_tools}
\rall{\stattools{}}

\begin{table}[t]
\vspace{-0.6cm}
\rall{
\small
\setlength{\tabcolsep}{2pt} 
\centering
\begin{tabular}{c c c}
\toprule
\textbf{Variant} & \textbf{Quality Metric} & \textbf{Primary Parameters} \\
\midrule
\name{}$_P$-A & Precision & $k, T, \delta$ \\\midrule
\name{}$_A$-A & Accuracy & $T, \delta$  \\
\name{}$_A$-M & Accuracy & $T, \delta$  \\\midrule
\name{}$_R$-A & Recall  & $k, T, \delta$  \\
\bottomrule
\end{tabular}
\caption{\rall{Main \name{} Variants}}\label{tab:prism_variants}
}
\vspace{-0.2cm}
\end{table}

\subsection{\rall{Overview and Outline}}\label{sec:overview_outline}
\rall{Armed with the statistical tools from Sec.~\ref{sec:stat_tools}, \name{} solves the cascade threshold problem for AT, PT, and RT queries. We present \name{} for PT queries in Sec.~\ref{sec:pt}, discussing alternatives and solution components  in depth. We extend \name{} to AT and RT queries, respectively, in Secs.~\ref{sec:at} and \ref{sec:rt_short}. \overview{}}

\if 0\textbf{Quality Constraints and Cascade Utility}. 

. Then, to decide when to use the proxy model, we sample a set $S$ from the dataset of size $|k|$ and use it to determine a \textit{cascade threshold} $\rho$ so that, for any data point $x_i$ such that $\mathcal{S}(x_i)\geq\rho$, we use $\mathcal{P}(x_i)$ as the answer, and use $\mathcal{O}(x_i)$ as the answer otherwise. \sep{show overview of procedure}

The problem, then, is to determine a cascade threshold. How to determine a cascade threshold depends on the exact problem setting. For example, if the user wants to ensure high recall, the threshold needs to be set to a smaller value than when the goal is to ensure high precision. Thus, we define three problem setting, depending on the quality metric the user is interested in. 

\textit{Precision/Recall Target Queries}. In these queries, the output of the oracle is a binary label. The user wants the proxy to label a subset of $D$ to ensure precision (reps. recall) is at least some desired target $T$ with probability at last $1-\delta$, while maximizing recall (resp. precision). The system is also provided an oracle budget, $k$, the the algorithm can use to obtain ground-truth labels. The labeling by the proxy is done using a cascade threshold $\rho$. To formalize this, define
\begin{align*}
\mathfrak{R}_D(\rho)&=\frac{\sum_{x\in D}\mathds{I}[\mathcal{O}(x)=1\, \&\, \mathcal{S}(x)\geq\rho]}{\sum_{x\in D}\mathds{I}[\mathcal{O}(x)=1]}, \\
\mathfrak{P}_D(\rho)&=\frac{\sum_{x\in D}\mathds{I}[\mathcal{O}(x)=1\, \&\, \mathcal{S}(x)\geq\rho]}{\sum_{x\in D}\mathds{I}[\mathcal{S}(x)\geq\rho]}.    
\end{align*}
We define\sep{need to somehow bring up why there is a pre-specified budget}
\begin{definition}
    Given a precision (resp. recall) target $T$ and failure probability $\delta$, and an oracle budget $k$, use at most $k$ oracle calls to find a cascade threshold $\rho$ such that  \sep{not sure this makes sense}
    \begin{align*}
        &\max_{\rho\in[0, 1]} \mathfrak{R}(\rho)\\
        \text{s.t.}\quad &\mathds{P}(\mathfrak{P}_D(\rho)< T)\leq \delta
    \end{align*}
    \begin{align*}
        &\max_{\rho\in[0, 1]} \mathfrak{P}(\rho)\\
        \text{s.t.}\quad &\mathds{P}(\mathfrak{R}_D(\rho)< T)\leq \delta
    \end{align*}
\end{definition}


\textit{Accuracy Target Queries}. In such queries, the user wants to processes the entire dataset $D$, but wants to use the oracle  as little as possible, while ensuring the answer is accurate. Define,\sep{change these, define problems as abstrat set retun, then say what cascade is}
\begin{align*}
\mathfrak{A}_D(\rho)&=\frac{\sum_{x\in D, \mathcal{S}(x)\geq\rho}\mathds{I}[\mathcal{O}(x)=\mathcal{P}(x)]+\sum_{x\in D, \mathcal{S}(x)<\rho}1}{n}.\\
\end{align*}

Formally,

\begin{definition}
    Given an accuracy target $T$ and failure probability $\delta$ finda cascade threshold $\rho$ such that  \sep{not sure this makes sense}
    \begin{align*}
        &\max_{\rho\in[0, 1]} \sum_{x\in D, \mathcal{S}(x)\geq\rho}1\\
        \text{s.t.}\quad &\mathds{A}(\mathfrak{P}_D(\rho)< T)\leq \delta
    \end{align*}
\end{definition}
\sep{Move down cascade stuff}
\textbf{Model Cascade}. 
\fi

%% file: newer_precision.tex
\section{\name{} for PT Queries}\label{sec:pt}
\begin{figure}
    \centering
    \includegraphics[width=1\linewidth]{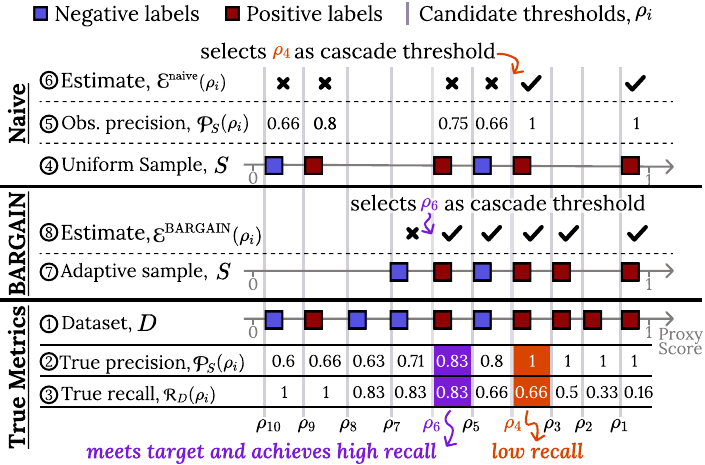}
    \caption{An example of cascade approaches for PT queries given precision target $T=0.75$.}
    \label{fig:overview}
\end{figure}

In this section, we discuss \name{} for solving PT queries. We use Fig.~\ref{fig:overview} (top half) as a running example, where we perform a PT query with target $T=0.75$ on the dataset, $D$. That is, our goal is to find a cascade threshold with precision at least $0.75$ w.h.p while maximizing recall. {\small \CircledText[]{1}} Red/blue squares show records $x \in D$, plotted at $\mathcal{S}(x)$ on the proxy score axis and colored based on $\mathcal{O}(x)$. The vertical lines $\rho_1,..., \rho_{10}$ are \textit{candidate thresholds} to choose the final cascade threshold from. Bottom of Fig.~\ref{fig:overview} {\small \CircledText[]{2}} shows the true precision, $\mathfrak{P}_D(\rho_i)$, and  {\small \CircledText[]{3}} true recall, $\mathfrak{R}_D(\rho_i)$, for the candidate thresholds. \textit{candidate thresholds} are precisely the proxy scores of records in $S$, $\mathscr{C}=\{\mathcal{S}(x); x\in S\}$, because $\mathfrak{P}_S(\rho)$ (i.e., the observed precision at threshold $\rho$) changes only at these values and is constant otherwise. 


\textbf{Overview of \name{} for PT Queries}. Our framework for PT queries consists of three main components, as detailed in Fig.~\ref{fig:cascade_overview} and discussed here at a high level: (1) A \textit{sampling} component samples a set of $k$ records, $S$, from $D$, to be labeled by the oracle. (2) An \textit{estimation} component estimates for any candidate threshold whether it is expected to meet the precision target. This is done by defining a boolean \textit{estimation function} $\mathscr{E}(S, T, \rho, \alpha)$, which given a sample set $S$, the target $T$, and a candidate threshold $\rho$, returns 1 if $\mathfrak{P}_D(\rho)\geq T$ is expected to hold and 0 otherwise. $\mathscr{E}$ depends on a \textit{confidence parameter} $\alpha$ that quantifies the probability that $\mathscr{E}$ makes wrong estimates, discussed later. (3) A \textit{selection} component selects, among the thresholds expected to meet the target, the one with the highest recall. This step repeatedly uses $\mathscr{E}$ to estimate if different candidate thresholds meet the requirement before choosing the one with the highest recall among ones where $\mathscr{E}$ returns true. 

Next, to illustrate the framework and associated challenges, we describe a naive approach following the framework. We then present \name{}$_P$-U, which keeps the uniform sampling procedure of the naive method but improves upon estimation and selection procedures. We then present  \name{}$_P$-A, our final algorithm that incorporates adaptive sampling to improve utility.

\subsection{Warm-up: Naive Algorithm with Guarantees}\label{sec:pt:naive}

\textbf{Sampling and Estimation}. A naive algorithm samples a set $S$ of $k$ records uniformly at random from $D$. The estimation function $\mathscr{E}$ uses this set to estimate if different candidate thresholds meet the precision target. We want $\mathscr{E}$  
to have a \textit{low false positive probability}:
\begin{align}\label{eq:est_guaranee}
    \mathds{P}_{S\sim D}(\mathscr{E}(S, T, \rho, \alpha)=1)\leq \alpha\;\text{if}\; \mathfrak{P}_D(\rho)< T,
\end{align}
for any confidence parameter $\alpha\in[0, 1]$. False positives must be avoided as they can lead to the selection of a threshold incorrectly estimated to have true precision above $T$. To design $\mathscr{E}$, we can use Hoeffding's inequality to relate the observed precision, $\mathfrak{P}_S(\rho)$ to the true precision, $\mathfrak{P}_D(\rho)$. By doing so, we show
\begin{align}\label{eq:naive_est}
\hspace*{-10pt}
\mathscr{E}^{\text{naive}}(S, T, \rho, \alpha)=\mathds{I}\big[\mathfrak{P}_S(\rho)\geq T+\Delta\big],\, \text{where}\, \Delta=\sqrt{\frac{\log(1/\alpha)}{2|S^\rho|}}
\end{align}
satisfies Eq.~\ref{eq:est_guaranee}, where $\mathds{I}$ is the indicator function and $S^\rho$ is the subset of $S$ with proxy scores more than $\rho$. $\mathscr{E}^{\text{naive}}$ estimates that a threshold $\rho$ meets the target whenever the observed precision, $\mathscr{P}_S(\rho)$ is more than $T+\Delta$. Note that $\Delta$ serves as an adjustment factor for sampling. That is, it is not sufficient for observed precision $\mathfrak{P}_S(\rho)\geq T$ to guarantee $\mathfrak{P}_D(\rho)\geq T$, but we instead need $\mathfrak{P}_S(\rho)\geq T+\Delta$. Fig.~\ref{fig:overview} shows an example at threshold $\rho_9$ where $\mathfrak{P}_S(\rho)\geq T$, but $\mathfrak{P}_D(\rho)$ is not. 
We can use Hoeffding's inequality to show the following. 
\begin{proposition}\label{prop:hoef_single}
    For any $\rho\in[0, 1]$ with $\mathfrak{P}_D(\rho)<T$, we have
$\mathds{P}(\mathfrak{P}_S(\rho)\geq T+\sqrt{\frac{\log(1/\alpha)}{2|S^\rho|}})\leq\alpha,$ 
where $S$ is i.i.d and uniformly sampled from $D$ and $\alpha\in[0, 1]$ is a confidence parameter. Consequently, $\mathds{P}(\mathscr{E}^\text{naive}(S, T, \rho, \alpha)=1)\leq \alpha$.
\end{proposition}

Fig.~\ref{fig:overview} shows an example of this procedure. The figure depicts a sample $S$ {\small \CircledText[]{4}} and the observed precision at different candidate thresholds $\rho_i$ {\small \CircledText[]{5}}. In the example, we see that $\mathscr{E}^{\text{naive}}$ estimates  $\rho_i\in\{\rho_1, ..., \rho_4\}$ meet the target while $\mathscr{E}^{\text{naive}}(S, \rho_i, T, \alpha)=0$ for all other thresholds {\small \CircledText[]{6}}. \rone{\chernoff{}}

\textbf{Selection}. The naive approach for threshold selection chooses the smallest threshold in the candidate set as the cascade threshold:
\begin{align}\label{eq:naive_sel}
    \rho^*=\min\{\rho; \rho\in \mathscr{C}, \mathscr{E^{\text{naive}}}(S, T,\rho,\alpha)=1\}.
\end{align} 
$\rho^*$ is the $\rho$ that maximizes recall (returns the most records as ``positive'') among all $\rho$ that satisfy $\mathscr{E^{\text{naive}}}$. Finding $\rho^*$ requires $|\mathscr{C}|$ applications of $\mathscr{E^{\text{naive}}}$. To guarantee $\rho^*$ meets the target, \textit{all} applications of  $\mathscr{E^{\text{naive}}}$ must return 0 for candidate thresholds with $\mathfrak{P}_D(\rho)<T$. Prop.~\ref{prop:hoef_single} shows this holds for a single application of $\mathscr{E^{\text{naive}}}$. We use union bound across all applications of $\mathscr{E^{\text{naive}}}$ to ensure the probability that any of $\mathscr{E^{\text{naive}}}$ return 0 for candidate thresholds with $\mathfrak{P}_D(\rho)<T$ is small. Doing so proves:
\begin{proposition}\label{prop:select_naive}
    Let $\mathscr{E}$ be a function with false positive probability bounded by $\alpha$ (as defined in Eq.~\ref{eq:est_guaranee}) when sampling a set $S$ from $D$. Setting $\rho^*$ as Eq.~\ref{eq:naive_sel}, we have $\mathds{P}(\mathfrak{P}_D(\rho^*)<T)\leq |\mathscr{C}|\alpha$.
\end{proposition}

Prop.~\ref{prop:select_naive} shows the probability of the selection method failing to meet the target is proportional to $|\mathscr{C}|$, the candidate set's size. 

\textbf{Final Algorithm and Guarantees}. Since we want $|\mathscr{C}|\alpha=\delta$, we set $\alpha=\frac{\delta}{|\mathscr{C}|}$. Therefore, we select the cascade threshold as
\begin{align}\label{eq:rho_S_naive}
    \rho_S^{\text{naive}}=\min\{\rho; \rho\in \mathscr{C}, \mathscr{E}^{\text{naive}}(S, T,\rho,\frac{\delta}{|\mathscr{C}|})=1\}.
\end{align} 
$\rho_S^{\text{naive}}$ uses $\mathscr{E^{\text{naive}}}$ for estimation with $\alpha=\frac{\delta}{|\mathscr{C}|}$. This ensures, using Prop.~\ref{prop:select_naive}, that the probability of $\rho_S^{\text{naive}}$ missing the target is bounded by $\delta$. Indeed, combining Props.~\ref{prop:hoef_single} and \ref{prop:select_naive} shows:
\begin{lemma}\label{lemma:prec_heof_union}
For cascade threshold $\rho_S^{\text{naive}}$ returned by the naive method, we have $\mathds{P}_{S\sim D}(\mathfrak{P}_D(\rho_S^{\text{naive}})<T)\leq \delta.$
\end{lemma}

\textbf{Discussion}. Consider our running example, in Fig.~\ref{fig:overview}, where the naive algorithm chooses $\rho_4$ as the cascade threshold, since it is the smallest threshold with $\mathscr{E}^{\text{naive}}(S, \rho_i, T, \frac{\delta}{|\mathscr{C}|})=1$. This cascade threshold leads to recall 66\%, well below the best possible of 83\% at $\rho_6$ while meeting the precision requirement. This low recall 
can be attributed to the \textit{high false negative rate} of $\mathscr{E}^{\text{naive}}$ which incorrectly precludes the algorithm from selecting thresholds that meet the target and have better utility; e.g., in Fig.~\ref{fig:overview}, $\mathscr{E}^{\text{naive}}$ incorrectly estimates that $\rho_5$ and $\rho_6$ don't meet the target. To improve on the naive approach, \name{} leverages novel sampling, estimation, and selection methods that significantly reduce the false negative rate and ultimately provide much better utility.\techreport{ As Fig.~\ref{fig:overview} shows, \name{} performs adaptive sampling and uses a more accurate estimation function and threshold selection mechanism. These significantly improve recall while providing the same theoretical guarantees. }

\subsection{\name{}$_P$-U: Data-Aware Estimation}\label{sec:prec:unif}
As described above, the low recall achieved by the Naive approach can be attributed to the estimator $\mathscr{E^{\text{naive}}}(S, T,\rho,\frac{\delta}{|\mathscr{C}|})$ frequently returning 0 when $\mathfrak{P}_D(\rho)\geq T$. 
\name{}$_P$-U uses a better estimation function $\mathscr{E}$ and selection algorithm to remedy this. This selection algorithm performs a tighter analysis across applications of $\mathscr{E}$ to avoid splitting the failure probability $\delta$ into $\frac{\delta}{|\mathscr{C}|}$ for each application of $\mathscr{E}$. Both improvements allow \name$_P$-U to identify better valid candidate thresholds, $\rho$, where $\mathfrak{P}_D(\rho)\geq T$, thus improving the recall. We describe these two components in more detail before presenting the final \name$_P$-U algorithm. \name{}$_P$-U performs uniform sampling; we incorporate adaptive sampling in Sec.~\ref{sec:prec:adap}. 

\subsubsection{Estimation.}\label{sec:pt:unif:est}
Observe that the true precision at a threshold $\rho$ is the mean of the random variables $\mathds{I}[\mathcal{O}(x)=1]$ for $x\in S^\rho$. That is, $\mathds{E}[\mathds{I}[\mathcal{O}(x)=1]]=\mathfrak{P}_D(\rho)$, because $x\in S^\rho$ is a uniform sample from $D^\rho$, so
$$\mathds{P}(\mathcal{O}(x)=1)=\sum_{x'\in D^\rho}\frac{\mathds{I}[\mathcal{O}(x')=1]}{|D^\rho|}=\mathfrak{P}_D(\rho).$$  Therefore, to estimate whether $\mathfrak{P}_D(\rho)\geq T$ or not, we can use the hypothesis test of whether the true mean of the random variables in $S^\rho$ is more than $T$ or not. 
\rall{This hypothesis testing formulation enables us to use Lemma~\ref{lemma:hypothesis_test} to design our estimation function. Our approach results in a new function $\mathscr{E}^{\text{\name{}}}$ that helps test whether $\mathscr{P}_D(\rho)\geq T$. $\mathscr{E}^{\text{\name{}}}$ performs the hypothesis test, $\mathscr{T}$, from Lemma~\ref{lemma:hypothesis_test}, to decide if the true precision is below the threshold $T$. The following Lemma shows the guarantees of $\mathscr{E}^{\text{\name{}}}$. Define the set of random variables $S_O^\rho=\{\mathds{I}[\mathcal{O}(x)=1]; x\in S^\rho\}$.}
\begin{lemma}\label{lemma:lb}
\rall{    For  a confidence parameter $\alpha\in[0, 1]$ and any $\rho\in[0, 1]$ where $\mathfrak{P}_D(\rho)<T$, 
    \begin{align}\label{eq:est_prism_bound}
        \mathds{P}(\mathscr{E}^{\textnormal{\name{}}}(S, T, \rho, \alpha)=1)\leq \alpha, \quad \text{where}
    \end{align}    
    \begin{align}\label{eq:est_prism_def}
        \mathscr{E}^{\textnormal{\name{}}}(S, T, \rho, \alpha)=\mathscr{T}(T, S_O^\rho, \alpha), 
    \end{align}
    and $\mathscr{T}$ is the hypothesis test defined in Lemma~\ref{lemma:hypothesis_test}.}
\end{lemma}
\textbf{Benefits}. Eq.~\ref{eq:est_prism_bound} shows that $\mathscr{E}^{\text{\name{}}}$ performs our desired hypothesis test of whether $\mathfrak{P}_D(\rho)\geq T$ or not with bounded false positive probability. That is, $\mathscr{E}^{\text{\name{}}}$ is unlikely to return 1 for thresholds where $\mathfrak{P}_D(\rho)<T$.  As such, we use $\mathscr{E}^{\text{\name{}}}$ instead of $\mathscr{E}^{\text{naive}}$, replacing Prop.~\ref{prop:hoef_single} in the naive approach with Lemma~\ref{lemma:lb} in \name{}$_P$-U. 

\begin{figure}
\vspace{-0.5cm}
    \begin{minipage}{0.49\columnwidth}
    \includegraphics[width=0.95\linewidth]{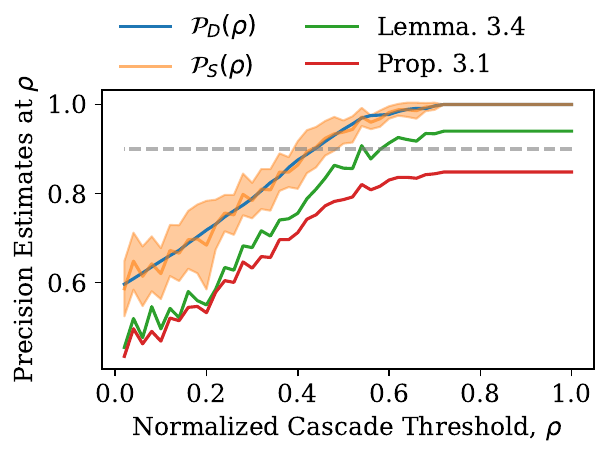}
    \caption{Comparison of different estimation methods}
    \label{fig:bounds}
    \end{minipage}
    \hfill
    \begin{minipage}{0.49\columnwidth}
    \includegraphics[width=0.95\linewidth]{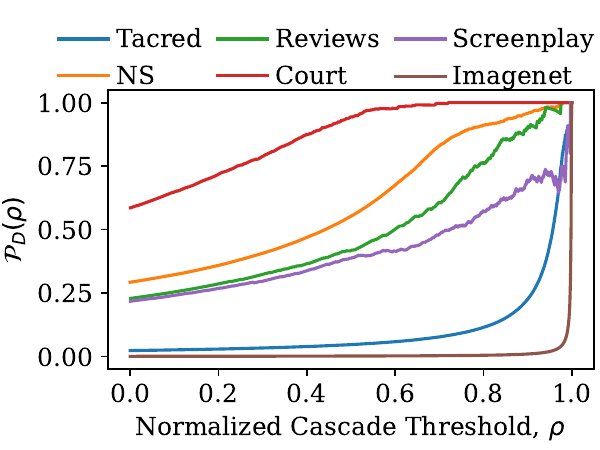}
    \caption{Ground-truth precision at different thresholds}
    \label{fig:prec_candidate}
    \end{minipage}
    \vspace{-0.1cm}
\end{figure}

The benefit of $\mathscr{E}^{\text{\name{}}}$ over $\mathscr{E}^{\text{naive}}$ can be attributed to the use of observed variance in $\mathscr{E}^{\text{\name{}}}$ to provide significantly better estimates, especially so when observations have low variance. $\mathscr{E}^{\text{\name{}}}$ takes both observed means and variances into account (via $\hat{\sigma_i}$ and $\hat{\mu_i}$ in Lemma~\ref{lemma:hypothesis_test}), but $\mathscr{E}^{\text{naive}}$ only depends on the observed mean (i.e., the observed precision in Eq.~\ref{eq:naive_est}). We empirically show the benefits of $\mathscr{E}^{\text{\name{}}}$ in Fig.~\ref{fig:bounds}, where we plot 
\begin{align}\label{eq:T_rho}
T_\rho^*=\max\{T; T\in[0, 1],\mathscr{E}(S, T, \rho, \alpha)=1\}
\end{align}
at different values of $\rho$ for fixed random samples, $S$ and for $\mathscr{E}$ as either $\mathscr{E}^{\text{\name{}}}$ or $\mathscr{E}^{\text{naive}}$. $T_\rho^*$ denotes the largest target $T$ that a threshold $\rho$ will be estimated to meet using $\mathscr{E}$, given $S$ and $\alpha$. The closer $T^*_\rho$ is to $\mathfrak{P}_D(\rho)$ the more accurate $\mathscr{E}$ is. In fact, for any target $T\in (T^*_\rho, \mathfrak{P}_D(\rho)]$, the estimate of $\mathscr{E}$ is a false negative, so how close $T^*_\rho$ is to $\mathfrak{P}_D(\rho)$ determines the accuracy of $\mathscr{E}$ at different targets. To plot $T_\rho^*$, we use one of our real-world datasets, Court (details in Sec.~\ref{sec:exp}), as $D$, set $\alpha=0.1$, and for each threshold, $\rho$, sample a (large enough) set, $S$ so that $S^\rho$ contains $50$ records. We keep the size of $S^\rho$ fixed to ensure the number of samples does not impact the trends across thresholds (e.g., $|S^\rho|=50$ in Eq.~\ref{eq:naive_est} for all $\rho$). Fig.~\ref{fig:bounds} also plots $\mathfrak{P}_D(\rho)$ and $\mathfrak{P}_S(\rho)$, and the shaded region around $\mathfrak{P}_S(\rho)$ shows its observed standard deviation across 10 runs. 

Fig.~\ref{fig:bounds} shows $\mathscr{E}^{\text{\name{}}}$ provides increasingly better estimates than $\mathscr{E}^{\text{naive}}$ as $\mathfrak{P}_D(\rho)$ increases. Indeed, if the user provides target $T=0.9$, $\mathscr{E}^{\text{naive}}$ estimates that no target meets the threshold. Instead, $\mathscr{E}^{\text{\name{}}}$ correctly identifies thresholds larger than $\sim$0.5 that meet the target, because it takes advantage of the lower variance in the observed samples when the true precision is high. For instance, when the true precision is 1, observed variance will be zero because all observed samples at the threshold will always be positive. Indeed, in \Cref{appx:variance} we show that the variance in observations decreases as the true precision increases; thus $\mathscr{E}^{\text{\name{}}}$ makes more accurate estimates when true precision is high. Meanwhile, $\mathscr{E}^{\text{naive}}$ provides the same estimates independent of the variance.

\subsubsection{Threshold Selection.}\label{sec:prism_pu:thresh}
The Naive algorithm selected the smallest threshold among all the candidate thresholds estimated to meet the precision target. Prop.~\ref{prop:select_naive} showed that this leads to the probability of missing the target increasing in the number of candidate thresholds, $|\mathscr{C}|$. Next, we show that we can incorporate data characteristics into the selection algorithm and analysis to avoid this dependence on $|\mathscr{C}|$. 

\textbf{Selection Procedure.} The following equation presents our selection method (used in place of Eq.~\ref{eq:naive_sel}) to improve utility. It uses a tolerance parameter $\eta$ that, as discussed later, helps us incorporate data characteristics into the algorithm: 
\if Denote by $$\gamma_\rho = \sum_{\rho'\in \mathscr{C}, \rho'\geq\rho}\mathds{I}[\mathscr{E}(S, T, \rho)=0],$$
the number of candidate thresholds more than or equal to $\rho$ that are estimated to miss the target by $\mathscr{E}$. $\gamma_\rho$ is computed based on the observed sample. Then, for a system parameter, $\eta$, define 
\begin{align}\label{eq:select_eta}
\rho^*=\min\{\rho; \rho\in\mathscr{C}, \gamma_{\rho}\geq\eta\}.    
\end{align}
\begin{align}\label{eq:select_eta}
\rho^*=\min\{\rho; \rho\in\mathscr{C}, \exists\,\text{at most} \eta\, \text{thresholds}, \rho'>\rho \text{s.t.} \mathscr{E}(S, T, \rho)=0\}.    
\end{align}
\fi
\begin{align}\label{eq:select_eta}
\rho^*&=\min\{\rho; \rho\in\mathscr{C}, \mathscr{E}(S, T, \rho, \alpha)=1, \gamma_\rho\leq \eta\}, \;\text{where}\\
\gamma_\rho&=\sum_{\rho'\in \mathscr{C}, \rho'\geq\rho}\mathds{I}[\mathscr{E}(S, T, \rho', \alpha)=0].\nonumber
\end{align}
$\gamma_\rho$ denotes the number of candidate thresholds greater than or equal to $\rho$ that $\mathscr{E}$ estimates to miss the target. Thus, $\rho^*$ is the smallest candidate threshold where at most $\eta$ larger thresholds are estimated to miss the target. When $\eta=|\mathscr{C}|$, $\rho^*$ is the smallest threshold that meets the target (same as Eq.~\ref{eq:naive_sel}), but when $\eta=0$, $\rho^*$ is the smallest candidate threshold such that both itself and all larger thresholds  are estimated to meet the target. The failure probability of Eq.~\ref{eq:select_eta} to meet the target now depends on $\eta$ not $|\mathscr{C}|$:
\begin{lemma}\label{lemma:select_eta}
    Let $\mathscr{E}$ be a function with false positive probability bounded by $\alpha$ (as defined in Eq.~\ref{eq:est_guaranee}) when sampling a set $S$ from $D$. Setting $\rho^*$ as Eq.~\ref{eq:select_eta}, we have $\mathds{P}(\mathfrak{P}_D(\rho^*)<T)\leq (\eta+1
    )\alpha$.
\end{lemma}

\textbf{Benefits}. The tolerance parameter $\eta$ allows us to capture data characteristics, and, in real-world settings, can be set to a small value to obtain low failure probability using Lemma~\ref{lemma:select_eta}.
A suitable value for $\eta$ depends on how often, in a dataset, the true precision oscillates around the target. For instance, if after the largest $\rho$ with true precision below $T$, $\mathfrak{P}_D(\rho)<T$, all thresholds $\rho'<\rho$ also have $\mathfrak{P}_D(\rho)<T$, then there is no benefit in considering thresholds less than $\rho$, and it is sufficient to set $\eta=0$. Fig.~\ref{fig:prec_candidate} that plots $\mathfrak{P}_D(\rho)$ for six different real-world datasets (from Sec.~\ref{sec:exp}) shows this is the case in real-world datasets, where true precision is often monotonically decreasing. Thus, the above argument implies setting $\eta=0$ is sufficient to obtain a good utility. We note that setting $\eta>0$ may be beneficial in cases where $\mathfrak{P}(D)$ drops below $T$  for $\eta$ different candidate thresholds, after which $\mathfrak{P}(D)$ rises up again to above $T$. 
Fig.~\ref{fig:prec_candidate} shows this is rarely the case in real-world datasets. We emphasize that the choice of $\eta$ only affects the utility of the methods. Our guarantees, as Lemma~\ref{lemma:select_eta} shows, hold for all possible datasets. Given the above observations, and to simplify our results, in the rest of this paper, we set $\eta=0$. Generalization to $\eta>0$ is straightforward and discussed in \Cref{sec:gen:eta}.

\begin{algorithm}[t]
\small
\begin{algorithmic}[1]
\State $S\leftarrow $ Sample $k$ records from $D$ uniformly at random
\State Sort $\mathscr{C}$ in descending order 
\For{$i$ \textbf{in} $|\mathscr{C}|$}
    \State $\rho\leftarrow \mathscr{C}[i]$
    \If{$\mathscr{E}^{\text{\name{}}}(S, T, \rho, \delta)=0$}\label{alg:iterative_select_precision:delta}
        \State \Return $\mathscr{C}[i-1]$\label{alg:iterative_select_precision:return}\label{line:prismpu:return}
    \EndIf
\EndFor
\State\Return $\mathscr{C}[|\mathscr{C}|]$
\caption{\name{}$_P$-U}\label{alg:iterative_select_precision}
\end{algorithmic}
\end{algorithm}

\subsubsection{\name{}$_P$-U Algorithm and Guarantees} 
\name{}$_P$-U performs uniform sampling, and uses the above estimation and threshold selection procedures, as stated in Alg.~\ref{alg:iterative_select_precision}. It iterates through candidate thresholds in decreasing order and stops after reaching the first threshold estimated by $\mathscr{E}^{\text{\name{}}}$ to miss the target. If such a threshold is $\mathscr{C}[i]$, it returns $\mathscr{C}[i-1]$, which was estimated to meet the target at the previous iteration. The early stopping is possible because $\eta=0$ in threshold selection. Combining Lemmas~\ref{lemma:lb} and ~\ref{lemma:select_eta} proves: 

\begin{lemma}\label{lemma:precision_unif}
    Let $\rho_S^{\text{\name{}$_P$-U}}$ be the cascade threshold found by Alg.~\ref{alg:iterative_select_precision}. We have $\mathds{P}_{S\sim D}(\mathfrak{P}_D(\rho_S^{\text{\name{}$_P$-U}})<T)\leq \delta.$
\end{lemma}

\subsubsection{Discussion} 
\begin{figure}
    \vspace{-0.05cm}
    \centering
    \begin{minipage}{0.22\textwidth}
        \includegraphics[width=\linewidth]{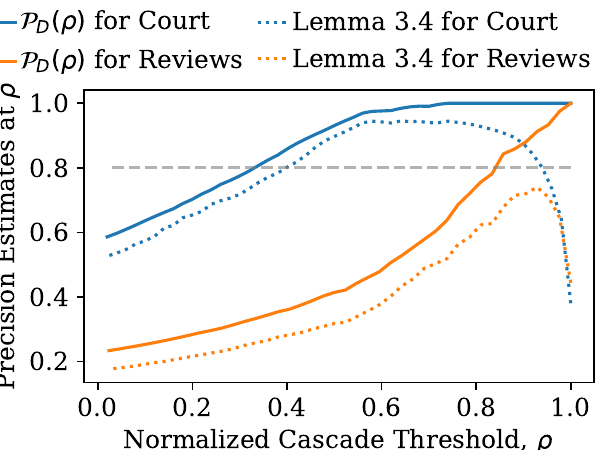}
        \caption{Estimation with a fixed uniform sample}
        \label{fig:lb_single_shot}
    \end{minipage}
    \hfill
    \begin{minipage}{0.23\textwidth}
    \includegraphics[width=1\linewidth]{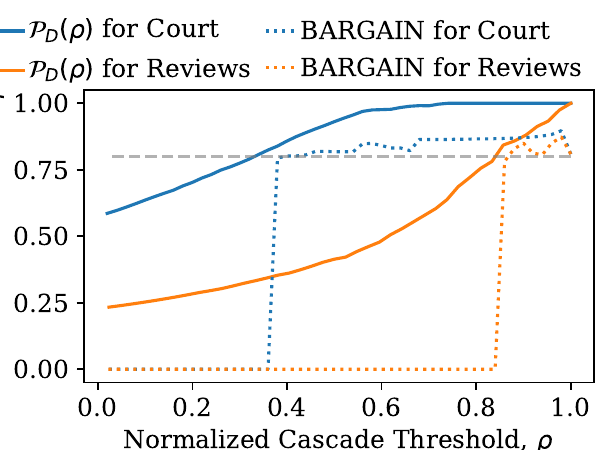}
    \caption{Estimation from sampling with \name{}$_P$-A}
    \label{fig:prism_pA_bounds}
    \end{minipage}
    \vspace{-0.15cm}
\end{figure}

The improved estimation and selection methods in \name{}$_P$-U improve upon the naive approach, but its utility is still hampered by uniform sampling. To see why, observe that applications of $\mathscr{E}^{\text{\name{}}}$ for different $\rho$ values use the same sample set $S$ but different subsets, $S^\rho$, of $S$ to make an estimate for each $\rho$. 
Given a fixed set, $S$, the size of $S^\rho$ decreases the larger $\rho$ gets. 
This causes two problems. First, for large $\rho$, $\mathscr{E}^{\text{\name{}}}$ has access to a smaller set to perform estimation. Using fewer observations, $\mathscr{E}^{\text{\name{}}}$ will be less confident about the true precision; thus, to provide the same low false positive guarantees (as defined in Eq.~\ref{eq:est_guaranee}) it has to err on the safe side and estimate that thresholds don't meet the target more frequently. This is plotted in Fig.~\ref{fig:lb_single_shot}, showing $T_\rho^*$ (as defined in  Eq.~\ref{eq:T_rho}) for different values of $\rho$, but for a fixed uniform sample from $D$ (for two real-world datasets, Court and Reviews). We see that Lemma~\ref{lemma:lb} provides poor estimates for large $\rho$, incorrectly estimating that larger thresholds don't meet many targets. Comparing Figs.~\ref{fig:lb_single_shot} and \ref{fig:bounds} shows that this is indeed due to decreasing sample sizes, where Fig.~\ref{fig:bounds} plots the same quantity for Court dataset, but when the sample size is fixed (with $k_\rho=50$) for all thresholds.
Second, even though uniform sampling leads to few samples for large candidate thresholds, it, on the other hand, \textit{wastes} many samples for estimation when the candidate threshold is small. In fact, when \name{}$_P$-U stops at a candidate threshold, $\rho$, any sample with proxy score less than $\rho$ is simply not used during estimation at all. 

\begin{algorithm}[t]
\small
\begin{algorithmic}[1]
\State Sort $\mathscr{C}_M$ in descending order 
\For{$i$ \textbf{in} $M$}
    \State $\rho\leftarrow \mathscr{C}_M[i]$
    \State $S\leftarrow \emptyset$ 
    \While{$\mathscr{E}^{\text{\name{}}}(S, T, \rho, \delta)=0$}\label{alg:precision_sample:prec_est2}
        \If{Out of sampling budget}\label{alg:precision_sample:outofbudget}
            \State \Return $\mathscr{C}_M[i-1]$
        \EndIf
        \State Sample a record uniformly from $D^\rho$ and add to $S$\label{alg:prism_pA:sample}
    \EndWhile
\EndFor
\State\Return $\mathscr{C}_M[M]$
\caption{\name{}$_P$-A}\label{alg:iterative_sample_precision}
\end{algorithmic}
\end{algorithm}

\subsection{\name{}$_P$-A: Task-Aware Sampling}\label{sec:prec:adap}
In this section, we present \name{}$_P$-A, our final algorithm for PT Queries that builds on \name{}$_P$-U, but additionally performs adaptive sampling to maximally utilize the oracle budget. 


\textbf{\name{}$_P$-A}. \name{}$_P$-A combines the sampling, estimation and selection steps to direct the sampling budget to both sample enough records when needed and avoid wasting samples when not. To do so, \name{}$_P$-A follows a similar algorithm as \name{}$_P$-U, but instead of sampling $S$ upfront, it samples records on the go and interleaved with the estimation and selection steps.

That is, for each candidate threshold $\rho$, in decreasing order, when checking whether $\mathscr{E}^{\text{\name{}}}$ is true or false, it \textit{continuously samples} new records in range $[\rho, 1]$ until it obtains $\mathscr{E}^{\text{\name{}}}(S, T, \rho,\delta)=1$. Only after it obtains $\mathscr{E}^{\text{\name{}}}(S, T, \rho, \delta)=1$ the algorithm considers the next (smaller) candidate threshold, for which the same process is repeated until it runs out of oracle budget. Note that for a candidate threshold, $\rho$ with $\mathfrak{P}_D(\rho)>T$, given enough records $\mathscr{E}^{\text{\name{}}}$ will eventually estimate that it meets the target. However, if $\mathfrak{P}_D(\rho)<T$, with high probability $\mathscr{E}^{\text{\name{}}}$ will return 0, in which case \name{}$_P$-A continues sampling until it runs out of budget. Since we do not sample a set $S$ apriori, we cannot use the candidate threshold set $\mathscr{C}$. Define, for a system parameter $M$,  
\begin{align}\label{eq:c_M_def}
    \mathscr{C}_M=\{\mathcal{S}(x_{i}); i=\lfloor\frac{j}{M}n\rfloor , j\in[M]\},
\end{align}
as the candidate threshold set of size $M$, where $x_1, .., x_n \in D$, are sorted according to their proxy scores, so that 
$\mathscr{C}_M$ contains every $\frac{j}{M}$-th percentile of proxy scores of $D$ as the candidate threshold set. We discuss the impact of $M$ as well as other potential choices for the candidate threshold set in \Cref{sec:candidate_set}. 

\name{}$_P$-A as presented in Alg.~\ref{alg:iterative_sample_precision}, iterates over candidates in $\mathscr{C}_M$ in descending order, and at the $i$-th iteration for $\rho=\mathscr{C}_M[i]$ checks if $\mathscr{E}^{\text{\name{}}}$ returns true (Line~\ref{alg:precision_sample:prec_est2}). If $\mathscr{E}^{\text{\name{}}}$ is false, it samples a new record from $D^\rho$ (i.e., from records with proxy score $>\rho$) and checks $\mathscr{E}^{\text{\name{}}}$ again with this new sample. This sampling and estimation continues until either (1) $\mathscr{E}^{\text{\name{}}}$ returns 1, so the algorithm is confident that $\rho$ meets the target and moves on to the next candidate threshold, or (2) the algorithm runs out of oracle budget, wherein the algorithm returns the last candidate threshold estimated to meet the target. Note that when $\mathfrak{P}_D(\rho)<T$, it is unlikely that $\mathscr{E}^{\text{\name{}}}$ will return true. Thus, when the algorithm reaches a candidate threshold where $\mathfrak{P}_D(\rho)<T$, it keeps sampling new records and eventually runs out of budget, thus returning the last candidate threshold for which it is estimated that $\mathfrak{P}_D(\rho)\geq T$.

\textbf{Benefits}. To see the benefits of \name{}$_P$-A, Fig~\ref{fig:prism_pA_bounds} plots the same quantity $T_\rho^*$ across thresholds, as defined in Eq.~\ref{eq:T_rho}, but now after the sampling procedure of \name{}$_P$-A. Note that $T_\rho^*$ is plotted for the sample set obtained at the end of every iteration. Comparing Fig~\ref{fig:prism_pA_bounds} and Fig.~\ref{fig:lb_single_shot}, we see significant benefits where the sampling method improves the estimates for larger values of $\rho$. Furthermore, the sharp drop in Fig~\ref{fig:prism_pA_bounds} for small $\rho$ is due to the algorithm not sampling any more records after it chooses the cascade threshold, so it avoids wasting samples for candidate thresholds that will not be selected. 

\rone{\textbf{Guarantees and Proof Overview.}  \name{}$_P$-A provide the same required theoretical guarantees as \name{}$_P$-U stated in Lemma~\ref{lemma:precision_unif}. We omit the theorem for \name{}$_P$-A for the sake of space but discuss overview of the analysis. \proofoverview{}}

\textbf{Sampling without Replacement and Reuse}. So far, the discussion uses sampling with replacement, additionally with $S$ being set to $\emptyset$ per new threshold considered. We show that we can sample without replacement and obtain the same theoretical guarantees but with a slightly different estimation function. Appx.~\ref{sample:wor} discusses how to modify $\mathscr{E}^{\text{\name{}}}$, and Appx.~\ref{lemma:_anytime_no_replace_lb} shows we obtain the same guarantees. \name{}$_P$-A uses sampling without replacement. Furthermore, the sampling process \techreport{in Line~\ref{alg:prism_pA:sample} of Alg.~\ref{alg:iterative_sample_precision}}can reuse the oracle labels from samples from previous iterations, discussed in detail in Appx.~\ref{sec:sampling_proc}.

%% file: extension.tex
\section{AT and RT Queries}\label{sec:discussion}
\name{} applies similar intuitions as PT queries to solve  AT and RT queries. Theoretical statements are deferred to \Cref{sec:statements}. 

\input{accuracy}

\subsection{\name{}$_R$ for RT Queries}\label{sec:rt_short}
Next, we discuss RT queries; we start with uniform samples and then extend the discussion to adaptive sampling.   

\subsubsection{\name{}$_R$-U with Uniform Samples.} Consider a uniform sample $S$, and denote by $S_+=\{x;x\in S, \mathcal{O}(x)=1\}$, subset of $S$ with positive labels, and define $D_+$ analogously for $D$. We next describe \name{}$_R$-U's estimation and threshold selection procedures.

\textbf{Estimation}. Here, we define an estimation function, $\mathscr{E}^{\text{\name{}}}_R$ to estimate whether a candidate threshold meets the recall target or not. To do so, note that a random variable $x\in S_+$, i.e., an observed sample with positive label, is a uniform sample from $D_+$ so  $$\mathds{E}[\mathds{I}[\mathcal{S}(x)\geq\rho]]=\sum_{x'\in D_+}\frac{\mathds{I}[\mathcal{S}(x')\geq\rho]}{|D^+|}=\mathfrak{R}_D(\rho).$$ Thus, the set
$
S_+^\rho=\{\mathds{I}[\mathcal{S}(x)\geq\rho]; x\in S_+\},
$
consists of i.i.d random variables with mean $\mathfrak{R}_D(\rho)$. We define $\mathscr{E}^{\text{\name{}}}_R$, analogous to $\mathscr{E}^{\text{\name{}}}$, to use observations in $S_+^\rho$ to estimate their true mean:
\rall{
\begin{align*}
 \mathscr{E}^{\text{\name{}}}_R(S, T, \rho, \alpha)&=\mathscr{T}(T, S_+^\rho, \alpha),
\end{align*}
where $\mathscr{T}$ is the hypothesis test from Lemma~\ref{lemma:hypothesis_test}.
$\mathscr{E}_R^{\textnormal{\name{}}}$ returns 1 if $\rho$ is estimated to meet the target, i.e., $\mathfrak{R}_D(\rho)\geq T$. It provides the same guarantees for RT (see \Cref{lemma:estimation_lb_r}) as $\mathscr{E}^{\text{\name{}}}$ for PT queries. }


\textbf{Threshold Selection}. \name{}$_R$-U selects cascade threshold
\begin{align}\label{eq:rho_prism_pu}
    \rho^{\text{\name{}}_R-U}_S=\max\{\rho; \rho\in \mathscr{C}, \mathscr{E}_R^{\text{\name{}}}(S, T, \rho, \delta)=1\},
\end{align}
the largest $\rho$ for which $\mathscr{E}^{\text{\name{}}}_R$ returns 1. To maximize precision, Eq.~\ref{eq:rho_prism_pu} selects the maximum over the thresholds since larger thresholds are expected to have higher precision (as Fig.~\ref{fig:prec_candidate} shows). 
$\rho^{\text{\name{}}}_S$ provides the required theoretical guarantees (see \Cref{lemma:prism_RU}). The proof shows that since recall is a monotonically decreasing function, we can take the minimum over $\rho\in \mathscr{C}$ without splitting the failure probability into $\frac{\delta}{|\mathscr{C}|}$ for each application of $\mathscr{E}^{\text{\name}}_R$. 

\subsubsection{\name{}$_R$-A with Adaptive Samples.} \techreport{When the total number of positives in $D$ is much smaller than $n$, a uniform sample will contain few, if any, positive labels, and \name{}$_R$-U will likely select a low-utility threshold. }\name{}$_R$-A complements \name{}$_R$-U with a pre-filtering step to focus sampling only on records expected to contain positive labels using the notion of \textit{positive density} that quantifies how positive labels are distributed. We first discuss \textit{positive density} before presenting \name$_R$-A.

\begin{figure}
    \begin{minipage}{0.49\linewidth}
        \includegraphics[width=1\linewidth]{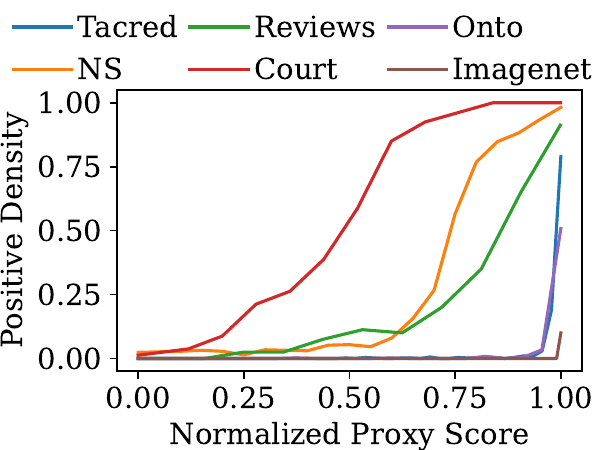}
        \caption{Positive density in real-world datasets}
        \label{fig:positive_scores_body}
    \end{minipage}
    \hfill
    \begin{minipage}{0.49\linewidth}
        \includegraphics[width=1\linewidth]{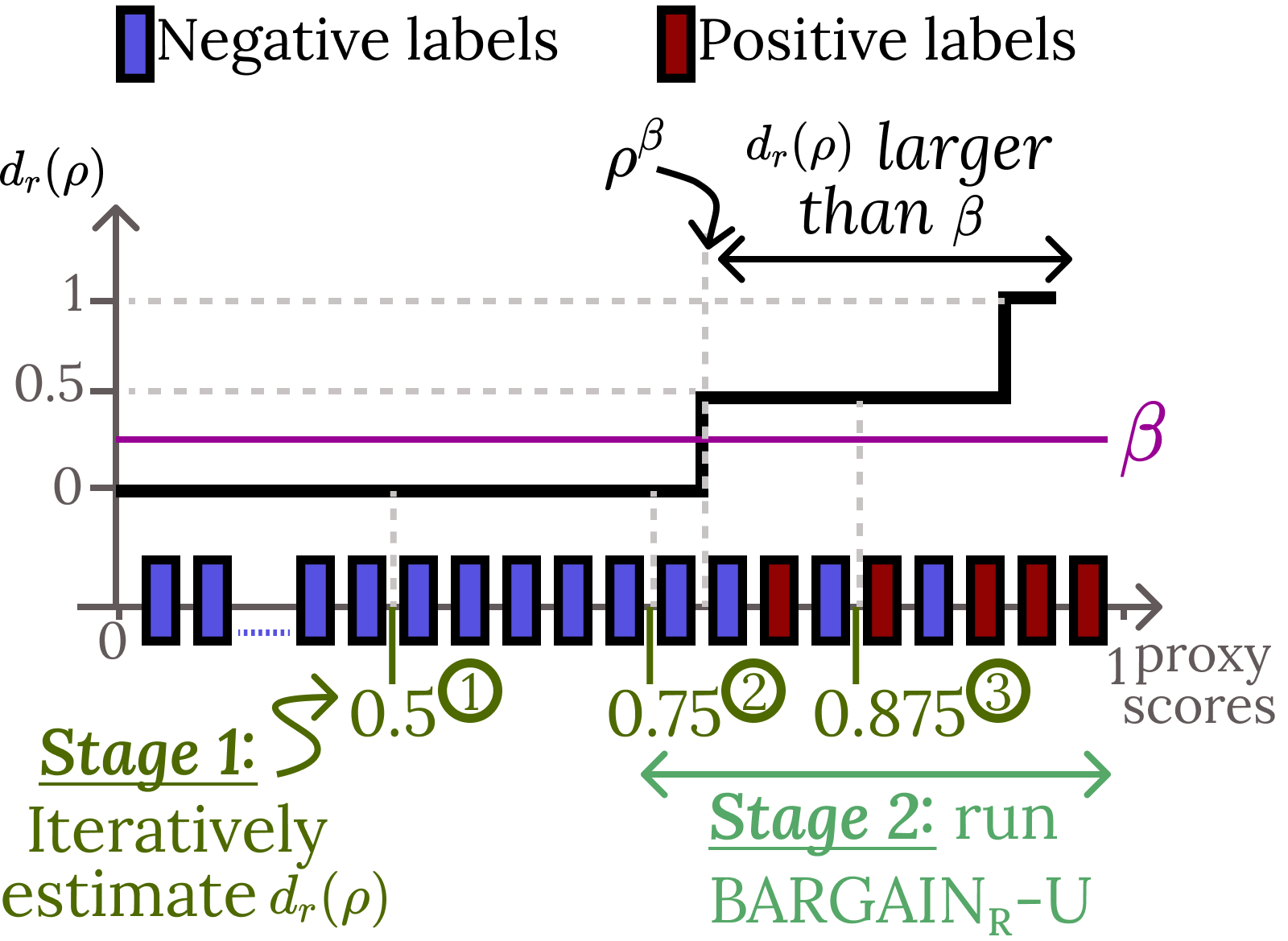}
        \caption{\name{}$_R$-A example}
        \label{fig:prism_RA}
    \end{minipage}
\end{figure}

\textbf{Positive Density}. To study how positive labels are distributed in our dataset, define \textit{positive density} at a threshold $\rho$, $d_r(\rho)$ as

$$d_r(\rho)=\frac{\sum_{x\in D^{\rho}_{r}}\mathds{I}[\mathcal{O}(x)=1]}{|D^{\rho}_{r}|},\, D^{\rho}_{r}=\{x;x\in D, \mathcal{S}(x)\in [\rho, \rho+r)\}.$$ 
where $D^{\rho}_{r}$ contains the records \textit{near} the threshold $\rho$ for a \textit{resolution parameter} $r$, and positive density denotes the fraction of records \textit{near} $\rho$ (i.e., in $D^{\rho}_{r}$) that are positive. Positive density can be seen as an approximation to $\mathds{P}(\mathcal{O}(x)=1|\mathcal{S}(x)=\rho)$ (known as correctness likelihood \cite{wang2023calibration}), the probability that a sample is positive given proxy score $\rho$, where the resolution $r$ is used to include enough records for this approximation. Fig.~\ref{fig:positive_scores_body} shows positive density for multiple real-world datasets. For datasets Onto and Imagenet, positive density for most proxy scores is zero, with non-zero density only at very large scores. \name{}$_R$-A uses this observation to improve utility.

\textbf{\name{}$_R$-A Algorithm}. \name{}$_R$-A takes advantage of this low positive density at small proxy scores to find a \textit{cutoff}, $\rho_P$, on proxy scores, such that positive labels are expected to have proxy scores only in $[\rho_P, 1]$. It then runs \name{}$_R$-U on $D^{\rho_P}$ (subset of $D$ with proxy score at least $\rho_P$) to obtain the cascade threshold. \name{}$_R$-A finds $\rho_P$, by estimating $d_r(\rho)$ at different thresholds to find the subset of $D$ with high positive density. It sets $\rho_P$ as the largest cutoff where any threshold $\rho$ with \textit{high positive density}, that is $d_r(\rho)\geq \beta$ for a  \textit{minimum positive density} parameter $\beta\geq 0$, is estimated to be in the range $\rho\in[\rho_P, 1]$. We discuss the role of $\beta$ later. 



\name{}$_R$-A is presented in Alg.~\ref{alg:prism_r_i}. It consists of two stages. The first stage performs a binary search on proxy scores to find $\rho_P$, and the second stage runs \name{}$_R$-U on $D^{\rho_P}$ to find the cascade threshold. To find the cutoff, $\rho_P$,  using binary search \name{}$_R$-A estimates whether $d_r(\rho)<\beta$ starting from the midpoint of possible values, $\rho=0.5$. To do so, it uses a function $\mathscr{E}_d^{\text{\name{}}}$, analogous to $\mathscr{E}^{\text{\name{}}}$ but now to estimate positive density, discussed later.
If at any iteration $\mathscr{E}_d^{\text{\name{}}}$ returns 1, the algorithm estimates that $\rho$ has low positive density, and it proceeds to check the density at $(1+\rho)/2$. This continues until the algorithm runs out of sampling budget or if it finds a threshold where it estimates that $d_r(\rho)\geq\beta$. Finally, the algorithm sets $\rho_P$ as the largest $\rho$ where it estimated $d_r(\rho)<\beta$. It runs \name{}$_R$-U on $D^{\rho_P}$ to find the final cascade threshold. \name{}$_R$-A assigns half of the sampling budget to the binary search and half to find the cascade threshold. It also divides $\delta$ into two and uses half for binary search for the cutoff and half for finding cascade threshold to bound the total probability of failure. 
An example of \name{}$_R$-A is presented in Fig.~\ref{fig:prism_RA} ($r$ is set so that $D^{\rho}_{r}$ contains 2 points at all $\rho$), where we see \name{}$_R$-A first estimates $d_r(\rho)$ at $\rho=0.5$, $\rho=0.75$ (where $d_r(\rho)<\beta$), and finally at $\rho=0.875$ where $d_r(\rho)\geq\beta$ so the search stops and  \name{}$_R$-A selects $\rho_P=0.75$. It then runs \name{}$_R$-U on points with proxy scores larger than 0.75. 

$d_r(\rho)$ can be seen as the precision of the set $D^{\rho}_{r}$, and we estimate whether $d_r(\rho)\geq \beta$ using the function $\mathscr{E}_d^{\textnormal{\name{}}}(S, \beta, \rho, \alpha)$, defined analogously to $\mathscr{E}^{\textnormal{\name{}}}$. We defer exact definition of $\mathscr{E}_d^{\textnormal{\name{}}}$ to \Cref{sec:density_est}. 
Here, we note that $\mathscr{E}_d^{\textnormal{\name{}}}$ guarantees that, if $d_r(\rho)\geq\beta$ and given a sample set $S$, $\mathds{P}(\mathscr{E}_d^{\textnormal{\name{}}}(S, \beta, \rho, \alpha)=1)\leq \alpha$, for any confidence parameter $\alpha$. That is, if $\rho$ meets the minimum positive density, $\mathscr{E}_d^{\textnormal{\name{}}}$ is unlikely to return 1.  

\begin{algorithm}[t]
\small
\begin{algorithmic}[1]
\State $k_1, k_2\leftarrow k/2$
\State $\delta_1, \delta_2\leftarrow \delta/2$
\State $\rho_{P}\leftarrow 0$, $\rho\leftarrow 0.5$
\While {$k_1\geq 0$}
    \State $S\leftarrow\emptyset$
    \While{$\mathscr{E}_d^{\text{\name{}}}(S, \beta, \rho, \delta_1)=0$ \textbf{and} $k_1\geq 0$}
        \State Sample a record uniformly from $D_r^\rho$ and add to $S$
        \State $k_{1} -= 1$
        \EndWhile
    \If{$\mathscr{E}_d^{\text{\name{}}}(S, \beta, \rho, \delta_1)=0$}
        \State \textbf{break}
    \EndIf
    \State $\rho_{P}\leftarrow \rho$, $\rho\leftarrow(1+\rho)/2$    
\EndWhile
\State\Return \name{}$_R$-U sampling $k_2$ points over $[\rho_{P}, 1]$ with $\delta_2$
\caption{\name{}$_R$-A}\label{alg:prism_r_i}
\end{algorithmic}
\end{algorithm}

\textbf{Guarantees and Role of $\beta$}. The guarantees \name{}$_R$-A provides depends on $\beta$. If $\beta=0$, \name{}$_R$-A provides the same guarantees as \name{}$_R$-U, but is also unlikely to improve on its utility, leading to poor utility when the total number of positive labels is low in a dataset. In fact, in \Cref{sec:utility_lb}, we show the following impossibility result: consider \textit{any approach} that samples records with probability monotonically increasing in the proxy scores and guarantees meeting the recall target on all possible datasets; such an approach \textit{must provide low precision for datasets with few positive records}. Due to this impossibility result, we allow users to optionally set $\beta>0$ to improve the utility of the solution but weaken the theoretical guarantees. To understand the role of $\beta$, for a dataset, $D$, let $\rho^\beta\in[0, 1]$ be the smallest proxy score such that $d_r(\rho)\geq\beta$ for all $\rho\in[\rho^\beta, 1]$ (Fig.~\ref{fig:prism_RA} shows an example). Define $\texttt{dense}_\beta(D)=\{x;x\in D, \mathcal{S}(x)\geq \rho^\beta\}$, and $\mathfrak{R}_{D}^\beta(\rho)=\mathfrak{R}_{\texttt{dense}_\beta(D)}(\rho)$, that is, $\mathfrak{R}_{D}^\beta$ is the recall on the set $\texttt{dense}_\beta(D)$, a subset of $D$ with minimum positive density $\beta$ at all proxy scores. \name{}$_R$-A guarantees $\mathfrak{R}_{D}^\beta(\rho)\geq T$ with high probability (see \Cref{lemma:prism_r_i}), but can potentially ignore any positive labels in $D$ but not in $\texttt{dense}_\beta(D)$, i.e., positive labels present where positive density is low. In real-world datasets, as Fig.~\ref{fig:positive_scores_body} shows, positive labels are densely distributed towards high proxy scores so that $\mathfrak{R}_{D}^\beta(\rho)$ and $\mathfrak{R}_{D}(\rho)$ are often very similar for small $\beta$. As such, our experiments show that setting $\beta$ as a fixed small value across all datasets allows us to obtain $\mathfrak{R}_{D}(\rho)\geq T$ in practice. Sec.~\ref{sec:exp:user_param} empirically evaluates the impact of $\beta$.

%% file: accuracy.tex
\begin{algorithm}[t]
\small
\begin{algorithmic}[1]
\State Sort $\mathscr{C}_M$ in descending order 
\For{$i$ \textbf{in} $|\mathscr{C}_M|$}
    \State $\rho\leftarrow \mathscr{C}_M[i]$
    \State $S\leftarrow \emptyset$ 
    \While{$\mathscr{E}_A^{\text{\name{}}}(S, T, \rho, \delta)=0$}\label{alg:precision_sample:prec_est}
        \If{$\texttt{avg}(S_A^\rho)-\texttt{std}(S_A^\rho)\geq T$ \textbf{and} $|S_A^\rho|\geq c$ }\label{alg:iterative_sample_accurcay:check_to_sample}
            \State \Return $\mathscr{C}_M[i-1]$
        \EndIf
        \State Sample a record uniformly from $D^\rho$ and add to $S$
    \EndWhile
\EndFor
\State\Return $\mathscr{C}_M[M]$
\caption{\name{}$_A$-A}\label{alg:iterative_sample_accurcay}
\end{algorithmic}
\end{algorithm}

\subsection{\name{}$_A$ for AT Queries}\label{sec:at}
\subsubsection{Model Cascade for AT Queries.}\label{sec:accuracy:cascade} We first introduce the model cascade framework for AT queries. For AT queries, because the output is not necessarily binary, the proxy score is defined more generally as the model's confidence in its output (not necessarily the positive class). Furthermore, AT queries do not come with a fixed predefined budget, but our goal is to determine the cascade threshold while minimizing cost. To perform AT queries following the model cascade framework, first, a subset $S$, $S\subseteq D$, is labeled and used to determine the cascade threshold, $\rho$. Given such a threshold, the set $D^\rho\setminus S$ is labeled by the proxy while the set $D\setminus(S\cup D^\rho)$ is labeled by the oracle. We denote the set of estimated labels by $\hat{Y}=\{\mathcal{P}(x_i); x_i\in D^\rho\setminus S\}\cup\{\mathcal{O}(x_i); x_i\in (D\setminus D^\rho)\cup S)\}$. The cost, i.e., the total number of oracle calls, is $C=|D|-|D^\rho\setminus S|$. Thus, under model cascade, for AT queries, we need to 
find a cascade threshold, $\rho$, by labeling a subset $S$ of $D$ with the oracle, such that $\hat{Y}$ as defined above meets $\mathds{P}(\mathfrak{A}(\hat{Y})\geq T)\geq 1-\delta$, while cost, $C=|D|-|D^\rho\setminus S|$, is minimized. 
For convenience of notation, for a set $S\subseteq D$, we define:
$$\mathfrak{A}_S(\rho)=\frac{\sum_{x\in S^\rho}\mathds{I}[\mathcal{O}(x)=\mathcal{P}(x)]}{|S^\rho|}.$$
$\mathfrak{A}_S(\rho)$ is the accuracy of the proxy model \textit{when processing records in }$S^\rho$ \textit{only}. We study $\mathfrak{A}_D(\rho)$ to provide our theoretical guarantees. 

\subsubsection{\name{}$_A$-A.} Our solution to AT queries is similar to PT queries. We use the same adaptive sampling procedure as \name{}$_P$-A, but our method differs from \name{}$_P$-A in two ways. First, we estimate accuracy at each iteration, not precision, by defining an estimation function $\mathscr{E}^{\text{\name{}}}_A$, analogous to $\mathscr{E}^{\text{\name{}}}$, but instead to estimate accuracy. Second, recall that for PT queries, we continuously sampled records until we ran out of oracle budget. For AT queries, we do not have a fixed budget. Instead, we stop sampling when we determine \textit{it's not worth sampling more to obtain a better estimate} at the threshold under consideration. We first discuss our estimation function before presenting the \name{}$_A$-A algorithm. 

\textbf{Estimation}. We present a new estimation function $\mathscr{E}_A^{\text{\name{}}}$ to estimate accuracy, in place of $\mathscr{E}^{\text{\name{}}}$ for precision. Recall that our adaptive sampling procedure for PT queries, when considering a threshold $\rho$, samples an $x\in S$ uniformly from $D^\rho$ (see line~\ref{alg:prism_pA:sample} in Alg.~\ref{alg:iterative_sample_precision}). Using the same sampling procedure and random variables $x\in S$, but considering $\mathds{I}[\mathcal{O}(x)=\mathcal{P}(x))]$,  we have $$\mathds{E}[\mathds{I}[\mathcal{O}(x)=\mathcal{P}(x))]]=\sum_{x'\in D^\rho}\frac{\mathds{I}[\mathcal{O}(x')=\mathcal{P}(x'))]}{|D^\rho|}=\mathfrak{A}_D(\rho).$$ Thus, defining the set, 
        $S_A^\rho=\{\mathds{I}[\mathcal{O}(x)=\mathcal{P}(x))]; x\in S^\rho\}$,
to estimate whether $\mathfrak{A}_D(\rho)\geq T$ at a threshold $\rho$ we use the observations in $S_A^\rho$ to estimate their true mean:
\rall{
\begin{align*}
        \mathscr{E}_A^{\textnormal{\name{}}}(S, T, \rho, \alpha)&=\mathscr{T}(T, S_A^\rho, \alpha),
\end{align*}
where $\mathscr{T}$ is the hypothesis test from Lemma~\ref{lemma:hypothesis_test}. $\mathscr{E}_A^{\textnormal{\name{}}}$ returns 1 if $\rho$ is estimated to meet the target. It provides the same guarantees for AT (see \Cref{lemma:_anytime_no_replace_lb_accuracy}) as $\mathscr{E}^{\text{\name{}}}$ for PT queries. 
}

\textbf{\name{}$_A$-A Algorithm}. We present \name{}$_A$-A in Alg.~\ref{alg:iterative_sample_accurcay}. The algorithm follows the same sampling procedure as \name{}$_P$-A, iteratively considering candidate thresholds in decreasing order and estimating whether each threshold meets the accuracy target. It now uses $\mathscr{E}_A^{\textnormal{\name{}}}$ as the estimation function instead of $\mathscr{E}^{\textnormal{\name{}}}$ to estimate accuracy. Furthermore, as the algorithm iterates through the thresholds, we stop sampling at a threshold $\rho$ if $\texttt{avg}(S_A^\rho)-\texttt{std}(S_A^\rho)\geq T$, but do so only after $|S_A^\rho|\geq c$ for some parameter $c$ so that the mean and standard deviation of $S_A^\rho$ are meaningful. 
This follows the intuition that if $T$ is within one standard deviation of the mean of $S_A^\rho$, it will be difficult to estimate if the true mean, $\mathfrak{A}_D(\rho)$, of the observation in $S_A^\rho$, meets the target without needing a large number of new samples. Thus, we terminate the algorithm and return the smallest threshold estimated to meet the target so far as the candidate threshold. This modification is done in Line~\ref{alg:iterative_sample_accurcay:check_to_sample} of Alg.~\ref{alg:iterative_sample_accurcay} to decide if the algorithm should continue sampling or not. \name{}$_A$-A provides the required theoretical guarantees (see \Cref{lemma:prism_AA}). 

\subsubsection{\name{}$_A$-M and Other Details.} We design \name{}$_A$-M, an extension to \name{}$_A$-A for classification tasks to \textit{set the cascade thresholds per class} which may help when proxy scores for some classes are more helpful than others. If there are $r$ classes, \name{}$_A$-M runs \name{}$_A$-A  $r$ times, and determines a different cascade threshold for each class. When the proxy predicts a class, the cascade threshold for that specific class is used for prediction. Details are discussed in \Cref{sec:prism_a_M}. Finally, \Cref{sec:acc:oracle}, discuss how to extend the analysis to take the data records labeled by the oracle into account. 

%% file: discussion.tex
\rone{\section{Parameter Setting}\label{sec:hyperparam}}
\vspace{-0.05cm}
\rone{\paramsettingmain{}}

%% file: exp.tex
\begin{table}[t]
\setlength{\tabcolsep}{3pt}
\hspace*{-5pt}
    \footnotesize
    \begin{tabular}{c c c c c c c c c}
    \toprule
& \textbf{Review}& \textbf{Court}& \textbf{Screen}& \textbf{Wiki} & \textbf{Onto}& \textbf{Imagenet}& \textbf{Tacred}& \textbf{NS}\\
\midrule
$\mathbf{n}$& 855& 1,000& 1,000& 1,000& 11,165& 50,000& 22,631& 973,085\\
$\mathbf{\frac{n^+}{n}}$& 0.23& 0.59& 0.22& 0.25& 0.02& 0.001& 0.02& 0.29\\
\bottomrule
\end{tabular}
    \caption{Dataset characteristics}
    \label{tab:datasets}
\end{table}
\vspace{-0.03cm}
\section{Experiments}\label{sec:exp}
\vspace{-0.05cm}
\techreport{
We present our experimental setup in Sec~\ref{sec:exp:setup}, comparison with baselines in Sec.~\ref{sec:exp:data}, analysis of sensitivity of approaches to user parameters in Sec.~\ref{sec:exp:user_param}, and robustness of the approaches to random noise and adversarial settings in Sec.~\ref{sec:exp:noise}. 
}
\vspace{-0.05cm}
\subsection{Setup}\label{sec:exp:setup}
\vspace{-0.05cm}
\textbf{Datasets and Tasks}. We perform experiments on 8 different datasets. We use all 4 of the datasets used in \cite{kang2020approximate} (obtained from \cite{supgdatasets}), Imagenet, night-street (NS), Tacred, and OntoNotes (Onto), with the first two datasets on image classification, and the latter two on text classification. These datasets use non-LLM deep learning models for image and text processing. To evaluate our approach for data processing using LLMs, we additionally consider 4 new datasets: Reviews (Steam game reviews, obtained from Kaggle \cite{gamereviews}), Court (US court opinions since 1970 \cite{court}), Screenplay (popular movie scripts from Kaggle \cite{Screenplay}), and Wiki (Wikipedia talk page discussions amongst editors \cite{chang2020convokit}). We randomly sampled 1,000 examples from each source, though Reviews contains only 855 examples because many reviews did not pass the OpenAI toxicity filter. The LLM tasks involve determining game reference comparisons (Reviews), court ruling reversals (Court), decisions based on false information (Screenplay), and whether discussions led to edit reversions (Wiki). The specific prompts used are described in \Cref{app:taskprompts}. Throughout, gpt4o-mini is used as the proxy model and gpt-4o as the oracle. \Cref{tab:datasets} shows the number of records, together with $\frac{n^+}{n}$ denoting the fraction of records that have positive labels in each dataset.

\begin{table*}[t]
\vspace{-0.3cm}
    \small
    \centering
    \begin{tabular}{c c c c c c c c c}
\toprule&\textbf{Reviews}&\textbf{Court Opinion}&\textbf{Screenplay}&\textbf{Wiki Talk}&\textbf{Onto}&\textbf{Imagenet}&\textbf{Tacred}&\textbf{NS}\\\midrule
\multicolumn{9}{c}{\underline{\textit{(a) Percentage of Oracle Calls Avoided for AT Queries}}}\\
\textbf{SUPG}&3.2&24.4&1.5&4.4&73.5&84.7&80.1&7.5\\ \textbf{Naive}&0.0 (\textcolor{burgundy}{-100.0})&0.0 (\textcolor{burgundy}{-100.0})&6.8 (\textcolor{cadmiumgreen}{+362})&0.0 (\textcolor{burgundy}{-100.0})&73.6 (\textcolor{cadmiumgreen}{+0.1})&84.6 (\textcolor{burgundy}{-0.1})&79.9 (\textcolor{burgundy}{-0.2})&0.0 (\textcolor{burgundy}{-100.0})\\ \textbf{\name{}$_A$-A}&\textbf{41.8} (\textbf{\textcolor{cadmiumgreen}{+1218}})&48.0 (\textcolor{cadmiumgreen}{+96.5})&0.0 (\textcolor{burgundy}{-100.0})&\textbf{48.1} (\textbf{\textcolor{cadmiumgreen}{+984}})&97.7 (\textcolor{cadmiumgreen}{+32.9})&\textbf{99.7} (\textbf{\textcolor{cadmiumgreen}{+17.8}})&97.0 (\textcolor{cadmiumgreen}{+21.2})&65.3 (\textcolor{cadmiumgreen}{+765})\\ \textbf{\name{}$_A$-M}&36.7 (\textcolor{cadmiumgreen}{+1060})&\textbf{58.6} (\textbf{\textcolor{cadmiumgreen}{+139}})&\textbf{11.1} (\textbf{\textcolor{cadmiumgreen}{+655}})&42.9 (\textcolor{cadmiumgreen}{+865})&\textbf{98.9} (\textbf{\textcolor{cadmiumgreen}{+34.6}})&\textbf{99.9} (\textbf{\textcolor{cadmiumgreen}{+18.0}})&\textbf{99.2} (\textbf{\textcolor{cadmiumgreen}{+24.0}})&\textbf{75.7} (\textbf{\textcolor{cadmiumgreen}{+903}})\\ 
\midrule
\multicolumn{9}{c}{\underline{\textit{(b) Observed Recall for PT Queries}}}\\
\textbf{SUPG}&59.4&75.6&44.2&54.9&49.1&89.7&33.5&10.2\\ \textbf{Naive}&47.0 (\textcolor{burgundy}{-20.9})&39.9 (\textcolor{burgundy}{-47.2})&40.2 (\textcolor{burgundy}{-9.1})&40.3 (\textcolor{burgundy}{-26.6})&3.4 (\textcolor{burgundy}{-93.1})&0.7 (\textcolor{burgundy}{-99.2})&1.7 (\textcolor{burgundy}{-95.0})&3.9 (\textcolor{burgundy}{-61.6})\\ \textbf{\name{}$_P$-U}&54.2 (\textcolor{burgundy}{-8.7})&\textbf{86.5} (\textbf{\textcolor{cadmiumgreen}{+14.4}})&40.2 (\textcolor{burgundy}{-9.2})&43.1 (\textcolor{burgundy}{-21.5})&3.5 (\textcolor{burgundy}{-92.8})&1.0 (\textcolor{burgundy}{-98.9})&1.7 (\textcolor{burgundy}{-95.1})&4.1 (\textcolor{burgundy}{-60.1})\\ \textbf{\name{}$_P$-A}&\textbf{89.0} (\textbf{\textcolor{cadmiumgreen}{+49.8}})&84.2 (\textcolor{cadmiumgreen}{+11.4})&\textbf{73.4} (\textbf{\textcolor{cadmiumgreen}{+65.9}})&\textbf{86.3} (\textbf{\textcolor{cadmiumgreen}{+57.2}})&\textbf{88.2} (\textbf{\textcolor{cadmiumgreen}{+79.8}})&\textbf{100} (\textbf{\textcolor{cadmiumgreen}{+11.5}})&\textbf{61.5} (\textbf{\textcolor{cadmiumgreen}{+83.3}})&\textbf{44.6} (\textbf{\textcolor{cadmiumgreen}{+337}})\\ 
\midrule
\multicolumn{9}{c}{\underline{\textit{(c) Observed Precision for RT Queries}}}\\
 \textbf{SUPG}&34.3&77.5&27.9&45.9&13.2 \textcolor{orange}{[MT]}&90.3&11.1&45.3\\ \textbf{Naive}&22.9 (\textcolor{burgundy}{-33.1})&74.8 (\textcolor{burgundy}{-3.5})&21.8 (\textcolor{burgundy}{-21.8})&24.8 (\textcolor{burgundy}{-46.0})&2.5 (\textcolor{burgundy}{-81.0})&0.1 (\textcolor{burgundy}{-99.9})&2.4 (\textcolor{burgundy}{-78.8})&31.3 (\textcolor{burgundy}{-31.0})\\ \textbf{\name{}$_R$-U}&\textbf{40.9} (\textbf{\textcolor{cadmiumgreen}{+19.2}})&\textbf{82.7} (\textbf{\textcolor{cadmiumgreen}{+6.7}})&\textbf{32.6} (\textbf{\textcolor{cadmiumgreen}{+16.8}})&\textbf{52.3} (\textbf{\textcolor{cadmiumgreen}{+14.0}})&2.5 (\textcolor{burgundy}{-81.0})&0.1 (\textcolor{burgundy}{-99.9})&2.4 (\textcolor{burgundy}{-78.8})&\textbf{59.1} (\textbf{\textcolor{cadmiumgreen}{+30.5}})\\ \textbf{\name{}$_R$-A}&39.7 (\textcolor{cadmiumgreen}{+15.9})&\textbf{82.0} (\textbf{\textcolor{cadmiumgreen}{+5.9}})&\textbf{33.0} (\textbf{\textcolor{cadmiumgreen}{+18.2}})&50.9 (\textcolor{cadmiumgreen}{+10.9})&\textbf{28.0} (\textbf{\textcolor{cadmiumgreen}{+112}})&\textbf{97.8} (\textbf{\textcolor{cadmiumgreen}{+8.3}})&\textbf{22.0} (\textbf{\textcolor{cadmiumgreen}{+97.8}})&56.1 (\textcolor{cadmiumgreen}{+23.8})\\ \bottomrule
    \end{tabular}
    \caption{Observed Utility for AT, PT and RT Queries with Target $T=0.9$. \textcolor{cadmiumgreen}{+}/\textcolor{burgundy}{-} shows percentage change over SUPG, \textcolor{orange}{[MT]} \rone{means the method missed the target, and all methods otherwise met the target}.}
    \label{tab:all_res}
\end{table*}

\textbf{Methods}. We compare \name{} against two baselines: SUPG~\cite{kang2020approximate}, the state-of-the-art model cascade method with \textit{asymptotic} guarantees, and Naive, a baseline we designed to provide the same guarantees as \name{}, but using simple statistical tools. We use the open-source implementation of SUPG \cite{supgdatasets}. SUPG was only designed for PT and RT queries, and uses importance sampling independent of target and data characteristics. To extend it to AT queries, we use the PT method to solve the AT query by changing the calculated metric during estimation. Since AT queries do not use an apriori budget while SUPG takes a fixed budget as input, we run SUPG for 10 different budgets 
and report the result with the best utility. Naive, for all queries, takes a uniform sample and uses Hoeffding's inequality to estimate quality, as in Sec.~\ref{sec:pt:naive}. For AT queries, we use the same method as for SUPG discussed above to determine the sample size for Naive. \rone{\naivemain{}} 


\textbf{Metrics}. To evaluate the methods for PT and RT queries, we respectively report the true recall and precision for the methods at the threshold returned by the algorithm. For AT queries, we count the total number of points, $n_{proxy}$ that only the proxy was evaluated on, and report the fraction $\frac{n_{proxy}}{n}$ as the \textit{percentage of oracle calls avoided}.  Reported results are averaged across 50 runs. \rone{We performed 50 runs to empirically evaluate whether the methods meet the quality target with the desired probability. We also report observed variances across the 50 runs in Appx.~\ref{appx:variance_exp}; our results show \name{} has lower variance than SUPG.}

\textbf{Default Parameters}. Unless otherwise stated, we set $T=0.9$, $\delta=0.1$ and $k=400$. For \name{}$_P$ variants and Naive $M=20$ for PT and AT queries, and for \name{}$_R$-A, we set $\beta=0.02$ and $r=150$. We show the impact of parameters in Sec.~\ref{sec:exp:user_param}, \rone{with further results in Appx.~\ref{appx:hyperparams}}. We use SUPG with default parameters  \cite{kang2020approximate}.

\subsection{Comparison across Datasets}\label{sec:exp:data}
Table~\ref{tab:all_res} shows baseline comparison across all datasets, showing significant benefits to \name{} over SUPG and the Naive method. First, consider AT queries. We see that both \name{}$_A$ variants significantly reduce oracle usage compared with SUPG. For many datasets, SUPG only uses the proxy model prediction less than 10\% of the time, while \textbf{\name{} increases proxy model usage up to 10$\times$ over SUPG}. Note that there is no clear winner between \name{}$_A$-A and \name{}$_A$-M, where \name{}$_A$-A chooses a single threshold for all classes while \name{}$_A$-M chooses different thresholds for different classes. \name{}$_A$-A performs better when a single cascade threshold for all classes provides sufficiently good utility, since this threshold can be determined with fewer labels. However,  if a single threshold cannot be applied to all classes (e.g., in Screenplay), \name{}$_A$-M, which chooses a per-class threshold, will perform better. 

For PT and RT queries, we see \name{}$_P$-A and \name{}$_R$-A significantly improve upon SUPG. We see that \name{}$_P$-U and \name{}$_R$-U, the \name{} variants with uniform sampling do improve upon the Naive approach but, depending on the dataset, often perform worse than SUPG which uses importance sampling. This is particularly visible on Onto, Imagenet and Tacred datasets that, as Table~\ref{tab:datasets} shows, have very low number of positive labels compared to the full dataset, and thus, uniform sampling is unlikely to find where the positives are located. Nonetheless, improved sampling procedures in \name{}$_P$-A and \name{}$_R$-A help \name{} provide the significant advantage over SUPG, \textbf{improving recall in PT queries by up to 80\% and precision in RT queries by up to 98\%.} 

\begin{figure}[t]
        \centering
\vspace{-5pt}
\hspace{-10pt}  
    \begin{minipage}{0.36\linewidth}
        \centering
        \includegraphics[width=0.9\linewidth]{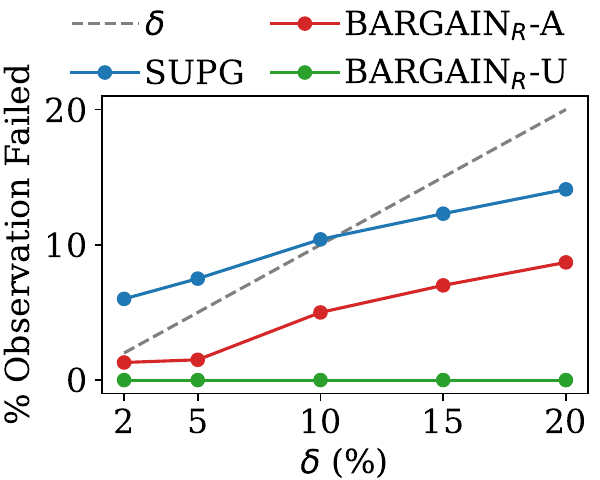}
        \caption{Meeting target in Onto Dataset}
        \label{fig:meeting_target_onto}
    \end{minipage}
    \begin{minipage}{0.66\linewidth}
        \centering
        \includegraphics[width=\linewidth]{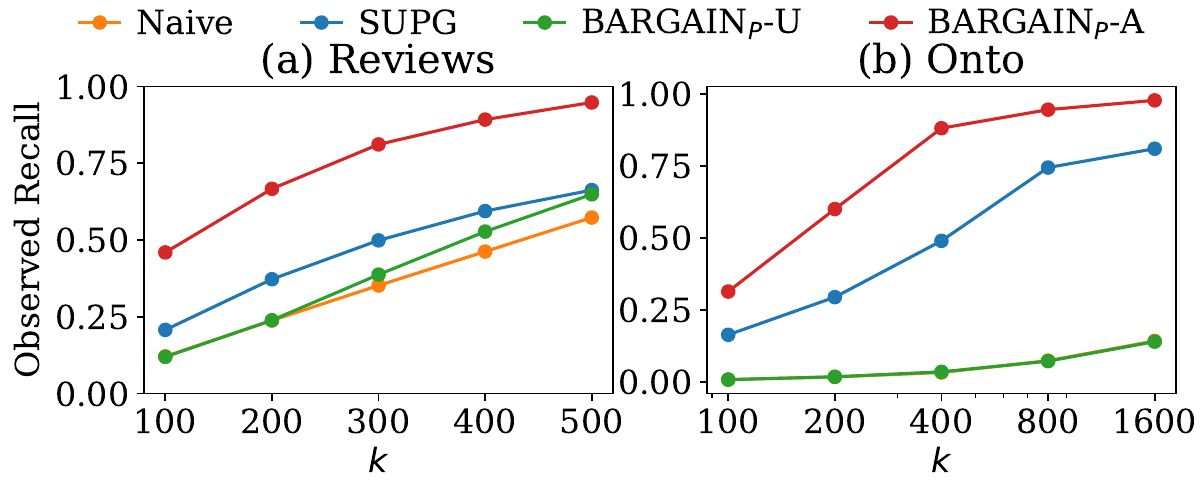}
        \caption{Impact of $k$ in PT Queries}
        \label{fig:pt_res_k}
    \end{minipage}
    \hfill
    \vspace{-0.2cm}
\end{figure}

All results in Table~\ref{tab:all_res} empirically meet the required target except for SUPG on Onto dataset for RT queries. Fig.~\ref{fig:meeting_target_onto} further illustrates this, showing percentage of runs across 1,000 runs where SUPG returned recall below the target, $T=0.9$. When $\delta$ is small, i.e., the allowed failure probability is low, SUPG fails to provide the desired guarantees. In Sec.~\ref{sec:exp:noise}, we show that this is not an isolated incident and construct datasets where SUPG frequently misses the target. 

\begin{figure*}[t]
\vspace{-0.2cm}
\centering
    \begin{minipage}{0.325\textwidth}
        \centering
        \includegraphics[width=1\linewidth]{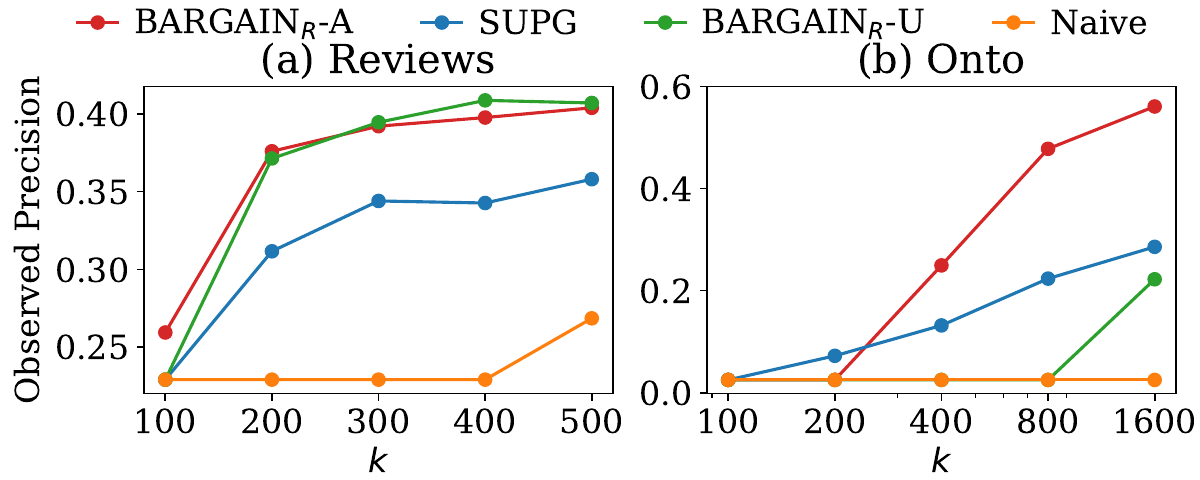}
        \caption{Impact of $k$ in RT Queries}
        \label{fig:rt_res_k}
    \end{minipage}
    \begin{minipage}{0.325\textwidth}
        \centering
        \includegraphics[width=1\linewidth]{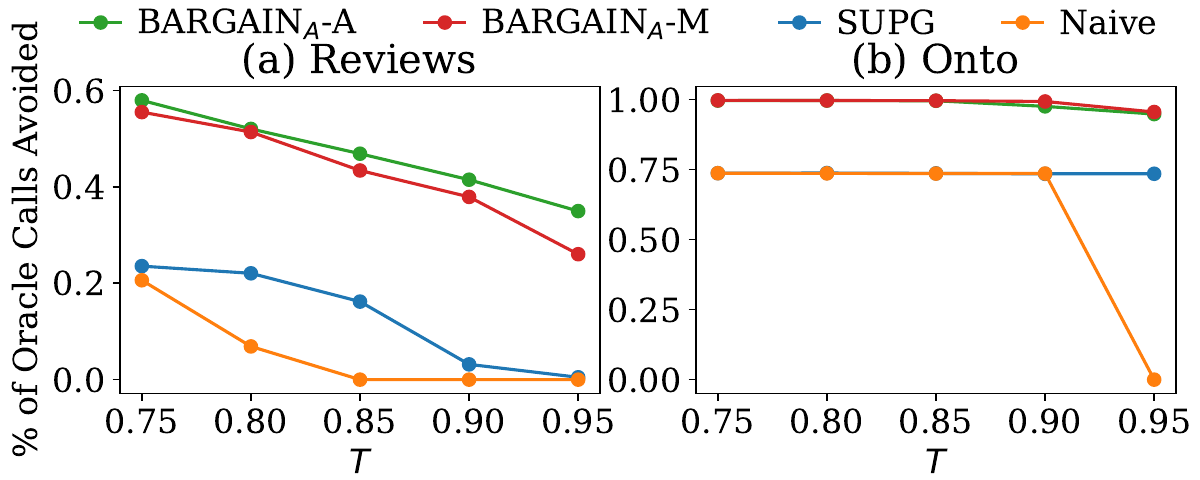}
        \caption{Impact of $T$ in AT Queries}
        \label{fig:at_res_T}
    \end{minipage}
    \begin{minipage}{0.325\textwidth}
        \centering
        \includegraphics[width=1\linewidth]{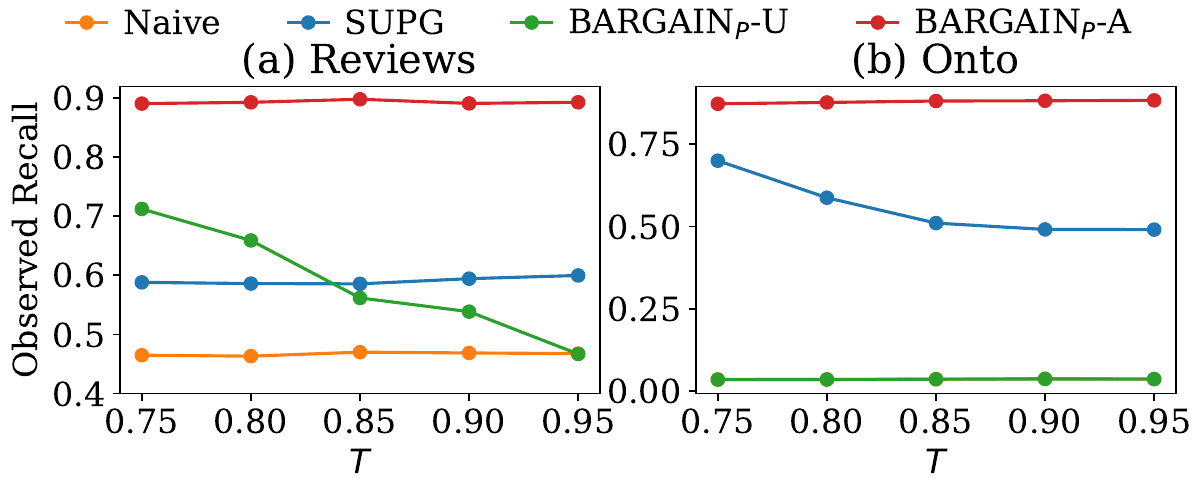}
        \caption{Impact of $T$ in PT Queries}
        \label{fig:pt_res_T}
    \end{minipage}
\end{figure*}

\begin{figure*}[t]
\vspace{-0.05cm}
    \centering
    \begin{minipage}{0.49\linewidth}
        \includegraphics[width=0.33\linewidth]{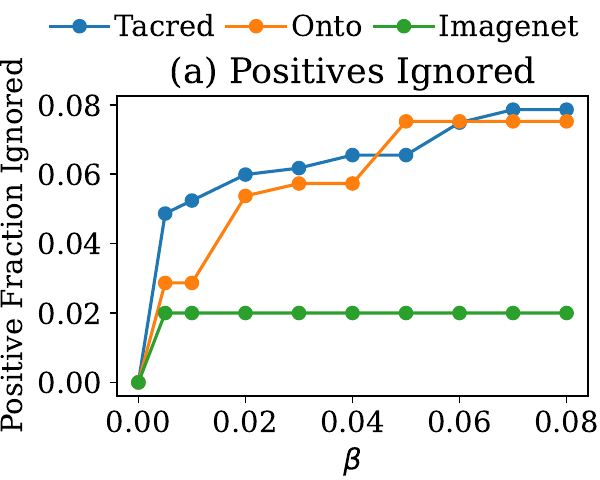}    \includegraphics[width=0.66\linewidth]{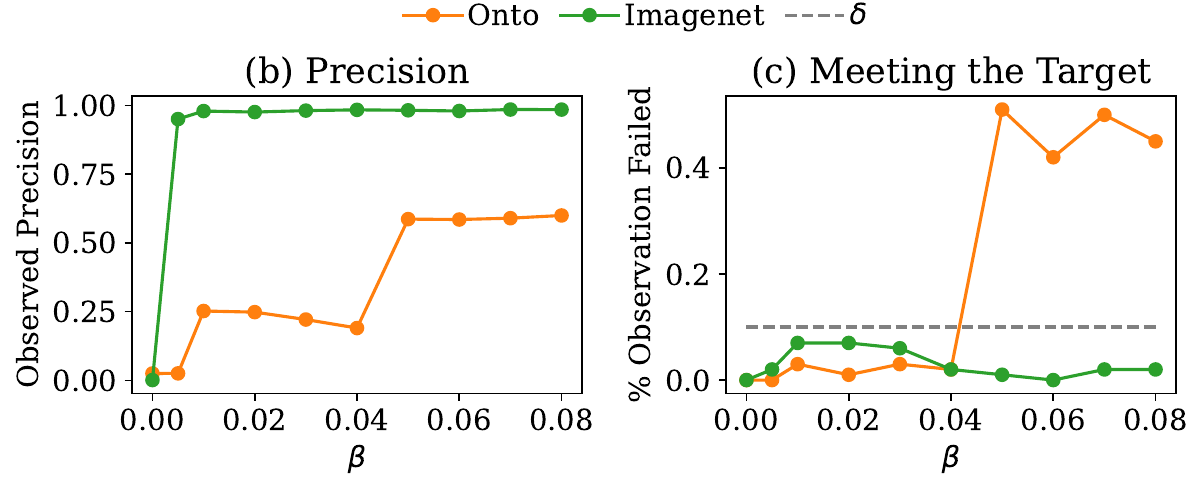}
        \caption{Impact of $\beta$}
        \label{fig:beta_exp}        
    \end{minipage}
    \hfill
    \begin{minipage}{0.16\linewidth}
        \centering
        \includegraphics[width=\linewidth]{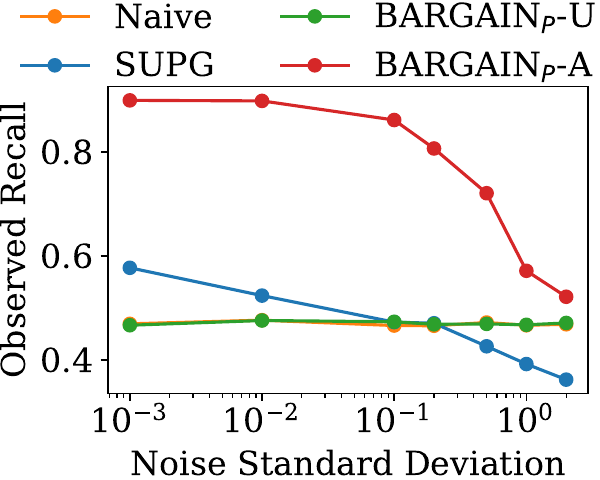}    
        \caption{Impact of noise in PT Queries}
        \label{fig:beta_noise_precision}
    \end{minipage}
    \hfill
    \begin{minipage}{0.16\linewidth}
        \centering
        \includegraphics[width=\linewidth]{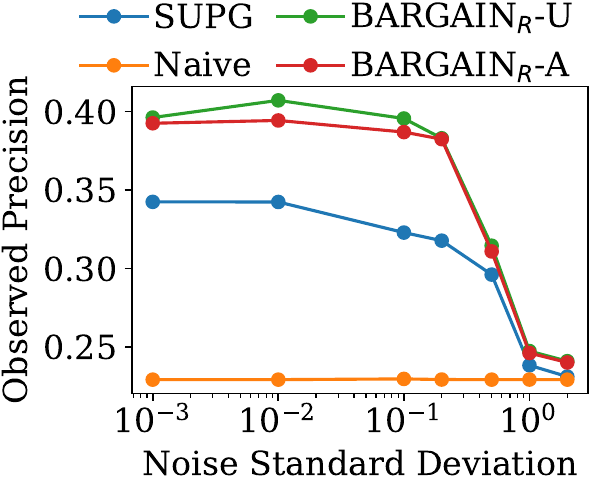}
        \caption{Impact of noise in RT Queries}
        \label{fig:beta_noise_recall}
    \end{minipage}
    \hfill
    \begin{minipage}{0.16\linewidth}
        \centering
        \includegraphics[width=\linewidth]{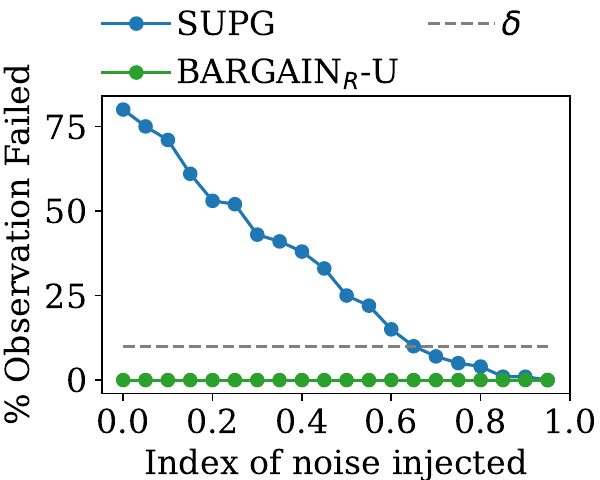}
        \caption{Meeting target in Noisy Data}
        \label{fig:meeting_target_adverserial}
    \end{minipage}
\end{figure*}
\vspace{-0.1cm}
\subsection{Sensitivity to Parameters}\label{sec:exp:user_param}
\vspace{-0.1cm}
Here, we use Review, our dataset for LLM-powered data processing, and Onto, a dataset from \cite{kang2020approximate} with a low positive rate (see \Cref{tab:datasets} to study the sensitivity of approaches to user parameters. 

\textbf{Varying $k$}. Fig.~\ref{fig:pt_res_k}-\ref{fig:rt_res_k} show the impact of budget size, $k$, for  AT and PT queries on two datasets, Reviews and Onto.  \name{}$_P$-A and \name{}$_R$-A outperform SUPG across budget sizes, although the gap between methods is smaller when the budget size is small, cases where  no method provide good utility due to limited budget.

\textbf{Varying $T$}. Figs.~\ref{fig:at_res_T}-\ref{fig:pt_res_T} show the sensitivity of \name{} to the target $T$ for AT and PT queries (results for RT queries are similar and omitted here for the sake of space) on Reviews and Onto datasets. \name{} outperforms SUPG across values of $T$.

\textbf{Varying $\beta$}. For RT queries, \name{}$_R$-A depends on the parameter $\beta$, the minimum positive density parameter. For datasets with large positive rates, i.e., all datasets except Tacred, Onto and Imagenet, any small value of $\beta$, e.g., $\beta<0.1$, does not impact the performance. The dataset's positive density at most thresholds is larger than $\beta$, so \name{}$_R$-A simply performs \name{}$_R$-U on the entire domain without reducing the range (i.e., sets $\rho^{\beta}=0$). Here, we investigate the impact of $\beta$ on datasets with low positive rates, i.e., Tacred, Onto, and Imagenet. First, in Fig.~\ref{fig:beta_exp} (a), we show what fraction of the positive records may be \textit{ignored} by \name{}$_R$-A at a particular $\beta$, that is the ratio between the number of positive labels in $D\setminus \texttt{dense}_\beta(D)$ and the total number of positives in $D$ (see Sec.~\ref{sec:rt_short} for definitions). 
As Fig.~\ref{fig:beta_exp} (a) shows, in real-world datasets, only a small fraction of the positive labels are not present in a dense subset of $D$, justifying use of dense positive labels to provide guarantees in real-world settings. Fig.~\ref{fig:beta_exp} (b) and (c) evaluate \name{}$_R$-A at different $\beta$ values (results for Tacred are similar to Onto are omitted for visual clarity), showing significant improvements in precision even at very low $\beta$ values. However, as $\beta$ increases, \name{}$_R$-A may stop meeting the target, $T$, as it starts to ignore too many positive points. For Onto, this happens at $\beta\geq0.05$. We note that even though \name{}$_R$-A stops to meet the target at the probability needed by $\delta$, the observed recall is often very close to the target since the deviation from the target is bounded based on the fraction of positive labels ignored by the algorithm, which as Fig.~\ref{fig:beta_exp} (a) shows, is small.

\rone{\textbf{Other parameters}. Additional parameters across \name{} variants are $M$, $c$ and $\eta$. Appx.~\ref{appx:hyperparams} shows experiments on the impact of these parameters on the utility. We observe that changing the parameters has little impact on the utility of \name{} and a large set of values perform well across datasets. }

\vspace{-0.1cm}
\subsection{Robustness}\label{sec:exp:noise}
\vspace{-0.1cm}
\textbf{Impact of Random Noise in Proxy Scores}. We add Gaussian noise to the proxy scores in the Review datasets to study the impact of model calibration. The larger the magnitude of the noise is, the less correlated the proxy scores will be with the correctness of proxy prediction. The results of this experiment are plotted in Figs.~\ref{fig:beta_noise_precision} and \ref{fig:beta_noise_recall}, where we measure utility as we increase the standard deviation of the noise injected. In general, \name{} outperforms other methods. When the noise magnitude becomes large, the proxy scores lose all correlation with proxy correctness, and all approaches perform similarly to Naive and provide little utility.

\textbf{Adversarial Setting for Meeting the Quality Target}. In this experiment, we adversarially modify the Imagenet dataset to show how frequently SUPG, which achieves only asymptotic guarantees, can fail to meet the target and make the case for strong non-asymptotic theoretical guarantees that \name{} achieves. Specifically, on the Imagenet dataset, we sort the records in the increasing order of proxy scores and denote the $i$-th record in this order as $x_i$. To modify the dataset, we change the label of records $x_i$,..., $x_{i+100}$ and set them to be positive, and vary $i$ in this experiment. For instance, when $i=0$, we set the label of the 100 records with the lowest proxy scores to be positive. Fig.~\ref{fig:meeting_target_adverserial} shows the result of this experiment. We see that SUPG misses the target more than 75\% of the times, significantly more than the 10\% allowed (since $\delta=0.1$). On the other hand, \name{}$_P$-U meets the target throughout. Putting this in the context of the results in Table~\ref{tab:all_res},  this lack of guarantees for SUPG explains why it can provide high utility on datasets with a low number of positives,  but \name{}$_P$-U that does provide guarantees provides poor utility. Our method \name{}$_R$-A strikes a balance between the two, by allowing the user to quantifiably (through the parameter $\beta$) relax the guarantee requirement to provide high utility on real-world datasets, unlike SUPG that may unpredictably fail to meet the target. We also note that although the results here are in a synthetically generated setting, existing work shows that LLMs may be uncalibrated in practice \cite{kapoor2024calibration,jiang2021can}, and thus rigorous theoretical guarantees are needed to ensure robustness.

%% file: related_work.tex
\vspace{-0.05cm}
\section{Related Work}\label{sec:rel_work}
\vspace{-0.05cm}

\textbf{LLM-Powered Data Management.} LLMs have revolutionized data management research and applications~\cite{fernandez2023large}, reshaping how our community approaches longstanding challenges. LLMs have been broadly used in two ways: (1), developing ``point'' solutions for challenging problems such as data discovery~\cite{freirelarge, wang2023solo}, data extraction and cleaning~\cite{arora2023language, narayan2022can, vos2022towards, naeem2024retclean, lee2025semantic}, query planning~\cite{urban2024demonstrating}, and text to SQL~\cite{pourreza2024chase}; and (2) creating flexible query processing frameworks that incorporate LLMs in open-ended ways~\cite{shankar2024docetl, patel2024lotus, liu2024declarative, anderson2024design, urban2024eleet, lin2024towards, wang2025aop, zeighami2025llm}. \name{} reduces costs while guaranteeing quality, making it applicable to any LLM-based component across these systems.

\textbf{Cost-Efficient ML-Powered Data Processing}. Many techniques have been used to reduce the cost of ML-powered data processing. Some strategies include optimizing for specific {\em types} of queries, like aggregations~\cite{kang13blazeit}, or building {\em indexes} to reduce online query processing time~\cite{kang2022tasti, bastani2022otif}. More recently, \cite{jo2024smart} performs profiling to estimate model accuracies for different tasks to decide which model to use, and \citet{huang2025thriftllm} ensembles various model answers to generate the final output. {\em Model cascade}, which routes queries through cheaper proxy models before using expensive oracle models, has been widely adopted in traditional ML and deep learning, particularly for video analytics~\cite{kang2017noscope, lu2018accelerating, kang13blazeit, anderson2019physical, cao2022figo, ding2022efficient}. In particular, recent work \cite{chen2023frugalgpt,patel2024lotus} has used this framework for LLM-powered data processing; \cite{chen2023frugalgpt} without providing guarantees while \cite{patel2024lotus} uses SUPG~\cite{kang2020approximate} to provide theoretical guarantees. 

\citet{kang2020approximate} demonstrated that model cascades approaches that lack theoretical guarantees frequently miss quality targets (e.g., achieving precision below 65\% when targeting 90\%). This motivated SUPG~\cite{kang2020approximate}, a method to set cascade thresholds with theoretical guarantees and improved utility over prior work \cite{kang2017noscope, lu2018accelerating}, but as already discussed, SUPG's guarantees only hold asymptotically. Our approach, \name{}, improves upon SUPG by presenting stronger theoretical guarantees, especially for LLM-powered data processing tasks, and significantly better empirical utility confirmed by our experiments. Current LLM-powered data processing frameworks seeking theoretical guarantees~\cite{patel2024lotus} rely on SUPG, underscoring the value of our improved methodology. \name{} can be integrated into any existing LLM-powered data processing frameworks to substantially reduce costs while maintaining quality guarantees.

%% file: conclusion.tex
\section{Conclusion}\label{sec:conc}
We studied the problem of low-cost LLM-powered data processing through model cascade while providing quality guarantees. We studied accuracy, precision, and recall quality guarantees, and presented \name{} to decide the model cascade threshold while providing tight theoretical guarantees and good utility. \name{} uses adaptive sampling to label records, hypothesis testing to estimate whether different cascade thresholds meet the target, and choose the cascade threshold based on the estimates with a tight theoretical analysis. We empirically showed \name{} significantly improves utility over the state-of-the-art. \rtwo{We plan to extend \name{} to open-ended tasks which requires further consideration on how to calculate proxy scores, as well as \textit{semantic join} \cite{patel2024lotus} and entity matching operations. The latter cases can be formulated as a filter on the cross product of two datasets. \name{} can be applied as is, but additional optimizations are possible by considering properties the operations (e.g., transitivity in entity matching).}

%% file: appendix/appendix_overview.tex
\section{Overview}
This appendix is organized as follows.
\begin{itemize}
    \item \Cref{sec:statements} presents formal statements and additional details for theoretical results discussed in the paper.
    \item \Cref{sec:proof} present proofs of all results, both results from the main body of the paper and results from \Cref{sec:statements}.  
    \item \Cref{sec:other_extesions} discusses extensions to multi-proxy settings and when considering proxy cost.
    \item \Cref{sec:candidate_set} discusses impact of the candidate threshold set on \name{}.
    \item \Cref{app:taskprompts} discusses details of prompts used in our datasets for evaluation.
    \item \Cref{appx:exp} discusses additional experiments.
\end{itemize}

%% file: appendix/appendix_additional_results.tex
\section{Technical Details and Formal Statements}\label{sec:statements}
This section includes additional theoretical results supporting the discussion in the paper, as well as formal statements of results deferred to here from the main body of the paper. Proofs are in \Cref{sec:proof}. 

\subsection{Mean Estimation through Hypothesis-Testing}
Here, we present simplified versions of the results by \citet{waudby2024estimating}, used to obtain our estimation functions. The results stated below are simplifications of the results by \cite{waudby2024estimating} to statements needed to prove the results in this paper. At a high level, the results here help test whether the mean, $\mu$, of a sequence of Bernoulli random variables is below or above a value $m$ that abstractly captures the estimation problems presented in this paper.  In definitions below, for convenience we let $\sum_1^0 f(x)=0$ for any $f(x)$.

\begin{lemma}[Simplified Version of Theorem 3 by \cite{waudby2024estimating}]\label{lemma:_anytime_replace_original}
    Consider a (potentially infinite) sequence of Bernoulli random variables $X=\langle X_1, X_2,... \rangle$ with mean $\mu$. Let $X[:i]$ be the subsequence of $X$ containing of the first $i$ random variables. For a confidence parameter $\alpha\in[0, 1]$, and any $\mu<m$, we have
    \begin{align}\label{lemma:betting_lb}
        \mathds{P}\big(\exists i\in\mathbb{N}, \mathds{I}\big[\exists j\in[i],\,\mathscr{K}(m, X[:j])\geq\frac{1}{\alpha}\big]=1\big)\leq \alpha
    \end{align}    
     where $\mathscr{K}(m, Y)$, for any sequence $Y=\langle Y_1, ..., Y_k\rangle$ is defined as 
    \begin{align}\label{eq:k}
        \mathscr{K}(m, Y)=\Pi_{j=1}^{i}(1+\min(\lambda_j, \frac{3}{4m})\times(Y_j-m)), 
    \end{align}
    \begin{align*}
        \hspace{-13pt}\lambda_i=\sqrt{\frac{2\log(2/\alpha)}{i\log(i+1)\hat{\sigma}^2_{i-1}}},\; \hat{\sigma}_i^2=\frac{1/4+\sum_{j=1}^i(Y_j-\hat{\mu}_j)^2}{i+1}, \;\hat{\mu}_i = \frac{1/2+\sum_{j=1}^iY_j}{i+1}.
    \end{align*} 
    Furthermore, for any $\mu>m$, 
    \begin{align}\label{lemma:betting_ub}
        \mathds{P}\big(\exists i\in\mathbb{N}, \mathds{I}\big[\exists j\in[i],\,\mathscr{K}^-(m, X[:j])\geq\frac{1}{\alpha}\big]=1\big)\leq \alpha
    \end{align}    
    Where $\mathscr{K}^-(m, Y)$, for any sequence $Y=\langle Y_1, ..., Y_k\rangle$ is defined as 
    \begin{align}\label{eq:k_minus}
        \mathscr{K}^{-}(m, Y)=\Pi_{j=1}^{i}(1-\min(\lambda_j, \frac{3}{4(1-m)})\times(Y_j-m))
    \end{align}    
    with $\lambda_j$ defined as above.
\end{lemma}

Eq.~\ref{lemma:betting_lb}  shows we can use $\mathds{I}[j\in[i],\,\mathscr{K}(m, X[:i])\geq\frac{1}{\alpha}]$ to test whether $m\geq \mu$ after sampling $i$ records, for any $i$, and obtain low false positive probability. That is, $\mathds{I}[j\in[i],\,\mathscr{K}(m, X[:i])\geq\frac{1}{\alpha}]$ is unlikely to be 1 if  $\mu<m$. Note that Eq.~\ref{lemma:betting_lb} shows the hypothesis test is \textit{anytime valid} and can be at any point during sampling. In other words, Eq.~\ref{lemma:betting_lb} allows us to check if $m$ is a suitable lower bound $\mu$ or not while sampling new points and still have the probability of making a wrong estimate bounded by $\alpha$. We use Eq.~\ref{lemma:betting_lb} to provide our estimation function for most of the discussion in the paper. 

Eq.~\ref{lemma:betting_ub} also allows us to check if $m$ is a suitable upper bound for $\mu$. That is, $\mathds{I}[j\in[i],\,\mathscr{K}^-(m, X[:i])\geq\frac{1}{\alpha}]$ is unlikely to be 1 if  $\mu>m$. We use this upper bound only for our density estimation function $\mathscr{E}_d^{\text{\name{}}}$.

We next present another result by \cite{waudby2024estimating} that provides the same guarantees as above but also allows us to perform sampling without replacement.

\begin{lemma}[Simplified Version of Theorem 4 by \cite{waudby2024estimating}]\label{lemma:_anytime_no_replace_original}
    Consider any sequence of $k$ Bernoulli random variables $X=\langle X_1, ..., X_k\rangle$ taken uniformly at random and without replacement from a population of size $N$ with mean $\mu$. Let with $X[:i]$ be the subsequence of $X$ containing of the first $i$ random variables. For a confidence parameter $\alpha\in[0, 1]$, and any $\mu<m$, we have
    \begin{align}\label{lemma:betting_lb_wor}
        \mathds{P}\big(\exists i\in[k], \mathds{I}\big[\exists j\in[i],\,\mathscr{K}_{WR}(m, X[:j])\geq\frac{1}{\alpha}\big]=1\big)\leq \alpha
    \end{align}    
    Where, $\mathscr{K}_{WR}(T, Y)$, for any sequence $Y=\langle Y_1, ..., Y_i\rangle$ is defined as 
    \begin{align}\label{eq:k_wr}
        \mathscr{K}_{WR}(T, Y)=\Pi_{j=1}^{i}(1+\min(\lambda_j, \frac{3}{4T_i^{WR}})\times(Y_j-T_i^{WR})), 
    \end{align}
    \begin{align*}
        T_i^{WR}=\frac{NT-\sum_{j=1}^{i-1}Y_j}{N-(i-1)},\,\lambda_i=\sqrt{\frac{2\log(2/\alpha)}{i\log(i+1)\hat{\sigma}^2_{i-1}}}.
    \end{align*}
    \begin{align*}
         \hat{\sigma}_i^2=\frac{1/4+\sum_{j=1}^i(Y_j-\hat{\mu}_j)^2}{i+1}, \;\hat{\mu}_i = \frac{1/2+\sum_{j=1}^iY_j}{i+1}.
    \end{align*}
\end{lemma}

Eq.~\ref{lemma:betting_lb_wor} is analogous to Eq.~\ref{lemma:betting_lb}, except that it proves that we can obtain the same guarantees while sampling without replacement. 

\subsection{\name{}$_P$-U for PT Queries}

\subsubsection{Variance of Observations}\label{appx:variance}
In Sec.~\ref{sec:pt:unif:est}, we noted that the variance of observed precision decreases as the true precision increases. Note that observed precision is $\mathscr{P}_S(\rho)=\frac{\sum_{x\in S_O^\rho}x}{|S_O^\rho|}$, where every random variable in $x\in S_O^\rho$ is a Bernoulli random variable with mean $\mathscr{P}_D(\rho)$. Thus, we have
\begin{align}\label{eq:observed_prec_var}
\mathtt{Var}(P_S(\rho))=\frac{1}{|S_O^\rho|}\mathfrak{P}_D(\rho)(1-\mathfrak{P}_D(\rho)).
\end{align}
When $\mathfrak{P}_D(\rho)$ is close to 1, the variance in observations will be very small, meaning the observed precision is likely closer to the true precision, compared with when $\mathfrak{P}_D(\rho)=0.5$. 

\subsubsection{Generalization to $\eta>0$}\label{sec:gen:eta} We presented both \name{}$_P$-U and \name{}$_P$-A in Sec.~\ref{sec:pt} with $\eta=0$ in our selection method (i.e., from Lemma~\ref{lemma:select_eta}). Here we note that for both \name{}$_P$-U and \name{}$_P$-A, when we have $\eta>0$ we need to make two algorithmic modifications. First, every application of $\mathscr{E}$ needs to be done with $\frac{\delta}{\eta+1}$ instead of $\delta$. Second, as we iterate through the candidate threshold in decreasing order, we keep a counter $c$, of the number of thresholds estimated to be below the target. After every estimate, we check if $c<\eta+1$ and stop the algorithm as soon as $c=\eta+1$. When $\eta=0$, we stop after the first threshold estimated to miss the target, as shown in Alg.~\ref{alg:iterative_select_precision}. We note that for \name{}$_P$-A and if $\eta>0$, we additionally deploy a strategy similar to Alg.~\ref{alg:iterative_sample_accurcay}, to estimate that a threshold does not meet the target if the target is within a standard deviation of the mean of observations. 

\subsection{\name{}$_P$-A for PT Queries}

\subsubsection{Anytime Valid Estimates while Sampling without Replacement}\label{sample:wor}
Here, we present the following lemma that generalizes Lemma~\ref{lemma:lb} in two ways: (1) it extends it to anytime valid estimation, i.e., the probability of returning a false positive is bounded at any time during sampling, and (2) uses sampling without replacement. 

\begin{lemma}[Corollary to Theorem 4 by \cite{waudby2024estimating}]\label{lemma:_anytime_no_replace_lb}
    For any $\rho\in[0, 1]$ with $\mathfrak{P}_D(\rho)<T$, let $X_1$, $X_2$,... $X_k$ be random samples from $D^\rho$ without replacement, and denote $N=|D^\rho|$. Let $S_i=\langle X_1, ..., X_i\rangle$ and  $S^\rho_O[:i]=\langle \mathds{I}[\mathcal{O}(X_1)=1], ..., \mathds{I}[\mathcal{O}(X_i)=1]\rangle$, i.e., sequence of Bernoulli random variables each denoting whether the $i$-th sample is a positive. Then, for a confidence parameter $\alpha\in[0, 1]$, 
    \begin{align}\label{eq:est_prism_guarantee_bound}
        \mathds{P}(\exists i\in[k],\,\mathscr{E}^{\textnormal{\name{}}}_{WR}(S_i, T, \rho, \alpha)=1)\leq \alpha, \quad \text{where}
    \end{align}    
    \begin{align}
        \mathscr{E}^{\textnormal{\name{}}}_{WR}(S, T, \rho, \alpha)=\mathds{I}[\exists i\in[|S|]\,\text{s.t.}\,\mathscr{K}_{WR}(T, S_O^\rho[:i])\geq\frac{1}{\alpha}], 
    \end{align}
    with $\mathscr{K}_{WR}$ as define in Eq.~\ref{eq:k_wr}
\end{lemma}

Note that Eq.~\ref{eq:est_prism_guarantee_bound} shows the stronger result that the estimate $\mathscr{E}^{\textnormal{\name{}}}_{WR}(S_i, T, \rho, \alpha)$ has low false positive probability throughout sampling and not just after taking a fixed number of samples. Furthermore, Lemma~\ref{lemma:_anytime_no_replace_lb} uses sampling without replacement, by defining the function $\mathscr{K}_{WR}(T, X)$ slightly differently than $\mathscr{K}(T, X)$ in Lemma~\ref{lemma:lb}.

In \name{}$_P$-A, we use $\mathscr{E}^{\textnormal{\name{}}}_{WR}$ instead of $\mathscr{E}^{\textnormal{\name{}}}$ and sample without replacement.

\subsubsection{Sampling Procedure}\label{sec:sampling_proc}
Recall that in \name{}$_P$-A, at each candidate threshold $\rho$ considered, we iteratively sample records in Line~\ref{alg:prism_pA:sample} of Alg.~\ref{alg:iterative_sample_precision}. Here, we provide further details of this sampling procedure. As discussed in \Cref{sample:wor}, we take these samples without replacement, which as Lemma~\ref{lemma:_anytime_no_replace_lb} showed provides the same guarantees as Lemma~\ref{lemma:lb}. We also sample records in a way that allows us to reuse samples across the thresholds, discussed next.

We use the following sampling procedure. Let $\mathring{D}$ be the sequence of records in $D$ in a random order. To sample records at a threshold $\rho$, consider the sequence $\mathring{D}^\rho$, defined as the subsequence of $\mathring{D}$ containing only records with a proxy score more than $\rho$. Let $c_\rho$ be a counter associated for each threshold, $\rho$, initially $c_\rho=0$ for all $\rho$. To sample a record when considering the threshold $\rho$, we take the $c_\rho$-th element from $\mathring{D}^\rho$ and increment $c_\rho$. This approach iterates over records in $D^\rho$---records in $D$ with a score more than $\rho$---in a random order and thus samples from $D^\rho$ without replacement. 

The above procedure also ensures that the samples are reused as the candidate threshold decreases. Note that for two thresholds $\rho,\rho'$ with $\rho'>\rho$, $\mathring{D}^{\rho'}$ is a subsequence of $\mathring{D}^\rho$ containing all the samples with score larger than $\rho'$, while $\mathring{D}^\rho$ additionally contains samples with score in $[\rho, \rho']$. As such, this procedure helps ensure the samples taken for larger candidate thresholds are reused for smaller thresholds as well. We show our theoretical guarantees still hold when performing such sample reuse. 

\subsubsection{Final Guarantees}
Let $\rho^{\text{\name{}}_P-A}_S$ be the output from \name{}$_P$-A when using the sampling procedure described in Sec.~\ref{sec:sampling_proc} and using the estimation function $\mathscr{E}_{WR}^{\text{\name{}}}$ instead of  $\mathscr{E}^{\text{\name{}}}$. We have:
\begin{lemma}\label{lemma:final_guarantees:prism_pa}
   $\rho^{\text{\name{}}_P-A}_S$ provides our desired theoretical guarantees, that is, $\mathds{P}_{S\sim D}(\mathfrak{P}_D(\rho_S^{\text{\name{}$_P$-A}})<T)\leq \delta.$ 
\end{lemma}

\subsection{\name{}$_A$-A and \name{}$_A$-M  for AT Queries}
\subsubsection{Estimation function}
We note that for \name{}$_A$-A and \name{}$_A$-M we also perform sampling without replacement. The following lemma shows the estimation function used and the resulting theoretical guarantees.

\begin{lemma}[Corollary to Theorem 4 by \cite{waudby2024estimating}]\label{lemma:_anytime_no_replace_lb_accuracy}
    For any $\rho\in[0, 1]$ with $\mathfrak{A}_D(\rho)<T$, let $X_1$, $X_2$,... $X_k$ be random samples from $D^\rho$ without replacement, and denote $N=|D^\rho|$. Let $S_i=\langle X_1, ..., X_i\rangle$ and  $S^\rho_A[:i]=\langle \mathds{I}[\mathcal{O}(X_1)=\mathcal{P}(X_1)], ..., \mathds{I}[\mathcal{O}(X_i)=\mathcal{P}(X_i)]\rangle$, i.e., sequence of Bernoulli random variables each denoting whether the $i$-th sample is correctly answered by the proxy. Then, for a confidence parameter $\alpha\in[0, 1]$, 
    \begin{align}
        \mathds{P}(\exists i\in[k],\,\mathscr{E}^{\textnormal{\name{}}}_{A}(S_i, T, \rho, \alpha)=1)\leq \alpha, \quad \text{where}
    \end{align}    
    \begin{align}
        \mathscr{E}^{\textnormal{\name{}}}_{A}(S, T, \rho, \alpha)=\mathds{I}[\exists i\in[|S|]\,\text{s.t.}\,\mathscr{K}_{WR}(T, S_A^\rho[:i])\geq\frac{1}{\alpha}].
    \end{align}
    $\mathscr{K}_{WR}(T, Y)$ is defined as in Eq.~\ref{eq:k_wr}.
\end{lemma}

\subsubsection{\name{}$_A$-M details}\label{sec:prism_a_M}
We present an extension to \name{}$_A$-A for classification tasks to \textit{set the cascade thresholds per class}. This improves \name{}$_A$-A whenever model inaccuracies differ for different classes. 
We find class-specific cascade thresholds, $\rho_i$ for all $i\in[r]$. Then, when $\mathcal{P}(x)=i$ for some $x\in D$ and $i\in [r]$, we use the cascade thresholds, $\rho_i$, to decide whether to use oracle or the proxy. 
We find such class-specific cascade thresholds by repeatedly running \name{}$_A$-A for different classes. Define $D_{i}=\{x;x\in D,\; \mathcal{P}(x)=i\}$, i.e., subset of $D$ predicted to belong to class $i$ by the proxy model. We run Alg.~\ref{alg:iterative_sample_accurcay} for each of $D_{i}$ to obtain the cascade thresholds, $\rho_{i}$. 
To ensure the same theoretical guarantees as before (see \Cref{lemma:prism_AM}), we use union bound to bound the probability of failure across $r$ classes so that each run of \name{}$_A$-A uses confidence parameter $\frac{\delta}{r}$.

\subsubsection{Taking Oracle Labels into Account}\label{sec:acc:oracle}
For a threshold $\rho$, since the records with proxy score less than $\rho$, i.e., $D\setminus D^\rho$, are processing with the oracle, we still can guarantee output accuracy $T$ if proxy accuracy is below $T$ on $D^\rho$. For any threshold $\rho$, let $N_\rho=|D^\rho|$, so that $N_\rho$ is the number of data points labeled by the proxy (ignoring points already labeled in $S$) and $N-N_\rho$ is the number of data points labeled by the oracle. If proxy achieves accuracy at least $T_\rho=\frac{ N_\rho-N(1-T)}{N_\rho}$ at a threshold $\rho$, the overall accuracy will be at least $T$. Thus, for \name{}$_A$-A, in Alg.~\ref{alg:iterative_sample_accurcay} we replace $T$ with $T_\rho$ for any candidate threshold $\rho$ considered. We do so similarly for \name{}$_A$-M, but every use of Alg.~\ref{alg:iterative_sample_accurcay} is on the subset $D_i$ of $D$, and $N_\rho$ and $N$ are computed for each $D_i$. 

\subsubsection{Guarantees for \name{}$_A$-A and \name{}$_A$-M}
Let $\rho^{\text{\name{}}_A-A}_S$ be the output from \name{}$_A$-A when using the sampling procedure described in Sec.~\ref{sec:sampling_proc} and the estimation function $\mathscr{E}_{A}^{\text{\name{}}}$ instead of  $\mathscr{E}^{\text{\name{}}}$. We have:

\begin{lemma}\label{lemma:prism_AA}
   $\rho^{\text{\name{}}_A-A}_S$ provides our desired theoretical guarantees, that is, $\mathds{P}_{S\sim D}(\mathfrak{A}_D(\rho_S^{\text{\name{}$_A$-A}})<T)\leq \delta.$ 
\end{lemma}

Building on the above lemma, we show that $\rho^{\text{\name{}}_A-M}_S$, the output of \name{}$_A$-M when running \name{}$_A$-A in a classification task with $r$ different classes also meets the accuracy constraints. 

\begin{lemma}\label{lemma:prism_AM}
   $\rho^{\text{\name{}}_A-M}_S$ provides our desired theoretical guarantees, that is, $\mathds{P}_{S\sim D}(\mathfrak{A}_D(\rho_S^{\text{\name{}$_A$-M}})<T)\leq \delta.$ 
\end{lemma}

\subsection{\name{}$_R$-U for RT Queries}
\subsubsection{Estimation function.} Note that for RT queries, we do not know the total number of positive labels in the entire dataset. Thus, we do not use sampling without replacement for RT queries. The following lemma shows the theoretical guarantees for $\mathscr{E}_R^{\text{\name{}}}$. 

\begin{lemma}[Corollary to Theorem 3 by \cite{waudby2024estimating}]\label{lemma:estimation_lb_r}
    For any $\rho\in[0, 1]$ with $\mathfrak{R}_D(\rho)<T$, let $X_1$, $X_2$,... $X_k$ be random samples from $D_+$ with replacement. Let $S_i=\langle X_1, ..., X_i\rangle$ and  $S^\rho_R[:i]=\langle \mathds{I}[\mathcal{S}(X_1)\geq\rho], ..., \mathds{I}[\mathcal{S}(X_i)\geq\rho]\rangle$, i.e., sequence of Bernoulli random variables each denoting whether the $i$-th positive sample has proxy score above the threshold. Then, for a confidence parameter $\alpha\in[0, 1]$, 
    \begin{align}
        \mathds{P}(\exists i\in[k],\,\mathscr{E}^{\textnormal{\name{}}}_{R}(S_i, T, \rho, \alpha)=1)\leq \alpha, \quad \text{where}
    \end{align}    
    \begin{align*}
        \mathscr{E}^{\textnormal{\name{}}}_{R}(S, T, \rho, \alpha)=\mathds{I}[\exists i\in[|S|]\,\text{s.t.}\,\mathscr{K}(T, S_R^\rho[:i])\geq\frac{1}{\alpha}].
    \end{align*}
    $\mathscr{K}(T, Y)$ is defined as in Eq.~\ref{eq:k}.
\end{lemma}

\subsubsection{Final Guarantees} The threshold chosen by \name{}$_R$-U as in Eq.~\ref{eq:rho_prism_pu} provides the required theoretical guarantees.

\begin{lemma}\label{lemma:prism_RU}
   Let $\rho^{\text{\name{}}_R-U}_S$ be the threshold selected by \name{}$_R$-U. We have $\mathds{P}_{S\sim D}(\mathfrak{R}_D(\rho_S^{\text{\name{}$_R$-U}})<T)\leq \delta.$ 
\end{lemma}

\subsection{\name{}$_R$-A for RT Queries}
\subsubsection{Estimation Function for Density}\label{sec:density_est}
Here, we describe $\mathscr{E}_d^{\text{\name{}}}$ for estimating whether a threshold has density higher than $\beta$, $d_r(\rho)\geq\beta$ or not.

\begin{lemma}[Corollary to Theorem 4 by \cite{waudby2024estimating}]\label{lemma:ub_density}
    For any $\rho\in[0, 1]$ and given the resolution parameter, $r$, with $\mathfrak{P}(D_r^{\rho})>\beta$, let $X_1$, $X_2$,... $X_k$ be random samples from $D_r^{\rho}$ without replacement, and denote $N=|D^{\rho, r}|$. Let $S_i=\langle X_1, ..., X_i\rangle$ and  $S^\rho_O[:i]=\langle \mathds{I}[\mathcal{O}(X_1)=1], ..., \mathds{I}[\mathcal{O}(X_i)=1]\rangle$, i.e., sequence of Bernoulli random variables each denoting whether the $i$-th sample is a positive. Then, for a confidence parameter $\alpha\in[0, 1]$, 
    \begin{align*}
        \mathds{P}(\mathscr{E}_d^{\textnormal{\name{}}}(S, \beta, \rho, \alpha)=1)\leq \alpha, \quad \text{where}
    \end{align*}  
    \begin{align*}
        \mathscr{E}_d^{\textnormal{\name{}}}(S, \beta, \rho, \alpha)=\mathds{I}[\exists i\in[k_\rho]\,\text{s.t.}\,\mathscr{K}^-(\beta, S_O^\rho[:i])\geq\frac{1}{\alpha}].
    \end{align*}
    Where $\mathscr{K}^-(T, X)$ is defined as in Eq.~\ref{lemma:betting_ub}.
\end{lemma}

\subsubsection{Lower Bound on Utility}\label{sec:utility_lb}
The lemma below shows that any algorithm that guarantees the target recall is met with probability $\delta$ and samples records with probability monotonically increasing in proxy score will have a precision upper bound based on the number of true positives in the dataset. 
We note that the lemma can be stated more generally to include other sampling classes as well---depending on the characteristic of the sampling class. We state only this special case for clarity. 

\begin{lemma}\label{lemma:recall_rho_must_be_zero_special}
Consider any algorithm that samples a set of $k$ points, $S$, i.i.d. where the probability of a point $x \in D$ is sampled is monotonically increasing in $\mathcal{S}(x)$. If the algorithm returns a cascade threshold, $\rho_{S}$, that meets the recall target $T$, $T\geq 0.5$, on all datasets of size $n$ and with $n^+$ positives, then, for any dataset $D$, it must have precision 
\begin{align}\label{eq:lower_bound}
\mathds{P}(\mathfrak{P}_D(\rho_{S})\leq \frac{n^+}{n})\geq (1-\frac{n^+}{n})^k-\delta. 
\end{align}
\end{lemma}

\if 0\begin{lemma}\label{lemma:recall_rho_must_be_zero_special}
Consider an algorithm, $A$, that samples a set of $k$ points, $S$, i.i.d. and with a p.m.f $\mathscr{W}(x)$ for $x\in D$, and returns a cascade threshold $\rho_{S}$. Assume algorithm $A$ meets the recall target $T$, $T\geq 0.5$, on all possible datasets of size $n$ and with $n^+$ positives. Then, for any dataset $D$, algorithm $A$ must have precision $$\mathds{P}(\mathfrak{P}_D(\rho_{S})\geq \frac{n^+}{n})\leq (1-\alpha)^k-\delta,$$
where $\alpha=\sum_{x\in D^{(1-T)n^+}}\mathscr{W}(x)$.
\end{lemma}\fi


Lemma~\ref{lemma:recall_rho_must_be_zero_special} is insightful when $n^+$ is much smaller than $n$, that is, when there are few true positives compared with total data size so that $\frac{n^+}{n}$ is small. In such cases, it shows that any algorithm returns very low precision with high probability, yielding low utility. 


\subsubsection{\name{}$_R$-A Guarantees}
The following shows that \name{} meets the required theoretical guarantees on datasets with dense positive labels. 

\begin{lemma}\label{lemma:prism_r_i}
    Let $\rho_S$ be the cascade threshold found by Alg.~\ref{alg:prism_r_i}. For any dataset, $D$, with dense positive labels, $\mathds{P}(\mathfrak{R}_D(\rho_S)<T)\leq \delta.$
\end{lemma}


\input{appendix/chernoff}

%% file: appendix/chernoff.tex
\subsection{\rone{Using Chernoff Bound}}\label{appx:chernoff}
\renewcommand{\tabchernoff}{tab:hoef_vs_chernoff}
\renewcommand{\chernoffall}{tab:avg_results}

\rone{\chernoffappx{}}

\begin{table}[t]
\hspace{-0.7cm}
\begin{minipage}{0.51\columnwidth}
    \rone{\centering
    \begin{tabular}{c c c}
    \toprule
        \textbf{Metric} & \textbf{Chern.} & \textbf{Hoeff.}\\\midrule
        AT@0.7 & 52.6 & 52.8\\
        AT@0.9 & 24.3 & 22.9\\
        \hline
        PT@0.7 & 45.3  & 46.9\\
        PT@0.9 & 22.6 & 22.1\\
        \hline
        RT@0.7 & 34.4 & 34.0\\
        RT@0.9 & 30.6 & 25.7  \\ \bottomrule
    \end{tabular}
    \caption{\rone{Obtained utility using Hoeffding \& Chernoff Bounds}}
    \label{tab:hoef_vs_chernoff}}
\end{minipage}
\hspace{0.1cm}
\begin{minipage}{0.48\columnwidth}
\rone{    \centering
    \begin{tabular}{c c c c}
    \toprule
        \textbf{Method} & \textbf{AT} & \textbf{PT}& \textbf{RT}\\\midrule
        \textbf{SUPG} & 27.8 & 46.7 & 36.5\\
        \textbf{Hoeff.} & 22.9 & 22.1 & 25.7\\
        \textbf{Chern.} & 24.3 & 22.6 & 30.6\\
        \textbf{\name{}$_*$-A} & \textbf{60.5} & \textbf{74.6} & \textbf{44.5}\\\bottomrule
    \end{tabular}
    \caption{\rone{Utility of different methods given quality target $T=0.9$ across queries}}
    \label{tab:avg_results}}
\end{minipage}
\end{table}
\vspace{-0.08cm}

%% file: appendix/appendix_proofs.tex
\section{Proofs}\label{sec:proof}

\subsection{Proofs of Estimation Function Guarantees}
We briefly discuss our lemmas that propose an estimation function using the results of \cite{waudby2024estimating}. \Cref{lemma:lb,lemma:estimation_lb_r,lemma:ub_density} are corollaries to Theorem 3 of \cite{waudby2024estimating}, whose simplified version was stated as Lemma~\ref{lemma:_anytime_replace_original}. \Cref{lemma:_anytime_no_replace_lb,lemma:_anytime_no_replace_lb_accuracy} are corollaries to Theorem 4 of \cite{waudby2024estimating}, whose simplified version was stated as Lemma~\ref{lemma:_anytime_replace_original}. It is easy to see that \Cref{lemma:_anytime_replace_original} proves \Cref{lemma:lb,lemma:estimation_lb_r,lemma:ub_density}, while \Cref{lemma:_anytime_no_replace_original} proves \Cref{lemma:lb,lemma:estimation_lb_r}. 

\subsection{Naive Algorithm for PT Queries}
\textit{Proof of Prop.~\ref{prop:hoef_single}.}  Observe that 
$$\mathds{E}[\mathfrak{P}_S(\rho)]=\sum_{x\in S^\rho}\frac{\mathds{E}[\mathds{I}[\mathcal{O}(x)=1]]}{|S^\rho|}=\mathfrak{P}_D(\rho),$$ so applying Hoeffding's inequality on the set of i.i.d random variables $S^\rho=\{\mathds{I}[\mathcal{O}(x)=1]; x\in S, \mathcal{S}(x)\geq\rho\}$ whose observed mean is $\mathfrak{P}_S(\rho)$ and have true mean $\mathfrak{P}_D(\rho)$, we have 
\begin{align*}
\mathds{P}(\mathfrak{P}_S(\rho)\geq \mathfrak{P}_D(\rho)+\sqrt{\frac{\log(1/\alpha)}{2|S^\rho|}})\leq\alpha.
\end{align*}
When additionally $\mathfrak{P}_D(\rho)<T$, combining $\mathfrak{P}_D(\rho)<T$ with the above proves the result.\qed

\textit{Proof of Prop.~\ref{prop:select_naive}.} Result follows applying union bound to sum the probability that $\mathscr{E}$ returns 1 for each of $|\mathscr{C}|$ applications of $\mathscr{E}$. \qed

\textit{Proof of Lemma.~\ref{lemma:prec_heof_union}.} Follows the application of Prop.~\ref{prop:select_naive} with $\delta=\frac{\alpha}{|\mathscr{C}|}$, using the fact that $\mathscr{E}^{\text{naive}}$ has bounded false positive probability, shown in Prop.~\ref{prop:hoef_single}.\qed

\subsection{\name{} for PT Queries}

\textit{Proof of Lemma.~\ref{lemma:select_eta}.}
Let $\bar{\rho}^1$, ..., $\bar{\rho}^{\eta+1}$ be the $\eta+1$ largest thresholds for which $\mathfrak{P}_D(\bar{\rho}^i)<T$ in decreasing order. Note that if $\mathscr{E}(S, T, \bar{\rho}^i, \alpha)=0$ for all $i\in[\eta+1]$, then $\rho^*$ will be a threshold larger than $\bar{\rho}^{\eta+1}$ and not any of $\bar{\rho}^1$, ..., $\bar{\rho}^{\eta}$ which implies $\mathfrak{P}_D(\rho^*)\geq T$. Thus, $\mathfrak{P}_D(\rho^*)<T$ only if $\mathscr{E}(S, T, \bar{\rho}^i,\alpha)\neq 0$ for some $i\in[\eta+1]$. Taking the union bound across the $\eta+1$ events, the probability that $\mathscr{E}(S, T, \bar{\rho}^i,\alpha)\neq 0$ for some $i\in[\eta+1]$ is at most $(\eta+1)\alpha$.\qed

\textit{Proof of Lemma.~\ref{lemma:precision_unif}.} The proof follows combining Lemma~\ref{lemma:lb} and \Cref{lemma:select_eta}.\qed

\textit{Proof of Lemma.~\ref{lemma:final_guarantees:prism_pa}.} Let $\bar{\rho}\in\mathscr{C}_M$ be the largest candidate threshold for which $\mathfrak{P}_D(\rho)< T$. \name{}$_P$-A fails to meet the target only if it estimates $\bar{\rho}$ to meet the target---it otherwise chooses a threshold larger than $\bar{\rho}$ which meets the target. Thus, we only need to analyze the probability of $\mathscr{E}_{WR}^{\text{\name{}}}(S, \bar{\rho}, T, \delta)=1$, where $\mathscr{E}_{WR}^{\text{\name{}}}$ is defined in  \Cref{lemma:_anytime_no_replace_lb}. Indeed, in the sampling procedure described in \Cref{sec:sampling_proc}, sampling elements of $\mathring{D}^\rho$ in order forms a uniform sample set without replacement from $D^\rho$, so that applying \Cref{lemma:_anytime_no_replace_lb} we have $\mathds{P}(\mathscr{E}_{WR}^{\text{\name{}}}(S, \bar{\rho}, T, \delta)=1)\leq \delta$, proving the desired result.\qed 

\subsection{\name{} for AT Queries}

\textit{Proof of Lemma.~\ref{lemma:prism_AA}.} The proof follows that of Lemma~\ref{lemma:final_guarantees:prism_pa}, except that we now use Lemma~\ref{lemma:_anytime_no_replace_lb_accuracy} instead of Lemma~\ref{lemma:_anytime_no_replace_lb}. Besides the estimation function, the only other difference between \name{}$_A$-A and \name{}$_P$-A is that \name{}$_A$-A has a different stopping condition. However, Lemma~\ref{lemma:_anytime_no_replace_lb_accuracy} shows our estimation is valid at any time during sampling and irrespective of the total number of samples taken, completing the proof.\qed

\textit{Proof of Lemma.~\ref{lemma:prism_AM}.} \name{}$_A$-M applies \name{}$_A$-A $r$ times, each time with probability of failure $\frac{\delta}{r}$. Thus, using union bound, the total probability of failure of \name{}$_A$-M is bounded by $\delta$.

\subsection{\name{} for RT Queries}

\textit{Proof of Lemma.~\ref{lemma:prism_RU}.} Let $\bar{\rho}_D=\min\{\rho; \mathfrak{R}_D(\rho)< T\}$. Observe that $\rho^{\text{\name{}}_R-U}_S$ fails to meet the target only if $\rho^{\text{\name{}}_R-U}_S\geq \bar{\rho}_D$. This is because $\rho^{\text{\name{}}_R-U}_S< \bar{\rho}_D$ implies $\mathfrak{R}_D(\rho^{\text{\name{}}_R-U}_S)\geq T$ by definition of $\bar{\rho}_D$. Thus, it remains to show $\mathds{P}(\rho^{\text{\name{}}_R-U}_S\geq \bar{\rho}_D)\leq \delta$. Note that since recall is monotonically decreasing in $\rho$, this only happens if $\mathscr{E}_R^{\text{\name{}}}$ returns 1 for threshold $\bar{\rho}_D$. By Lemma~\ref{lemma:estimation_lb_r}, this probability is bounded by $\delta$. \qed

\textit{Proof of Lemma.~\ref{lemma:recall_rho_must_be_zero_special}.} 
Recall that $D=\{x_1, ..., x_n\}$, and $x_1$, ..., $x_n$ are in sorted order based on proxy scores. Let $D^{\leq i}=\{x_1, ..., x_i\}\subseteq D$, $i\in[n]$ be the subset of $D$ consisting of records with the $i$ lowest proxy scores $D$. Let $\alpha=(1-\frac{n^+}{n})^k$, and note that when samples are selected i.i.d, with the probability increasing in proxy scores, then the probability that any sample is in the set  $D^{\leq n^+}$ data points is at most $\frac{n^+}{n}$. Thus, $\alpha$ is a lower bound on the probability of not selecting any of the first $n^+$ data points in $k$ samples. 

First, consider the case that $\sum_{x\in D^{\leq n^{+}}}\mathds{I}[\mathcal{O}(x)=1]\geq n^+\times(1-T)$. Any threshold that yields recall at least $T$ is at most $x_{n^+}$. Since $\rho_{S}$ meets the recall target with probability $1-\delta$, then it must have $\rho_{S}\leq x_{n^+}$ every time it satisfies the target, and therefore $\mathds{P}(\rho_{S}\leq x_{n^+})\geq 1-\delta$. Note that precision when $\rho_{S}\leq x_{n^+}$ is at most $\frac{n^+}{n-n^++(1-T)n^+}=\frac{n^+}{n-Tn^+}$, so that $\mathds{P}(\mathfrak{P}_D(\rho_{S})\leq \frac{n^+}{n-Tn^+})\geq 1-\delta$, proving the result for this case. 

Next consider the case where $\sum_{x\in D^{\leq n^+}}\mathds{I}[\mathcal{O}(x)=1]< n^+\times(1-T)$.
Assume for the purpose of contradiction that $$\mathds{P}(\rho_{S}\leq x_{n^+})< \alpha-\delta,$$ that is the algorithm returns a threshold at most $x_{n^+}$ with probability less than $\alpha-\delta$. This, by definition, means that $$\mathds{P}(S\cap D^{\leq x_{n^+}}=\emptyset,\; \rho_{S}\leq x_{n^+})< \alpha-\delta,$$
where $S\cap D^{\leq {n^+}}=\emptyset$ is the event that the algorithm does not sample any point with proxy score less than $x_{n^+}$.

We construct a dataset, $\bar{D}$, on which the algorithm fails with probability more than $\delta$. Consider the dataset $\bar{D}$ with the same proxy score and labels as $D$, except that all records with proxy scores in $[0, x_{n^+}]$ have a positive label. Note that the total number of positives in $\bar{D}$ is at most $2n^+$ with at least $n^+$ of the positives in $[0, x_{n^+}]$. Thus, a cascade threshold more than $x_{n^+}$ leads to a recall at most 0.5. 
Therefore, for $\bar{\rho}_{S}$ defined as the cascade threshold selected on $\bar{D}$ using $S$, we have
\begin{equation}\label{eq:error_if_more_than_t}
\mathds{P}(\mathfrak{R}_{\bar{D}}(\bar{\rho}_{S})<T)\geq\mathds{P}(\bar{\rho}_{S}> x_{n^+})
\end{equation}

Since $\rho_{S}$ is chosen deterministically given the observed samples, $\rho_{S}$ is identically selected for datasets with the same proxy score and the same observed labels. $D$ and $\bar{D}$ are identical except for labels in $[0, x_{n^+}]$ so whenever running the algorithm on $\bar{D}$ but for any $S$ where $S\cap\bar{D}^{\leq n^+}=\emptyset$, we must have $\bar{\rho}_{S}=\rho_{S}$. Thus,
\begin{align*}
    \mathds{P}(\bar{\rho}_{S}> x_{n^+})&\geq \mathds{P}(S\cap\bar{D}^{\leq {n^+}}=\emptyset,\;\bar{\rho}_{S}> x_{n^+})\\
    &=\mathds{P}(S\cap\bar{D}^{\leq {n^+}}=\emptyset)-\mathds{P}(S\cap\bar{D}^{\leq n^+}=\emptyset,\;\bar{\rho}_{S}\leq x_{n^+})\\
    &=\mathds{P}(S\cap D^{\leq {n^+}}=\emptyset)-\mathds{P}(S\cap D^{\leq {n^+}}=\emptyset,\;\rho_{S}\leq x_{n^+})\\
    &>\alpha-(\alpha-\delta)=\delta.
\end{align*}
Thus, combining the above with Eq.~\ref{eq:error_if_more_than_t}, we see that the algorithm fails with probability more than $\delta$ on $\bar{D}$, causing a contradiction. Thus, we must have $\mathds{P}(\rho_{S}\leq x_{n^+})\geq \alpha-\delta$, which as discussed before, implies $\mathds{P}(\mathfrak{P}_D(\rho_{S})\leq \frac{n^+}{n+Tn^+})\geq \alpha-\delta$, proving the result. \qed

\textit{Proof of Lemma.~\ref{lemma:prism_r_i}.} For a dataset $D$ with dense positive labels, let $\rho_P^*$ be the proxy score of a positive record in $D$ with the smallest proxy score. Note that if the first stage of \name{}$_P$-A finds a $\rho_P$ smaller than $\rho_P^*$, i.e., $\rho_P<\rho_P^*$, and if \name{}$_R$-U finds a threshold that meets the target on the set $\rho_P$, then \name{}$_P$-A meets the target. The probability that \name{}$_R$-U finds a threshold that doesn't meet the target is $\delta_2$, we next show that the probability that \name{}$_P$-A finds a $\rho_P$ larger than $\rho_P^*$ is bounded by $\delta_2$ so using union bound, the total probability of failure by \name{}$_P$-A is $\delta_1+\delta_2<\delta$. To show $\mathds{P}(\rho_P>\rho_P^*)\leq \delta_1$, note that $\rho_P>\rho_P^*$ happens only if for the first considered threshold larger than $\rho_P^*$, we have $\mathscr{E}_d^{\text{\name{}}}$ returns 1. Applying Lemma~\ref{lemma:ub_density} shows this probability is bounded by $\delta_1$.\qed

\subsubsection{\rone{Chernoff Bound}}\label{appx:chernoff_prof}

\rone{We use the Chernoff bound from \cite{mitzenmacher2017probability}, which shows, given i.i.d Bernoulli random variables, $X_1$, ..., $X_k$, with true mean $\mu$, we have
\begin{align}\label{eq:chernoff}
    \mathds{P}(\frac{1}{k}\sum_i X_i\geq \mu+\sqrt{\frac{2(1-\mu)\log(1/\alpha)}{k}})\leq \alpha,
\end{align}    
Where the bound above is obtained from applying the Chernoff's bound on the random variables $1-X_i$ (which gives a tighter bound when $\mu>0.5$).}

\rone{
Given additionally $\mu<T$, but since we do not know $\mu$, we need to find $$\max_{\mu\in[0, T)}\mu+\sqrt{\frac{2(1-\mu)\log(1/\alpha)}{k}}.$$
Let $f(\mu)=\mu+\sqrt{\frac{2(1-\mu)\log(1/\alpha)}{k}}$. We have $\derivative{\mu}{f}=1-\frac{1}{2}\sqrt{\frac{2\log(1/\alpha)}{k(1-\mu)}}$, and setting $\derivative{\mu}{f}=0$ we obtain 
$$
\mu^*=1-\frac{\log(1/\alpha)}{2k}
$$
Furthermore, we have $\derivative[2]{\mu}{f}<0$, which means $f$ is increasing and peaking at $\mu^*$ and decreasing after. Thus,
$$
\max_{\mu \in [0, T)} f(\mu) =
\begin{cases}
f(0), & \text{if } \mu^* \leq 0\iff k\leq\frac{\log(1/\alpha)}{2}, \\
f(\mu^*), & \text{if } 0 < \mu^* \leq T \iff 0 < 1-\frac{\log(1/\alpha)}{2k} \leq T , \\
f(T), & \text{if } \mu^* > T \iff T<1-\frac{\log(1/\alpha)}{2k}.
\end{cases}
$$
Now observe when $T\geq 1-\frac{\log(1/\alpha)}{2k}$ (i.e., not the last branch) we have $f(T)>1$. In such a case, $\mathds{P}(\frac{1}{k}\sum_i X_i\geq f(T))=0$ trivially, since the range of $X_1$, ...,  $X_k$ is $\{0, 1\}$. Combining this with the Chernoff bound above, we have for any $T$ and whenever $\mu<T$
\begin{align}\label{eq:chernoff_max}
    \mathds{P}(\frac{1}{k}\sum_i X_i\geq f(T))\leq \alpha,
\end{align}   
}

\rone{We then use Eq.~\ref{eq:chernoff_max} to define our estimation function
\begin{align}\label{eq:chernoff_special_max}
    \mathds{P}(\frac{1}{k}\sum_i X_i\geq T+\sqrt{\frac{2(1-T)\log(1/\alpha)}{k}})\leq \alpha.
\end{align}    
To conclude, the above argument shows that for the estimation function 
\begin{align*}
\hspace*{-10pt}
\mathscr{E}^{\text{Chernoff}}(S, T, \rho, \alpha)=\mathds{I}\big[\mathfrak{P}_S(\rho)\geq T+\sqrt{\frac{2(1-T)\log(1/\alpha)}{|S^\rho|}}\big], 
\end{align*}
and for any $\mu<T$ we have 
\begin{align*}
\hspace*{-10pt}
\mathds{P}(\mathscr{E}^{\text{Chernoff}}(S, T, \rho, \alpha)=1)\leq \alpha,
\end{align*}
the same guarantee as for $\mathscr{E}^{\text{naive}}$ that uses Hoeffding's inequality in Prop.~\ref{prop:hoef_single}.}

%% file: appendix/multi_proxy.tex
\section{Multi-Proxy Settings}\label{sec:other_extesions}
So far, we have discussed the setting where a single proxy model is used for inference. In such a setting, our approach helps determine whether to use the output of the cheap proxy model for a data point or whether to use the expensive oracle. As discussed in Sec.~\ref{sec:background}, this setting models many real-world scenarios where many service providers (e.g., OpenAI or Anthropic) offer a cheap model (e.g., GPT4o-mini, Claude Haiku) and an expensive model (e.g., GPT4o, Claude Sonnet), and \name{} helps decide which of the two models to use when processing a collection of documents. 

Additionally, the users may be interested in settings where a cascade system needs to consider more than two models to choose from, e.g., if the user wants to consider all four of  GPT4o-mini, GPT4o, Claude Haiku or Claude Sonnet and decide which one to use to process each data with, where different models have different costs. In such cases, a model cascade system needs to decide which proxy models to use and in what order, a problem known as \textit{proxy routing} \cite{chen2023frugalgpt}. Then, given a routing method, for each proxy model, a cascade threshold needs to be determined to decide whether to use a proxy model output when a data point is routed to the proxy model. For any proxy routing algorithm, \name{} can be used to find a cascade threshold for each model. Thus, \name{} can be combined with existing routing methods to perform high utility threshold selection and provide theoretical guarantees when a proxy model is used. 

%% file: appendix/candidate_set.tex
\section{Impact of Candidate Set}\label{sec:candidate_set} 
\rone{In Sec.~\ref{sec:hyperparam} we discussed the impact of $M$ on \name{}$_A$, showing that if we set $M=M'$ for some value $M'$,  increasing $M'$ to any larger value increases utility by at most $\frac{1}{M'}$, provided that accuracy monotonically decreases as proxy score decreases. A similar statement also holds for precision. This is because setting $M>M'$, \name{}$_P$ at most estimates $\frac{n}{M'}$ new records are estimated to be positive. E.g., if $M'=20$ increasing $M'$ can at most lead to an additional 5\% of the data being labeled positive. However, for PT queries, and in cases when the number of true positives is small, a small portion of the data can significantly affect the recall. Thus, in cases when the total number of true positives is small compared with data size, one can set $M$ to larger values. We see this in our results in Appx.~\ref{appx:exp}}

Here, we additionally note that, other candidate threshold sets are possible. For example, we can use exponentially spaced thresholds  
$$
\mathscr{C}_M=\{\mathcal{S}(x_{i}); i=\lfloor\frac{2^j}{M}n\rfloor , j\in[M]\}.
$$
Such non-uniform selection can be useful because precision, $\mathfrak{P}_D(\rho)$ depends cumulatively on, $D^\rho$, the records with score more than $\rho$. For smaller $\rho$, $|D^\rho|$ is large so small change in $\rho$ is likely to only marginally impact $\mathfrak{P}_D(\rho)$. While when $\rho$ is large $|D^\rho|$ is small and very small changes to $\rho$ can significantly impact $\mathfrak{P}_D(\rho)$. Thus, considering non-uniformly spaced candidate thresholds can be beneficial in practice. Furthermore, we can even modify the candidate threshold set considered as we sample more records. For instance, if the thresholds considered so far have quality much higher than the target, we can decide to skip some thresholds. We leave an in depth study of how the candidate threshold set should be designed to the future work.

%% file: appendix/taskprompts.tex
\section{LLM Task Prompts}
\label{app:taskprompts}

This section details the prompts for the tasks described in \Cref{sec:exp:setup}. The data is inserted at the \texttt{{text}} placeholder in each prompt. We prompt the LLM (gpt-4o-mini or gpt-4o) with temperature 0.

\subsection*{Game Review Classification}
\begin{spverbatim}
I will give you a review for a game.

Your task is to determine if this review references any other games in a more positive way than the game itself.

- True if the review mentions other games in a more positive light than the current game
- False if the review doesn't mention other games or doesn't reference them more positively

Here is the review: {text}

You must respond with ONLY True or False:
\end{spverbatim}

\subsection*{Court Opinion Classification}
\begin{spverbatim}
I will give you a Supreme Court opinion.

Your task is to determine if this opinion reverses a lower court's ruling.
Note that the opinion may not be an appeal, but rather a new ruling.

- True if the Supreme Court reverses the lower court ruling
- False otherwise

Here is the opinion: {text}

You must respond with ONLY True or False:
\end{spverbatim}

\subsection*{Screenplay Classification}
\begin{spverbatim}
I will give you a screenplay of a movie.

Your task is to determine if the protagonist makes a critical decision based on false information.

- True if the protagonist makes an important decision based on information that is incorrect or misleading
- False if the protagonist's key decisions are based on accurate information or no major decisions are made based on false information

Here is the screenplay: {text}

You must respond with ONLY True or False:
\end{spverbatim}

\subsection*{Wikipedia Talk Page Classification}
\begin{spverbatim}
I will give you a Wikipedia Talk page discussion.

Your task is to determine if this discussion led to a reversion (rollback of edits) rather than a stable change.

- True if the discussion resulted in reverting/rolling back changes
- False if the discussion led to stable changes or no changes

Here is the discussion: {text}

You must respond with ONLY True or False:
\end{spverbatim}

%% file: appendix/additional_exp.tex
\begin{figure}[t]
    \centering
    \includegraphics[width=\linewidth]{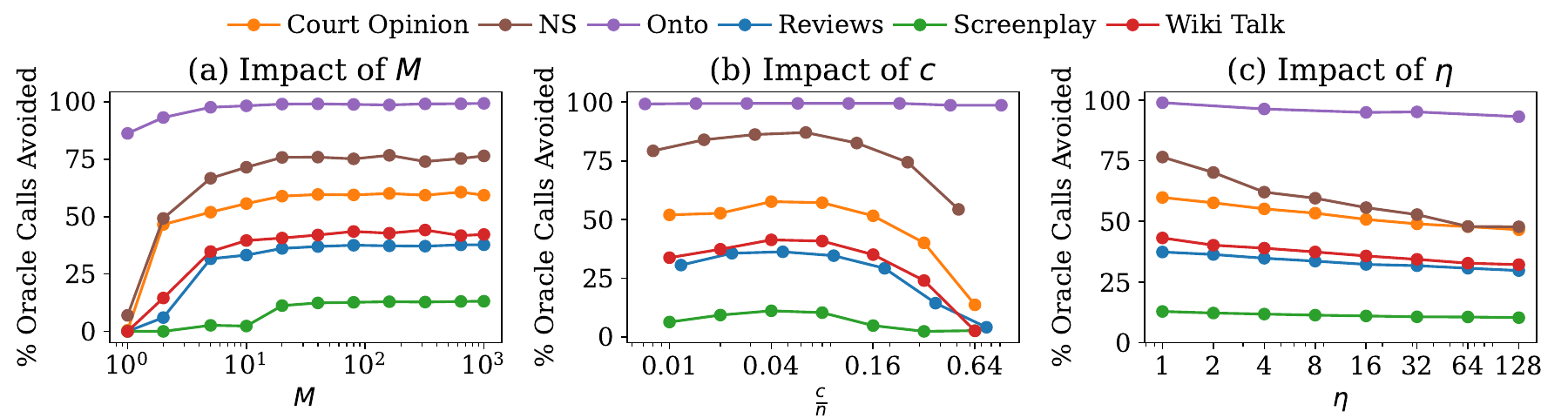}
    \caption{\rone{Impact of $\eta$ and $c$ on AT Queries}}
    \label{fig:hyperparam_AT}
\end{figure}

\section{\rone{Additional Experiments}}\label{appx:exp}
\renewcommand{\variance}{tab:avg_variance}
\renewcommand{\naivevariant}{tab:naive_vs_naive_witheq9}

\subsection{\rone{Alternative Naive approach}}\label{exp:additional_naive}
\rone{\naiveappx}
\begin{table}[t]
\rone{    \centering
    \begin{tabular}{c c}
        \toprule\textbf{Method} & \textbf{Observed Recall} \\\midrule
         Naive & 22.098 \\
         Naive with Eq.~\ref{eq:select_eta} &  22.149\\
        \bottomrule
    \end{tabular}
    \caption{\rone{PT Query with $T=0.9$}}
    \label{tab:naive_vs_naive_witheq9}}
\end{table}

\rone{\subsection{\rone{Variance of methods}}\label{appx:variance_exp}
\naivevariance{}}

\begin{table}[t]
\rone{    \centering
    \begin{tabular}{c c c c}
    \toprule
        \textbf{Method} & \textbf{AT} & \textbf{PT}& \textbf{RT}\\\midrule
        \textbf{SUPG}  & 4.7 & 3.3 & 13.7\\
        \textbf{Naive}  & 1.9 & 1.6& 0.7\\
        \textbf{\name{}$_*$-A}  & 3.7 & 2.1& 8.0\\\bottomrule
    \end{tabular}
    \caption{\rone{Standard deviation of utility of different methods given target $T=0.9$ averaged across datasets}}
    \label{tab:avg_variance}}
\end{table}

\rone{\begin{algorithm}[t]
\small
\begin{algorithmic}[1]
\State Sort $\mathscr{C}_M$ in descending order 
\State $r\leftarrow 0$ 
\State $\rho^*\leftarrow 1$
\For{$i$ \textbf{in} $|\mathscr{C}_M|$}
    \State $\rho\leftarrow \mathscr{C}_M[i]$
    \State $S\leftarrow \emptyset$ 
    \While{$\mathscr{E}_A^{\text{\name{}}}(S, T, \rho, \delta)=0$}\label{alg:precision_sample:prec_est}
        \If{$\texttt{avg}(S_A^\rho)-\texttt{std}(S_A^\rho)\geq T$ \textbf{and} $|S_A^\rho|\geq c$ }\label{alg:iterative_sample_accurcay:check_to_sample}
            \If{$r=\eta$}
                \State \Return $\rho^*$
            \Else
                \State $r\leftarrow r+1$
            \EndIf
        \EndIf
        \State Sample a record uniformly from $D^\rho$ and add to $S$
    \EndWhile
    \State Set $\rho^*\leftarrow\rho$ if $\mathscr{E}_A^{\text{\name{}}}(S, T, \rho, \delta)=1$
\EndFor
\State\Return $\mathscr{C}_M[M]$
\caption{\name{}$_A$-A}\label{alg:iterative_sample_accurcay_w_etha}
\end{algorithmic}
\end{algorithm}}

\subsection{\rone{System Parameters}}\label{appx:hyperparams}
\renewcommand{\fighyperparam}{fig:hyperparam_AT}
\rone{\systemparamsexp{}}

\begin{figure}
    \centering
    \includegraphics[width=0.4\linewidth]{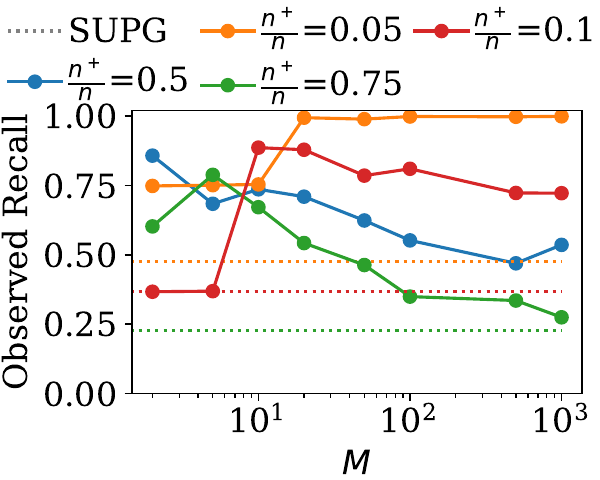}
    \caption{Impact of $M$}
    \label{fig:m_exp}
\end{figure}

\textbf{Varying $M$ on Synthetic Dataset}. We next empirically study the impact of the system parameter $M$ on \name{}$_P$-A. To isolate the impact of $M$, and show its interplay with the distribution of labels and proxy scores, we perform this experiment on synthetic datasets. Note that, for PT queries, our method only depend on proxy scores and oracle labels. We synthetically generate proxy scores and oracle labels for 4 different datasets each with 10,000 records but with a different number of positive labels, $\frac{n^+}{n}\in\{0.05, 0.1, 0.5, 0.75\}$. For all datasets, we generate uniformly distributed proxy scores in [0, 1]. To set the oracle labels for each dataset based on its given number of positive labels, we iterate over the records in descending order of proxy score and label each record as a positive with probability 0.95 until we reach the desired number of positive records. 
We run \name{}$_P$-A on these 4 datasets varying the parameter $M$ between 2 to 1000. The result is shown in Fig.~\ref{fig:m_exp}, which also includes the recall of SUPG as a reference.

Fig.~\ref{fig:m_exp} (b) shows that the recall achieved by \name{}$_P$-A initially improves but then gradually decreases. To see why, note that recall of \name{}$_P$-A is limited by the recall of the candidate thresholds considered. When $M$ is small, none of the few candidate thresholds considered may have good recall. $\frac{n^+}{n}=0.5$ is an exception where $M=2$ achieves good recall because all the positive labels are concentrated around the top half of the proxy scores, and thus the median proxy score achieves high recall while meeting the target precision. 
%
%
As the number of candidate thresholds increases, the likelihood that some candidate threshold has good recall and meets the precision target increases, leading to improved recall across methods. However, as this number further increases, the recall gets worse because the algorithm has to spend sampling budget to evaluate a large number of thresholds, and thus, may run out of sampling budget before it finds the best threshold. Nonetheless, across choices of $M$, we see that \name{}$_P$-A still outperforms SUPG. 